\documentclass[twoside,a4paper]{article}

\usepackage[a4paper,inner=2.5cm,outer=2.5cm,top=2.5cm,bottom=2.5cm]{geometry}
\usepackage[english]{babel}
\usepackage{amsmath,amsthm,amstext,amssymb}
\usepackage{pifont}
\usepackage{array}
\usepackage{upgreek}
\usepackage[normalem]{ulem}
\usepackage{enumitem}
\usepackage{pgf}
\newtheorem{remark}{Remark}
\usepackage{mathrsfs}
\usepackage{siunitx}
\usepackage{inconsolata}
\usepackage{multirow}
\usepackage{scrextend}
\usepackage{mathtools}
\usepackage{stmaryrd}
\usepackage{tikz}
\usepackage[absolute,overlay]{textpos}
\usepackage{hyperref}
\usepackage{subcaption}
\usepackage{graphicx}
\usepackage{cleveref}
\usepackage[affil-it]{authblk}
\usepackage{fancyhdr}

\usepackage{cite}

\makeatletter
\newcommand{\theauthor}{L. Scholz, Y. Ou, B. Grabowski, F. Fritzen}
\makeatother

\newcommand{\pd}[2]{\displaystyle\frac{\displaystyle\partial #1}{\displaystyle\partial #2}}
\newcommand{\lb}{\left(}
\newcommand{\rb}{\right)}
\newcommand{\sV}{\begin{bmatrix}}
\newcommand{\eV}{\end{bmatrix}}
\newcommand{\sty}[1]{{\boldsymbol{#1}}}
\newcommand{\styy}[1]{{\mathbb{#1}}}
\newcommand{\fa}{\sty{ a}}

\newcommand{\fe}{\sty{ e}}
\newcommand{\ff}{\sty{ f}}
\newcommand{\fg}{\sty{ g}}
\newcommand{\fn}{\sty{ n}}
\newcommand{\fq}{\sty{ q}}
\newcommand{\ft}{\sty{ t}}
\newcommand{\fv}{\sty{ v}}
\newcommand{\fx}{\sty{ x}}
\newcommand{\fA}{\sty{ A}}
\newcommand{\fB}{\sty{ B}}
\newcommand{\fD}{\sty{ D}}
\newcommand{\fE}{\sty{ E}}
\newcommand{\fG}{\sty{ G}}
\newcommand{\fI}{\sty{ I}}
\newcommand{\fQ}{\sty{ Q}}
\newcommand{\ffG}{\styy{ G}}
\newcommand{\ffR}{\styy{ R}}
\newcommand{\fdelta}{\mbox{\boldmath $\delta$}}
\newcommand{\cG}{{\cal G}}
\newcommand{\cL}{{\cal L}}
\newcommand{\WT}[1]{\widetilde{#1}}
\newcommand{\ol}[1]{\overline{#1}}
\newcommand{\ul}[1]{\underline{#1}}
\newcommand{\ulWT}[1]{\ul{\widetilde{#1}}}
\newcommand{\ull}[1]{\ul{\ul{#1}}}
\newcommand{\Argyro}{{Li\textsubscript{6}PS\textsubscript{5}Cl}}
\newcommand{\Kbulk}{{\ull{K}_\Omega}}
\newcommand{\Kebulk}{{\ull{K}^{\rm e}_\Omega}}
\newcommand{\KebulkO}{{\ull{K}^{\rm e,0}_\Omega}}
\newcommand{\Kpara}{{\ull{K}_\Vert}}
\newcommand{\Kepara}{{\ull{K}^{\rm e}_\Vert}}
\newcommand{\KeparaO}{{\ull{K}^{\rm e,0}_\Vert}}
\newcommand{\Kperp}{{\ull{K}_\perp}}
\newcommand{\Keperp}{{\ull{K}^{\rm e}_\perp}}
\newcommand{\KeperpO}{{\ull{K}^{\rm e,0}_\perp}}
\newcommand{\KbulkO}{{\ull{K}_\Omega^0}}
\newcommand{\KparaO}{{\ull{K}_\Vert^0}}
\newcommand{\KperpO}{{\ull{K}_\perp^0}}
\newcommand{\KGB}{\ull{K}_{\rm GB}}
\newcommand{\DGBzero}{\fD_{\rm GB,0}}

\newcommand{\NGB}{\ul{N}_{\rm GB}}
\newcommand{\BGB}{\ull{B}_{\rm GB}}
\newcommand{\BGBperp}{\ull{B}_{\rm GB, \perp}}
\newcommand{\BGBperpRef}{\ul{B}_{\rm GB, \perp}^{\rm ref}}
\newcommand{\BGBpara}{\ull{B}_{\rm GB, \Vert}}
\newcommand{\Pipara}{{\Pi_\Vert}}
\newcommand{\Piperp}{{\Pi_\perp}}
\newcommand{\cO}{\mathcal{O}}
\newcommand{\Gbulk}{{\ffG_\Omega}}
\newcommand{\Gpara}{{\ffG_\Vert}}
\newcommand{\Gperp}{{\ffG_\perp}}
\newcommand{\GbulkO}{{\ffG_\Omega^0}}
\newcommand{\GparaO}{{\ffG_\Vert^0}}
\newcommand{\GperpO}{{\ffG_\perp^0}}
\newcommand{\vfgb}{{f_{\rm GB}}}
\newcommand{\vfbulk}{{f_\Omega}}

\usetikzlibrary{shapes.arrows}
\usetikzlibrary{arrows}
\usetikzlibrary{backgrounds}
\usetikzlibrary{arrows.meta}
\usetikzlibrary{shapes.misc, positioning, arrows, decorations.markings, calc}

\definecolor{rev}{rgb}{0, 0, 0}

\title{\color{rev} A collapsed interface approach to resolve grain boundaries in finite element simulations of polycrystalline diffusion}

\author[a]{Lena Scholz\thanks{\href{mailto:scholz@mib.uni-stuttgart.de}{scholz@mib.uni-stuttgart.de}}}
\author[b,c]{Yongliang Ou\thanks{\href{mailto:yongliang.ou@imw.uni-stuttgart.de}{yongliang.ou@imw.uni-stuttgart.de}}}
\author[b]{Blazej Grabowski\thanks{\href{mailto:blazej.grabowski@imw.uni-stuttgart.de}{blazej.grabowski@imw.uni-stuttgart.de}}}
\author[a]{Felix Fritzen\thanks{\href{mailto:fritzen@mib.uni-stuttgart.de}{fritzen@mib.uni-stuttgart.de} (corresponding author)}}

\affil[a]{Institute of Applied Mechanics, University of Stuttgart, Universitätsstraße 32, 70569 Stuttgart, Germany}
\affil[b]{Institute for Materials Science, University of Stuttgart, Pfaffenwaldring 55, 70569 Stuttgart, Germany}
\affil[c]{Department of Materials Science and Engineering, Massachusetts Institute of Technology, 77 Massachusetts Ave, Cambridge, 02139, MA, USA}

\begin{document}
\maketitle

\textbf{Abstract}

{\color{rev}
Atomic diffusion affects the properties of various engineering materials, which predominantly occur in the polycrystalline state. A rigorous description of polycrystalline diffusion must therefore account for crystallographic defects, especially grain boundaries (GBs), whose structure and volume fraction - and hence the effective grain size - govern mass transport. Experiments and atomistic simulations consistently show that GBs can accelerate diffusion by up to several orders of magnitude and that fluxes along and across the interface are generally anisotropic.

Conventional mesoscale models either neglect GBs or invoke idealized analytical corrections. Fully resolved finite-element meshes are accurate but computationally infeasible when nanometer-thin GB layers are involved. We introduce a collapsed-interface finite element that integrates the GB thickness analytically and embeds the result in a two-dimensional surface element. The formulation (i) treats in-plane and through-plane diffusivity independently, (ii) couples to the surrounding grain matrix without the need for mesh manipulations, and (iii) parametrizes both grain size and GB volume fraction via simple affine scalings, allowing systematic variation without remeshing. Effective diffusivity tensors are extracted by linear computational homogenization.

The new finite element reproduces three-dimensional GB transport phenomena - channeled fluxes, concentration discontinuities - at a fraction of the computational cost of explicit models. Parametric studies spanning multiple orders of magnitude in GB diffusivity reveal four distinct diffusion regimes and quantify their impact on the overall response. The framework thus connects atomistic data and continuum predictions, providing an efficient tool for diffusion-driven design and optimization of polycrystalline materials.
}



\section{Introduction}
\label{Sec:introduction}

{\color{rev}
Atomic diffusion is a fundamental mechanism that governs the performance of a wide range of functional materials. For example, solid-state electrolytes require high ionic diffusivity to transport mobile ions between cathodes and anodes in all-solid-state batteries~\cite{Jun2024}. Electrodes in fuel cells utilize mixed ionic-electronic conductors, which combine high ionic diffusivity with good electronic conductivity~\cite{Riess1992May}. In hydrogen storage materials, hydrogen diffusion is deliberately enhanced to enable efficient absorption and release~\cite{Majer2003Apr}, whereas in structural metallic materials, it must be suppressed to mitigate the risk of hydrogen embrittlement~\cite{Song2013Feb}. In catalysts, diffusivity requirements vary across reactants and intermediates, and these diffusion characteristics influence the overall catalytic performance~\cite{Chmelik2010Nov}. The ability to tailor atomic diffusivity to meet application-specific requirements remains a central challenge in materials design and engineering. Therefore, investigating atomic diffusion processes at device-relevant scales is essential for both mechanistic understanding and technological advancement. 

Because of limitations in synthesis techniques and the intrinsic behavior of materials, crystallographic defects are invariably present in real solids. These defects span several length scales---zero-dimensional (0D) defects (e.g., vacancies, dopants), one-dimensional (1D) defects (e.g., dislocations), and two-dimensional (2D) defects (e.g., grain boundaries (GBs))---all of which can influence material properties. In particular, experimental studies have demonstrated that GBs and the grain size can substantially alter atomic diffusivity, thereby impacting the overall functionality of the material~\cite{Hasegawa2024, Saranya2018Aug, Lin2015Apr}. For solids with polycrystalline microstructures, theoretical predictions of diffusivity based on simulations should account for the influence of GBs to enable meaningful results. 

Molecular dynamics simulations based on interatomic potentials have been used to quantitatively study the influence of GBs on atomic diffusion~\cite{Ou2024, Lu2018Oct, Dawson2018, Yu2017Nov}. These simulations confirmed the significant impact of GBs on atomic diffusivity. Furthermore, they revealed that GBs modify the local structural and chemical environments, leading to distinct diffusion characteristics for different mobile species~\cite{Ou2024, Dawson2019, Dawson2018}. Anisotropic diffusivity in GB regions was also observed~\cite{Ou2024}. Due to the high computational cost, molecular dynamics simulations are generally limited in accessible length and time scales on the order of a few nanometers and nanoseconds. Therefore, simulations of polycrystals with micrometer-scale grains and investigations of the long-term impact of GBs remain out of reach. These limitations may contribute to discrepancies between simulated and experimentally observed diffusion behavior. 

Effective diffusivity of polycrystals with GBs was estimated by analytical or empirical volumetric averages based on local diffusivities from molecular dynamics simulations~\cite{Yoon2023, Dawson2018}. These approaches trace back to the early analytical insights~\cite{Hart1957} related to the Wiener bounds~\cite{Wiener1912} for permittivity or corresponding bounds in elasticity described by Voigt~\cite{Voigt1889} and Reuss~\cite{Reuss1929}. More accurate predictions were derived for idealized geometries, such as the Maxwell--Garnett equation for spherical~\cite{Garnett1904} or 2D square grains~\cite{Belova2003}, assuming isotropic GBs. For realistic polycrystalline microstructures featuring 3D grain geometries, establishing explicit relations between local and effective diffusivities remains an open challenge. 

Continuum-level simulations provide a rigorous macroscopic framework for modeling diffusion in polycrystals with complex microstructures. Phase-field simulations were applied at varying levels of GB description~\cite{Daubner2023, Lvov2023, Heo2021, Hoffrogge2021}. At the device-relevant scale, finite element (FE) simulations coupled with homogenization techniques offer a more computationally efficient alternative. Effective diffusivities can be obtained by numerically solving Fick's law of diffusion using spatially varying local diffusivities as inputs. In practice, however, FE simulations that fully resolve both grains and GBs are limited to simplified geometries designed to capture specific diffusivity characteristics~\cite{Chepak-Gizbrekht2020, Gryaznov2008}, due to the significant length-scale separation between GB width (typically a few nanometers) and grain size (on the order of micrometers). 

In the field of continuum mechanics and FE simulations, different approaches were proposed to address the length-scale separation challenge in modeling a wide variety of interface phenomena. Usually, the full resolution of GBs is avoided by reducing them to ``zero-thickness'' 2D manifolds and introducing a parametrization that accounts for application-specific characteristics~(e.g., Ref.~\cite{Sinzig2024} for polycrystals). For a detailed overview of specialized and generalized approaches, we refer to Refs.~\cite{Javili2025,Firooz2021}. In the context of polycrystalline microstructures, many studies represent GBs as interface elements, assuming constant concentration in the direction normal to the GB plane~\cite{Sarkar2021, Lacaille2014, Han2013}. GB diffusion was modeled by a single diffusion coefficient with a GB width parameter. This approach follows the Fisher model~\cite{Gibbs1966, Fisher1951} and matrix diffusion theory~\cite{Carrera1998Jun}, applicable only in the kinetic regime where GB diffusivity greatly exceeds bulk diffusivity. In contrast, Ref.~\cite{Peng2024} represents GBs as cohesive elements to model concentration jumps due to GB resistance, while neglecting diffusion along the GBs. This approach is valid only in the kinetic regime where GB diffusivity is much smaller than bulk diffusivity. To our knowledge, no existing study provides a GB representation that captures anisotropic GB diffusivity applicable to a wide range of kinetic regimes, i.e., transversely isotropic GB diffusion with both in-plane and through-plane transport being active has not been considered.

Here, we introduce a novel collapsed interface element to effectively represent such anisotropic GBs in FE simulations of polycrystals. Thin GB layers are modeled using an analytical formulation along the (virtual) GB thickness direction, eliminating the need for explicit geometric resolution. The mesoscale diffusivity response is computed with linear homogenization techniques. This approach facilitates efficient FE simulations of polycrystalline microstructures with anisotropic GB diffusivities, without imposing assumptions on kinetic regimes. Further, effective diffusivities for varying grain sizes can be obtained without modifying the model geometry or mesh, facilitating accelerated parametric studies. By utilizing grain and GB diffusivities of the promising solid-state electrolyte \Argyro{} from atomistic simulations as inputs, this study demonstrates the feasibility of simulating realistic polycrystalline microstructures at the mesoscale with high accuracy. 

The remainder of this article is structured as follows: In \Cref{Sec:constitutive_model}, a concise overview of the notation is given first (\Cref{Sec:notation}). Then, the proposed diffusion model is introduced together with a suitable FE discretization: Starting from the Nernst-Planck equation the diffusion in the bulk is recapped (\Cref{Sec:bulk:diff}) before introducing a dimensionality reduced ``zero-thickness'' ansatz for the GB diffusion (\Cref{Sec:surface:diffusion}), followed by the FE discretization (\Cref{Sec:fem:model}) and the computational homogenization techniques (\Cref{Sec:homogenization}). To reduce computation times without any loss of accuracy, we exploit the affine structure of the problem in \Cref{Sec:affine}. In \Cref{Sec:2d_validation}, a 2D example is introduced for validation and to prove the consistency of the model assumptions. The captured effects on the diffusion behavior in 3D polycrystals are then discussed in \Cref{Sec:results} in extensive parametric studies. In \Cref{Sec:case_study}, the proposed FE diffusion model is applied to \Argyro{} with a realistically parametrized polycrystalline microstructure.
Finally, the summary and outlook are given in \Cref{Sec:conclusion}.
}

\section{Constitutive model and discretization}
\label{Sec:constitutive_model}

{\color{rev}
\subsection{Notation}
\label{Sec:notation}

In the following, we represent vectors by boldface lowercase letters (e.g., atomic flux $\fq$) and 2-tensors by boldface uppercase letters (e.g., diffusivity tensor $\fD$) to distinguish them from scalar quantities (e.g., concentration $c$). The coefficient matrix $\ull{A}$ is equivalent to the tensor $\fA$ in a fixed orthonormal coordinate system. It allows for a sleek notation of the algebraic operations for assembly of the FE system.

Given an orthonormal basis (with basis vectors $\fe_i$, $i\in \{1,2,3\}$ or $\{x, y, z\}$) and based on the Einstein summation over repeated indices, a single contraction~$\cdot$ and the tensor product~$\otimes$ are defined as ($\fA$: order $n$ tensor; $\fB$: order $m$ tensor)
\begin{align}
    \fA &= A_{i_1 i_2\cdots i_{n-1}i_n} \fe_{i_1} \otimes \cdots \otimes \fe_{i_n}, \qquad
    \fB = B_{j_1 j_2\cdots j_{m-1}j_m} \fe_{j_1} \otimes \cdots \otimes \fe_{j_m}, \\
    \fA \cdot \fB &= A_{i_1 i_2 \cdots i_{n-1} k}B_{k j_2 \cdots j_m}
    \fe_{i_1} \otimes \cdots \fe_{i_{n-1}} \otimes \fe_{j_2} \otimes \cdots \fe_{j_m}, \\
    \fA \otimes \fB &= A_{i_1 \cdots i_n} B_{j_1 \cdots j_m} \fe_{i_1} \otimes \cdots \otimes \fe_{i_n} \otimes \fe_{j_1} \otimes \cdots \fe_{j_m}\, .
\end{align}
This yields the (right) divergence and the (right) gradient of a general tensor $\fA$,
\begin{align}
    \fA \cdot \nabla &= A_{i_1 \cdots i_{n-1} i_n,i_n}\,\fe_{i_1} \otimes \cdots \fe_{i_{n-1}} \, , &
    \fA \otimes \nabla &= A_{i_1 \cdots i_{n},k}\,\fe_{i_1} \otimes \cdots \fe_{i_{n}} \otimes \fe_k\, , 
\end{align}
with the operator
\begin{align}
    \nabla = \partial_x \fe_x + \partial_y \fe_y + \partial_z \fe_z \, .
\end{align}
Note that applying the operator $\nabla$ from the right refers to the standard notation in continuum mechanics (see, e.g., Ref.~\cite{Bertram2015}) since the operator $\nabla$ is, in general, applicable to tensors of arbitrary order. In case of a scalar~$a$ or a vector~$\fa$, the following equivalences hold:
\begin{align}
    a\otimes\nabla &= a_{,j} \fe_j = \nabla a \, , &
    \fa\cdot\nabla &= a_{i,i} = \nabla\cdot\fa\, .
\end{align}

In the context of computational homogenization (see \Cref{Sec:eff}) phase-wise averages will be used. We define the spatial average for an arbitrary quantity $[\cdot]$ over a domain $\mathcal{A}$ with measure $A=|\mathcal{A}|$ as
\begin{align}
    \langle [\cdot] \rangle_\mathcal{A} = \frac{1}{A} \int_{\mathcal{A}} [\cdot]\,{\rm{d}}A.
\end{align}
}

\subsection{Bulk diffusion}
\label{Sec:bulk:diff}
{\color{rev}To simulate atomic diffusion, a model describing the spatial transport of mobile species with concentration~$ c (\fx)$ is required. } The Nernst-Planck equation~\cite{Raijmakers2020} is used to describe the flux~$\fq$ consisting of an advection term~$\fq_{\rm v}$ due to a velocity $\fv$ and the flux~$\fq_{\rm f}$ due to a driving force~$\ff$, 
\begin{align}
    \fq = \fq_{\rm v} + \fq_{\rm f} && \text{with } \fq_{\rm v} = c \fv , \quad \fq_{\rm f} =  \fD_\Omega \ff\, . \label{eq:NP}
\end{align}
For atomic diffusion in solids, the velocity~$\fv$ is zero, i.e., no advection needs to be considered, independently of the concentration $c$ ($\fq_{\rm v}=\boldsymbol{0}$).
The diffusivity tensor $\fD_\Omega \in \mathrm{Sym}_+(\ffR^3)$ is symmetric and positive definite. It captures material symmetries (if any), i.e., given a symmetry group $\cG \subseteq {\rm{SO}}(3)$, the relation
\begin{align}
    \fD_\Omega &= \fG \, \fD_\Omega \, \fG^\mathsf{T}, & \forall \fG & \in \cG \, ,
\end{align}
holds. If $\mathcal{G} \neq \mathrm{SO}(3)$, then the actual diffusivity tensor depends on the local crystal orientation~$\fQ$ via
\begin{align}
    \fD_\Omega &= \fQ \, \fD_0 \, \fQ^\mathsf{T}\, , \label{eq:D:CODF}
\end{align}
where $\fD_0$ is the diffusivity tensor in the reference coordinate system.

\begin{figure}[h]
    \centering
    \includegraphics[width=0.4\textwidth]{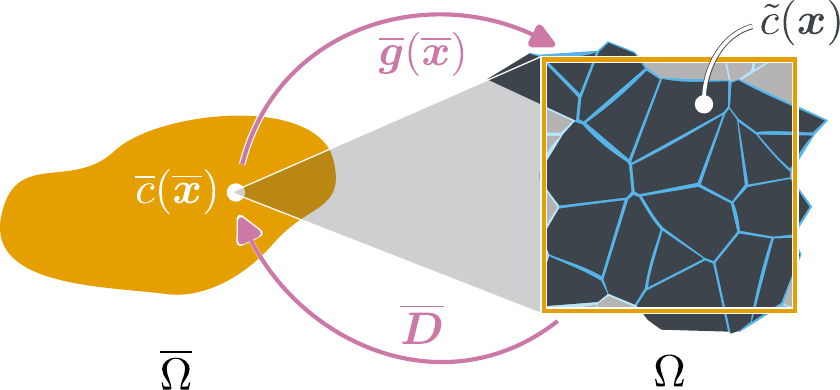}
    \caption{Relation between macroscopic position $\overline{\fx}\in\overline{\Omega}$ (mesoscale) and corresponding periodic microscopic polycrystalline sample $\Omega$ (microscale) in the context of linear computational homogenization. The quantities introduced in \cref{eq:homogenization}, as well as the macroscopic diffusion tensor $\ol{\fD}$, are indicated for reference.}
    \label{fig:homogenization_framework}
\end{figure}

In the sequel, $\Omega\subset\ffR^3$ is a representative volume element of the material as shown in \Cref{fig:homogenization_framework}, and the coordinate $\fx\in\Omega$ is used to emphasize potential spatial dependencies. Time dependencies are not stated explicitly in favor of a simplified notation. The driving force $\ff$ in \cref{eq:NP} is composed of two additive contributions:
\begin{align}
    \ff(\fx) = \ff_{\rm c}(\fx) + \ff_{\rm E}(\fx)\,  && \text{with } \ff_{\rm c}(\fx) = - \nabla c(\fx), \quad \ff_{\rm E}(\fx)  = \frac{z \, e}{k_{\rm B} T} c(\fx) \fE(\fx).
\end{align}
The first part~$\ff_{\rm c}(\fx)$ contains the driving force due to the gradient of the chemical concentration~$c(\fx)$ according to Fick's first law. The second part~$\ff_{\rm E}(\fx)$ accounts for the electro-migration due to an external electric field~$\fE(\fx)$. Here $z$~denotes the valence of the atomic species, $e$~is the elementary charge, $k_{\rm B}$~denotes the Boltzmann constant, and $T$~is the absolute temperature. Although including $\ff_{\rm E}(\fx)$ is of interest, we confine attention to the driving force $\ff_{\rm c}$ according to Fick's law for determining the effective diffusivity. 
{\color{rev} The investigation of contributions due to $\ff_{\rm E}$ is deferred to future studies. }

{\color{rev} Within an isolated polycrystal undergoing atomic diffusion}, mobile species are neither generated nor annihilated. In this region and under certain external conditions, their  concentration settles at a steady state, fulfilling the balance equation
\begin{align}
    \fq(\fx) \cdot \nabla = 0 \quad\text{in }\Omega, &&\text{+ boundary conditions.} \label{eq:balance}
\end{align}
Note that, due to the small size of~$\Omega$, the variations of the absolute value of the concentration $c(\fx)$ are small, matching results of, e.g., Ref.~\cite{Ou2024}, on the atomistic scale of the material. At the same time, the gradient~$\nabla c(\fx)$ can take substantial values, and relevant driving forces can emerge from these small fluctuations.

Regarding the boundary conditions, we limit attention to periodic fluctuation conditions for the concentration~$c(\fx)$. This choice is motivated by the established results for linear computational homogenization~\cite{Ostoja-Starzewski2011} and follows the schematic in~\Cref{fig:homogenization_framework}:
\begin{align}
    c(\fx) &= \ol{c}(\ol{\fx}) + \ol{\fg}(\ol{\fx}) \cdot \fx + \WT{c} (\fx)\, . \label{eq:homogenization}
\end{align}
The overall concentration~$c(\fx)$ is composed of the macroscopic concentration~$\ol{c}(\ol{\fx})$ at the macroscopic position $\ol{\fx}\in\ol{\Omega}\subset \ffR^3$, a contribution of the imposed macroscopic concentration gradient~$\ol{\fg}$, and the periodic concentration fluctuation field~$\WT{c}(\fx)$. Note that $\ol{\fg}$ could be replaced by an equivalent effective driving force, e.g., due to an applied electric field~$\ol{\fE}$ (cf. \Cref{r:electric}). 

Solving \cref{eq:balance} with the unit gradients $\ol{\fg}_k = \fe_k$ for $k\in\{1,2,3\}$ applied as boundary conditions yields three linear problems. The macroscopic diffusion tensor~$\ol{\fD}$ is derived from the corresponding averaged gradient and flux fields, see~\Cref{Sec:eff}.

\begin{remark}
    The presented steady-state diffusion problem is equivalent to a steady-state heat conduction problem or to electrostatics. Therefore, established discretizations, e.g., using FEs, can readily be applied to the elliptic problem.
\end{remark}

\begin{remark}
    The macroscopic gradient $\ol{\fg}$ introduces a spatially constant driving force $\ff_{\rm c, 0}=-\ol{\fg}$. This is equivalent to an external electric field (assumed constant inside of the microstructure) $\overline{\fE} = - \frac{k_{\rm B} T}{z \, e} \frac{1}{\ol{c}} \ol{\fg}$. Therefore, the model presented herein also calculates the sought-after diffusion tensor~$\ol{\fD}$ in the presence of electric fields.
    \label{r:electric}
\end{remark}

\subsection{Interface diffusion with in-plane transport}
\label{Sec:surface:diffusion}
\subsubsection{Assumptions and definitions}
In a recent study focusing on solid-state electrolyte \Argyro{}~\cite{Ou2024}, the diffusivities of both bulk and GB structures were assessed through molecular dynamics simulations. It was found that (i) the diffusivity of the GBs varied considerably from that of the bulk, and (ii) the diffusivities along and across the GBs are different. Further, the dependency of the diffusion coefficient $D$ on the temperature $T$ introduced via the Arrhenius relation~\cite{Arrhenius1889},
\begin{align}
    D (T) = D_{0}\exp\lb - \frac{E_{{\rm a}}}{k_{\rm B}T} \rb,
    \label{eq:arrhenius}
\end{align}
potentially differs due to the structure-specific fitted constant $D_{0}$ and activation energy $E_{{\rm a}}$. The Boltzmann constant is denoted with $k_{\rm B}$ here. Note that $D$ is treated as concentration-independent in its derivation from molecular dynamics simulations.

To model this behavior, we propose a special-purpose FE type inspired by cohesive zone elements used in mechanical problems (see Ref.~\cite{Firooz2021} for a comprehensive review). The model is put into the context of GBs \cite{Wulfinghoff2017,Pezzotta2008,Mori1987} to deal with the aforementioned aspects. In a mechanical setting, cohesive elements allow for a displacement jump~$\fdelta$ across the interface, defining the traction vector~$\ft(\fdelta)$ (e.g., Refs.~\cite{Leuschner2015, vandenBosch2006, Needleman1990}). We adopt the concept of a discontinuity between adjacent bulk domains for the concentration~$c(\fx)$ by connecting them with a continuous function through a virtual finite-thickness interface.

\begin{figure}[h]
    \centering
    \begin{subfigure}[b]{0.23\textwidth}
        \centering
        \includegraphics[width=\textwidth]{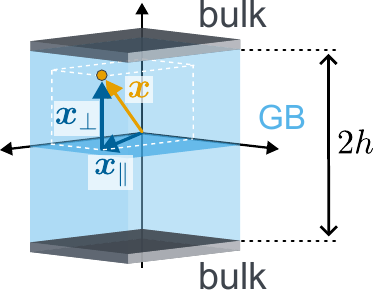}
        \caption{}
        \label{fig:interface}
    \end{subfigure}
    \hfill
    \begin{subfigure}[b]{0.23\textwidth}
        \centering
        \includegraphics[width=\textwidth]{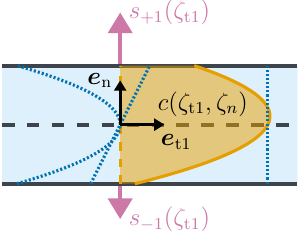}
        \caption{}
        \label{fig:channel_GB}
    \end{subfigure}
    \hfill
    \begin{subfigure}[b]{0.3\textwidth}
        \centering
        \includegraphics[width=\textwidth]{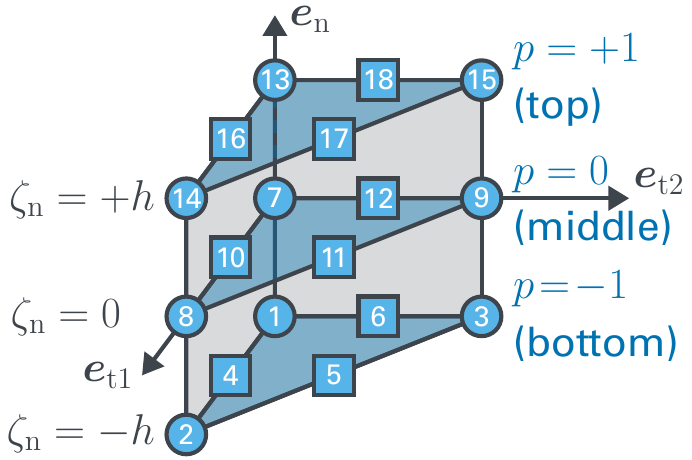}
        \caption{}
        \label{fig:iface_element}
    \end{subfigure}
    \caption{(a) Interface of thickness $2h$ and the split into tangential and normal coordinates (b): Atomic flux in a 2D channel example with in-/outflux on top and bottom. The concentration profile $c(\zeta_{\rm t1}, \zeta_{\rm n})$ as well as the required constant, linear and quadratic modes (dotted lines) are given. (c): Exploded view of the $\zeta_{\rm n}$-collapsed quadratic prism element: Nodes 1-6 constitute the \textit{bottom} ($p=-1$), nodes 7-12 the \textit{middle} ($p=0$), and nodes 13-18 the \textit{top} layer ($p=1$).}
    \label{fig:iface:element:assumptions}
\end{figure}

We systematically construct the interfacial model and the related special-purpose element building on the following assumptions:
\begin{enumerate}[label=\textbf{[A\arabic*]}]
    \item The variable to be modeled is the concentration $c(\fx)$. It is assumed to be continuous in the bulk as well as within and throughout the interface, i.e., the GBs.
    \label{A:continuous}
    \item The surrounding mesh, representing a polycrystal based on the Voronoi tessellation \cite{Quey2011,Fritzen2009,Aurenhammer1991}, is assumed to be tetrahedral.
    \label{a:voronoi}
    \item The interface width $2h$ is much smaller than the grain size $L_{\rm grain}$. Therefore, the interface element is collapsed along the out-of-plane direction~$\fn$. In the FE mesh, the elements will appear to have zero thickness, which aligns with the assumed separation of length scales. The (usually small) GB volume $v_{\rm GB} = 2hA$ added to the representative volume element will partially be corrected during post-processing of the results (see \Cref{Sec:geo_quant}). The vector of a point within the interface is decomposed based on the interface normal $\fn$ (see also \Cref{fig:interface}) according to:
    \begin{align}
        \fx & = \fx_\Vert + \fx_\perp\, , &
        \fx_\Vert &= \lb \fI - \fn \otimes \fn \rb \fx =  x_\Vert\, \fe_{\rm t}\, , &
        \fx_\perp &= \lb \fn \cdot \fx \rb \,\fn \, = x_\perp\,\fn \, . \label{eq:x:split}
    \end{align}
    Here, $\fe_{\rm t}$ denotes a unit vector in the interface plane. The same decomposition applies to vector-valued quantities such as the flux $\fq$ (see \Cref{Sec:surface:properties}).
    \item Within the GB region, a transversely isotropic diffusion tensor $\DGBzero\in \mathrm{Sym}_+$ is assumed. The GB-attached local coordinate system is defined as $\{ \fe_{\rm t1}, \fe_{\rm t2}, \fn \}$, i.e., the two orthonormal tangential basis vectors followed by the normal orientation of the GB, with coordinates $\ul{\zeta} = [\zeta_{\rm t1}, \zeta_{\rm t2}, \zeta_{\rm n}]$ (see also \Cref{fig:channel_GB,fig:iface_element}). With this simplification, the matrix representation of the diffusion tensor reduces to
    \begin{align}
    \fD_{\rm GB,0} &= \sV D_\Vert & 0 & 0 \\ 0 & D_\Vert & 0 \\ 0 & 0  & D_\perp \eV_{\rm (t1, t2, n)} ,
    \end{align}
    where $D_\Vert$ and $D_\perp$ are the in-plane and through-plane diffusion coefficients, respectively. 
    \label{A:D_GB}
    \item In the local coordinate system we assert a quadratic ansatz in the through-thickness direction for each $\ul{\zeta}_{\rm t}=[\zeta_{\rm t1}, \zeta_{\rm t2}]$, i.e.,
    \label{A:quadratic}
    \begin{align}
        c(\zeta_{\rm t1}, \zeta_{\rm t2}, \zeta_{\rm n}) & = \sum_{j=0}^2 \alpha_j(\ul{\zeta}_{\rm t}) \lb \frac{\zeta_{\rm n}}{h} \rb^j ,
        \label{eq:c_ansatz}
    \end{align}
    which is motivated by the following model problem: When investigating a 2D planar channel-like problem in the system $\{\fe_{\rm t1}, \fe_{\rm n}\}$ with atomic in-/outflux $s_{+1}(\zeta_{\rm t1})$ on top and $s_{-1}(\zeta_{\rm t1})$ on bottom (cf. \Cref{fig:channel_GB}) the in-plane flux must depend on $\zeta_{\rm t1}$. By combining the analytical expression for the flux at top and bottom, one finds:
    \begin{align}
        \begin{rcases}
            q(\zeta_{\rm n}=h) = -D_\perp\lb\frac{\alpha_1(\zeta_{\rm t1})}{h} + \frac{2 \alpha_2(\zeta_{\rm t1})}{h}\rb \overset{!} = s_{+1}(\zeta_{\rm t1}) \\
            -q(\zeta_{\rm n}=-h) = -D_\perp\lb\frac{\alpha_1(\zeta_{\rm t1})}{h} - \frac{2 \alpha_2(\zeta_{\rm t1})}{h}\rb \overset{!} = -s_{-1}(\zeta_{\rm t1})
        \end{rcases}
        \alpha_2(\zeta_{\rm t1}) = \frac{h}{4D_\perp} \lb s_{+1}(\zeta_{\rm t1})+s_{-1}(\zeta_{\rm t1})\rb \, .
    \end{align}
    This implies that {\color{rev} only quadratic modes (or, more generally, modes of even order)} enable a net flux into and out of the GB, which is infeasible when considering only linear concentration profiles in the thickness direction. Further, this mechanism enables diffusion channeling along the GBs. {\color{rev} Note that this aligns with findings for general imperfect interface elements in the context of heat transport \cite{Javili2014} despite the differing parametrization: Opposed to mechanical interfaces the admissible concentration at the center plane of the interface is not bounded by the concentration levels on both sides of the interface in case of in-plane transport or storage.}
    \item The mechanical separation of the interface \textcolor{rev}{and the inhomogeneous charge distribution in the interface} are not considered.
\end{enumerate}

\subsubsection{Model properties}
\label{Sec:surface:properties}
The diffusion coefficients $D_\Vert>0 $ and $D_\perp > 0$ characterize the behavior of the GB alongside the GB half-width~$h$. Regarding the amplitude of the atomic flux parallel~$q_\Vert$ and perpendicular~$q_\perp$ to the interface, we obtain the relation
\begin{align}
    q_\perp &  = \fq \cdot \fn = - D_\perp \pd{c(\ul{\zeta})}{\zeta_{\rm n}} {\color{rev} =} \frac{D_\perp}{h} \llbracket c \rrbracket , \\
    q_\Vert & = \vert \lb \fI - \fn \otimes \fn \rb \fq \vert = D_\Vert \sqrt{ \vert c \otimes \nabla \vert^2 - \lb \pd{c(\ul{\zeta})}{\zeta_{\rm n}} \rb^2 } {\color{rev} =
    D_\Vert \vert \lb \fI - \fn \otimes \fn \rb  \lb c \otimes\nabla \rb \vert
    }\, ,
\end{align}
following the decomposition in \cref{eq:x:split}, where \textcolor{rev}{$\llbracket c \rrbracket=c(\ul{\zeta}_{\rm t}, \zeta_{\rm n}=h) - c(\ul{\zeta}_{\rm t}, \zeta_{\rm n}=-h)$} denotes the concentration jump across the GB. In particular, $q_\Vert$ does not explicitly depend on $h$, while $q_\perp$ does. In the limit case where $h\to 0$ or $D_\perp\to\infty$ (or both), the diffusion~$q_\perp$ in the through-thickness direction also tends to infinity, which recovers the case of a classical bulk-only simulation, i.e., it renders the proposed GB model obsolete. Inversely, this observation illustrates implicit assumptions of such ``classical'' simulations, i.e., assumed zero atomic flow resistance of the GBs.

In the case where $D_\Omega=D_\perp=D_\Vert$, the GB should not influence the response at all. This case will be used to motivate the volume fraction compensation strategy for the computation of effective properties in~\Cref{Sec:eff}.

\subsection{FE implementation}
\label{Sec:fem:model}
\subsubsection{Bulk}
\label{Sec:fem:bulk}
As for the bulk material, we use isoparametric, quadratic tetrahedra (commonly referred to as \texttt{TET10} {\color{rev} element}) with the {\color{rev} row} vector of shape functions $\ul{N}_\Omega(\ul{\eta}) \in\ffR^{1 \times 10}$ for the reference coordinates $\ul{\eta}=[\eta_1,\eta_2,\eta_3]$ (see \Cref{app:T10ansatz}). Using isoparametric finite elements results in the gradient operator $\ull{B}_\Omega$ to depend on the gradient operator~ $\ull{B}^{\rm ref}_\Omega\in\ffR^{3\times 10}$ in the reference element~$\Omega^{\rm e}_{\rm ref}$ and on the {\color{rev} element} Jacobian
\begin{align}
    \ull{J}_\Omega^{(i)}(\ul{\eta}) &= \ull{B}^{\rm ref}_\Omega(\ul{\eta}) \ull{X}_\Omega^{(i)} \, ,
\end{align}
where $\ull{X}_\Omega^{(i)} \in \ffR^{10 \times 3}$ is a matrix containing the coordinates of the 10~nodes of the $i$-th bulk element. The gradient operator then reads $\ull{B}_\Omega(\ul{\eta}) = \ull{J}_\Omega^{-1}(\ul{\eta}) \ull{B}^{\rm ref}_\Omega(\ul{\eta})$ where the superscript $i$ is omitted for readability. As per the deployed cubature scheme, we have resorted to $n^{\rm T10}_{\rm GP}=4$ integration points $\ul{\eta}_j$ with corresponding integration weights $\omega_j=v^{\rm e}/4$ ($j=1, \dots, 4=n^{\rm T10}_{\rm GP}$) given the element volume~$v^{\rm e}$.

In the following, an isotropic bulk material $\ull{D}_\Omega \equiv \fD_\Omega = D_\Omega\fI$ is assumed\footnote{\textcolor{rev}{We relate the coefficient matrix $\ull{D}_\Omega$ (defined in a fixed coordinate system) to the equivalent tensor $\fD_\Omega$ here, as pointed out in \Cref{Sec:notation}. In the following, we stick with the coefficient matrices for a sleek notation of the FE system assembly.}}. This assumption is valid for materials exhibiting isotropic or cubic symmetry. The resulting element stiffness matrix $\Kebulk$ in the bulk then reads
\begin{align}
    \Kebulk = \sum\limits_{j=1}^{n_{\rm GP}^\Omega} \omega_j \ull{B}_\Omega^\mathsf{T}(\ul{\eta}_j) \ull{D}_\Omega \, \ull{B}_\Omega(\ul{\eta}_j)
    = D_\Omega \sum\limits_{j=1}^{n_{\rm GP}^\Omega} \omega_j \ull{B}_\Omega^\mathsf{T}(\ul{\eta}_j) \ull{B}_\Omega(\ul{\eta}_j)\, .
    \label{eq:Kebulk}
\end{align}

\subsubsection{Special purpose interface diffusion element}
\label{Sec:fem:interface}
Building on the considerations of \Cref{Sec:surface:diffusion}, we suggest a shape function based on the following tensor-product ansatz: On each of the three planes ($p\in\{-1, 0, 1\}$) in \Cref{fig:iface_element}, we introduce classical 2D quadratic shape functions $\ul{N}_{\rm P2}(\zeta_{\rm t1},\zeta_{\rm t2})\in\ffR^{1 \times 6}$ through a six node triangular element (see \Cref{app:T6ansatz}) with integration weights $\nu_i$. In conjunction with the used quadratic tetrahedral elements in the bulk, this enforces the assumed continuity of~$c(\fx)$ in Assumption \labelcref{A:continuous}. Then, on each of the three $\zeta_{\rm n}$ levels, a planar interpolation is pursued, leading to a total of 3$\times$6=18~nodes per element. The element degree of freedom vector $\ul{c}^{\rm e}$ is partitioned along the three layers according to:
\begin{align}
    \ul{c}^{\rm e} &= \begin{bmatrix*}[l] \ul{c}^{\rm e}_{-1} \\ \ul{c}^{\rm e}_{0} \\ \ul{c}^{\rm e}_{1} \end{bmatrix*} \in \ffR^{18}\, .
\end{align}
The surface interpolated data on the three layers
\begin{align}
    c^{\rm e}_p(\zeta_{\rm t1}, \zeta_{\rm t2}) = \ul{N}_{\rm P2}(\zeta_{\rm t1}, \zeta_{\rm t2}) \ul{c}^{\rm e}_p, && p  \in \{ -1, 0, 1 \}\, ,
\end{align}
is further interpolated {\color{rev} along} the through-thickness coordinate $\zeta_{\rm n}\in[-h, h]$ by a quadratic Lagrange basis~$\{\cL_{-1}, \cL_0, \cL_1 \}$, i.e.,
\begin{align}
    \cL_p(\zeta_{\rm n}) &= \prod_{\beta \in \{-1, 0, 1 \} \setminus \{p\}} \frac{\zeta_{\rm n}-\beta h}{\lb p - \beta \rb h}\, ,  && p  \in \{ -1, 0, 1 \}\, .
\end{align}
Putting the staggered interpolation together, the shape function within the prismatic GB element $\NGB{}$ gets
\begin{align}
        c^{\rm e}(\zeta_{\rm t1}, \zeta_{\rm t2}, \zeta_{\rm n}) &= \sum_{p=-1}^{1} \cL_p(\zeta_{\rm n}) c^{\rm e}_p(\zeta_{\rm t1}, \zeta_{\rm t2}) = \sum_{p=-1}^{1} \cL_p(\zeta_{\rm n}) \ul{N}_{\rm P2}(\zeta_{\rm t1}, \zeta_{\rm t2}) \ul{c}^{\rm e}_p = \NGB{}(\zeta_{\rm t1}, \zeta_{\rm t2}, \zeta_{\rm n}) \ul{c}^{\rm e} \, ,
\label{eq:tensor_prod}
\end{align}
i.e., $\NGB{} = \begin{bmatrix}
\cL_{-1}(\zeta_{\rm n}) \ul{N}_{\rm P2}(\ul{\zeta}_{\rm t}) & \cL_{0}(\zeta_{\rm n}) \ul{N}_{\rm P2}(\ul{\zeta}_{\rm t}) & \cL_{+1}(\zeta_{\rm n}) \ul{N}_{\rm P2}(\ul{\zeta}_{\rm t})
\end{bmatrix}\in\ffR^{1\times 18}$. The gradient of the concentration within the GB results from applying the product rule
\begin{align}
    \nabla c^{\rm e} (\zeta_{\rm t1}, \zeta_{\rm t2}, \zeta_{\rm n}) &=  \sV  \pd{\NGB}{\zeta_{\rm t1}} \\[6pt]  \pd{\NGB}{\zeta_{\rm t2}} \\[6pt] \pd{\NGB}{\zeta_{\rm n}} \eV \ul{c}^{\rm e}
    = \sum_{p=-1}^1 \lb \sV 1 & 0 \\ 0 & 1 \\ 0 & 0 \eV \ull{B}_{\rm P2}^{\rm ref}(\ul{\zeta}_{\rm t}) \cL_p(\zeta_{\rm n}) + \sV 0 \\0 \\ 1\eV \pd{\cL_p(\zeta_{\rm n})}{\zeta_{\rm n}} \ul{N}_{\rm P2}(\ul{\zeta}_{\rm t}) \rb \ul{c}^{\rm e}_p\\
    &= \sV 1 & 0 \\ 0 & 1 \\ 0 & 0 \eV \BGBpara^{\rm ref}(\ul{\zeta}_{\rm t}, \zeta_{\rm n}) \ul{c}^{\rm e} + \sV 0 \\ 0 \\ 1 \eV \BGBperpRef(\ul{\zeta}_{\rm t}, \zeta_{\rm n}) \ul{c}^{\rm e} = \BGB^{\rm ref}(\zeta_{\rm t1}, \zeta_{\rm t2}, \zeta_{\rm n}) \ul{c}^{\rm e} \, ,
\end{align}
matching the GB coordinate system in \Cref{fig:iface:element:assumptions}.

In the following, the element-specific gradient operators $\BGBpara{}$ and $\BGBperp{}$ are derived. The operator $\BGBpara^{\rm ref}\in\ffR^{2\times 18}$ defined in the 2D reference system $\{\fe_{t1}, \fe_{t2}\}$ needs to be related to the 3D coordinate system of the element. The nodal positions of the element ${\color{rev}\ull{\widehat{X}}_{\rm GB}^{(i)}}\in\ffR^{6\times 3}$ in the global coordinate system are identical on all three layers due to the zero-thickness of the element. We construct a transformation matrix ${\color{rev}\ull{W}^{(i)}}\in\ffR^{3\times 2}$, which consists of two vectors that form a 2D orthonormal basis {\color{rev} spanning the planar interface element}. The 2D representation of the nodal positions in this system is then given by
\begin{align}
    \WT{\ull{X}}_{\rm GB}^{{\color{rev}(i)}} = {\color{rev}\ull{\widehat{X}}_{\rm GB}^{(i)}}\, \ull{W}^{{\color{rev}(i)}} \in\ffR^{6 \times 2}\, .
\end{align}
Using the in-plane gradient operator $\ull{B}_{\rm P2}^{\rm ref}\in\ffR^{2\times 6}$, the inverse of the Jacobian $\ull{J}_{\rm GB}^{{\color{rev}(i)}}$ yields the gradient operator which is then projected back to the global 3D system, i.e.,
\begin{align}
    \BGBpara^{{\color{rev}(i)}}(\ul{\zeta}_{\rm t},\zeta_{\rm n}) = \ull{W}^{{\color{rev}(i)}}\,\lb\ull{J}_{\rm GB}^{{\color{rev}(i)}}\rb^{-1}\BGBpara^{\rm ref} \in\ffR^{3\times 18} \qquad \text{with } \ull{J}_{\rm GB}^{{\color{rev}(i)}} = \ull{B}_{\rm P2}^{\rm ref}\,\WT{\ull{X}}_{\rm GB}^{{\color{rev}(i)}} \in \ffR^{2\times 2}\, .
\end{align}
Note the dependence on $\zeta_{\rm n}$ due to the implicit representation of the normal direction. The contribution in the normal direction is directly given by multiplication with the interface normal:
\begin{align}
    \BGBperp^{{\color{rev}(i)}}(\ul{\zeta}_{\rm t}, \zeta_{\rm n}) = \ul{n}^{{\color{rev}(i)}} \, \BGBperpRef\, .
\end{align}
In the following, the superscript $i$ is omitted for readability. For the assembly of the element stiffness matrices, the orthogonality of normal and in-plane contributions is exploited:
\begin{align}
        \KGB^{\rm e}
        = \sum_{j=1}^{n_{\rm GP}^{\rm C18}} \nu_j \int_{-h}^h  D_\Vert \, \BGBpara^\mathsf{T}(\ul{\zeta}_{\rm{t}(j)}, \zeta_{\rm n}) \BGBpara(\ul{\zeta}_{\rm{t}(j)}, \zeta_{\rm n}) + D_\perp \, \BGBperp^\mathsf{T}(\ul{\zeta}_{\rm{t}(j)}, \zeta_{\rm n}) \BGBperp(\ul{\zeta}_{\rm{t}(j)}, \zeta_{\rm n}) \,\mathrm{d}\zeta_{\rm n}\, .
\end{align}
The integration over the interface plane is performed with the respective cubature scheme based on integration weights $\nu_j$ ($j=1,\dots,7=n_{\rm GP}^{C18}$) and in-plane integration points $\ul{\zeta}_{\rm{t}(j)}$. Due to the tensor product ansatz, the integration over the thickness of the element can be performed analytically, yielding constant prefactors independent of the specific element shape in the plane. The element stiffness matrix can be decomposed in an in-plane contribution $\Kepara$ and a normal contribution $\Keperp$:
\begin{align}
    \Kepara
    &= \frac{D_\Vert h}{15} 
    \begin{bmatrix*}[r]
        4[\ast] & 2[\ast] & -1[\ast] \\ 
        2[\ast] & 16[\ast] & 2[\ast] \\ 
        -1[\ast] & 2[\ast] & 4[\ast] 
    \end{bmatrix*} \in\ffR^{18\times 18} &&\text{with } [\ast] = \sum\limits_{j=1}^{n_{\rm GP}^{\rm C18}}\nu_j \ull{B}_{\rm P2}^\mathsf{T}(\ul{\zeta}_{\rm{t}(j)})\ull{B}_{P2}(\ul{\zeta}_{\rm{t}(j)}) \in\ffR^{6\times 6}\, , 
    \label{eq:Kepara} \\
    \Keperp
    &= \frac{D_\perp}{6h}
    \begin{bmatrix*}[r]
        7[\ast\ast] & -8[\ast\ast] & 1[\ast\ast] \\   
        -8[\ast\ast] & 16[\ast\ast] & -8[\ast\ast] \\ 
        1[\ast\ast] & -8[\ast\ast] & 7[\ast\ast]
    \end{bmatrix*} \in\ffR^{18\times 18} &&\text{with } [\ast\ast] = \sum\limits_{j=1}^{n_{\rm GP}^{\rm C18}}\nu_j \ul{N}_{\rm P2}^\mathsf{T}(\ul{\zeta}_{\rm{t}(j)})\ul{N}_{P2}(\ul{\zeta}_{\rm{t}(j)}) \in\ffR^{6 \times 6}.
    \label{eq:Keperp}
\end{align}
Note that the element stiffness matrices consist of nine identical blocks with different prefactors, which form a symmetric pattern. This allows for an efficient assembly of the matrices under consideration of the material- and microstructure-specific parameters $D_\Vert$, $D_\perp$, and $h$.

\begin{remark}
    The notation for the block-symmetric structure of the stiffness matrices $\Kepara$ and $\Keperp$ could be condensed by the Kronecker product, i.e., 
    \begin{align}
        \Kepara = \frac{D_\Vert h}{15} 
        \begin{bmatrix*}[r]
            4 & 2 & -1 \\ 
            2 & 16 & 2 \\ 
            -1 & 2 & 4 
        \end{bmatrix*}
        \otimes [\ast]\, , &&
        \Keperp = \frac{D_\perp}{6h}
        \begin{bmatrix*}[r]
            7 & -8 & 1 \\   
            -8 & 16 & -8 \\ 
            1 & -8 & 7
        \end{bmatrix*}
        \otimes [\ast\ast]\, .
    \end{align}
    However, we stick to the given notation in order to avoid ambiguity with respect to the tensor product, commonly denoted with $\otimes$ as well (see \Cref{Sec:notation}).
\end{remark}

\begin{remark}
    In general, the material parameters of a GB depend on the orientation of its interface and the neighboring grains. The structure of the global FE system allows for such a direct interface-specific parametrization without any overhead. For simplicity, however, we assume identical GB properties for all interfaces in the geometry for now. This will allow for an in-depth analysis of the transport mechanisms in \Cref{Sec:results}.
\end{remark}

\subsubsection{Periodicity constraints and overall system}
Following the homogenization setting in \cref{eq:homogenization}, the imposed macroscopic driving force $\ol{\ul{g}}$ enters the FE system via the right-hand side. The assembled global system to be solved comprises contributions from the volume elements representing the bulk $[\cdot]_\Omega$ as well as the in-plane $[\cdot]_\Vert$ and normal $[\cdot]_\perp$ contributions from the suggested interface elements. \textcolor{rev}{Here $[\cdot]$ denotes the stiffness matrices or right-hand side vectors.} The resulting linear system to be solved for the degrees of freedom (DOFs) of the fluctuation field $\ulWT{c}$ reads
\begin{align}
    \lb\Kbulk + \Kpara + \Kperp\rb \ulWT{c} = \ul{r}_\Omega + \ul{r}_\Vert + \ul{r}_\perp \, ,
\end{align}
with the respective contributions for bulk and GB,
\begin{align}
    \ull{K}_{[\cdot]} = \sum\limits_{i=1}^{n_{\rm el}^{[\cdot]}} \lb\ull{L}_{[\cdot]}^{(i)}\rb^\mathsf{T} \ull{K}_{[\cdot]}^{(i)} \ull{L}_{[\cdot]}^{(i)}\, , && \ul{r}_{[\cdot]} = -\sum\limits_{i=1}^{n_{\rm el}^{[\cdot]}}\lb\ull{L}_{[\cdot]}^{(i)}\rb^\mathsf{T} \ull{K}_{[\cdot]}^{(i)}\ull{X}_{[\cdot]}^{(i)}\ol{\ul{g}}\, ,
    \label{eq:global_K_r}
\end{align}
for given scatter matrices $\ull{L}_{[\cdot]}^{(i)}$ referring to the individual elements. The positions of the nodes within an element are denoted by $\ull{X}_\Omega^{(i)}\in\ffR^{10\times 3}$ for the bulk and $\ull{X}_{\rm GB}^{(i)}{\color{rev}=\left[\lb\ull{\widehat{X}}_{\rm GB}^{(i)}\rb^\mathsf{T}, \lb\ull{\widehat{X}}_{\rm GB}^{(i)}\rb^\mathsf{T}, \lb\ull{\widehat{X}}_{\rm GB}^{(i)}\rb^\mathsf{T}\right]}\in\ffR^{18\times 3}$ for the GB domain. To enforce periodicity between primary and replica DOFs $\ulWT{c}_{\rm p}$ and $\ulWT{c}_{\rm r}$ on the boundary we introduce two projectors $\ull{P}_{\rm per}\in\ffR^{d\times d}$ and $\ull{P}_{\rm f}\in\ffR^{d\times d_{\rm f}}$: The $d_{\rm f}$ free DOFs in $\ulWT{c}_{\rm f}$ are first scattered to the full system size $d$ by using $\ull{P}_{\rm f}$. With $\ull{P}_{\rm per}$, the entire set of DOFs is then obtained by keeping the free DOFs and reconstructing the constrained DOFs $\ulWT{c}_{\rm r}$ on the boundary from $\ulWT{c}_{\rm f}$. This can be expressed in the following symbolic notation
\begin{align}
    \ulWT{c} = 
    \begin{bmatrix}
        \ulWT{c}_{\rm p} \\
        \ulWT{c}_{\rm r} \\
        \ulWT{c}_{\rm i}
    \end{bmatrix}
    = 
    \begin{bmatrix}
        1 & 0 & 0 \\
        \ast & 0 & 0 \\
        0 & 0 & 1
    \end{bmatrix}
    \begin{bmatrix}
        1 & 0 \\
        0 & 0 \\
        0 & 1
    \end{bmatrix}
    \begin{bmatrix}
        \ulWT{c}_{\rm p} \\
        \ulWT{c}_{\rm i}
    \end{bmatrix}
    =\vcentcolon \ull{P}_{\rm per} \, \ull{P}_{\rm f} \, \ulWT{c}_{\rm f}\, ,
    \label{eq:c:reconstruction}
\end{align}
where $\ast$ encodes the periodicity constraints and $\ulWT{c}_i$ denotes the non-boundary DOFs inside the domain. Hence, one needs to solve the following system
\begin{align}
    \ull{P}_{\rm f}^{\mathsf{T}}\ull{P}_{\rm per}^{\mathsf{T}} \lb \Kbulk + \Kpara + \Kperp \rb \ull{P}_{\rm per} \ull{P}_{\rm f} \, \ulWT{c}_{\rm f} = \ull{P}_{\rm f}^{\mathsf{T}} \ull{P}_{\rm per}^{\mathsf{T}} \lb \ul{r}_\Omega + \ul{r}_\Vert + \ul{r}_\perp \rb
    \label{eq:FOM}
\end{align}
for the vector of free DOFs $\ulWT{c}_{\rm f}=\ull{P}_{\rm f}^\mathsf{T} \ulWT{c}\in\ffR^{d_{\rm f}}$. The full solution vector of the fluctuation field $\ulWT{c}$ is then recovered by reapplying $\ull{P}_{\rm f}$ and $\ull{P}_{\rm per}$ again using \cref{eq:c:reconstruction}.

\subsection{Linear homogenization: Effective properties and analytical bounds}
\label{Sec:homogenization}
\subsubsection{Geometry-specific characteristic quantities}
\label{Sec:geo_quant}
Following Assumption \labelcref{a:voronoi}, the considered spatial domain is given as a Voronoi tessellation with $N$ grains and characterized by its (bulk) volume $v_\Omega$ as well as the $n_{\rm F}$ GB facets with surface areas $A_i$ and normals $\fn_i$. In our model, the total volume $V$ is given as the sum of bulk volume $v_\Omega$ and the implicitly represented GB volume $v_{\rm GB}$. This yields the volume fractions of bulk $\vfbulk$ and GB $\vfgb$:
\begin{align}\color{rev}
    V &= v_\Omega + v_{\rm GB}, &
    A &= {\color{rev} \intop_{\rm GB} {\rm d} A} = \sum\limits_{i=1}^{n_{\rm F}} A_i, &
   v_{\rm GB} &= {\color{rev}\intop_{\rm GB} \intop_{-h}^h {\rm d} z\, {\rm d} A } =2hA\, ,&
   \vfbulk & = \frac{v_\Omega}{V}\, , &
   \vfgb & = \frac{v_{\rm GB}}{V}\, .
   \label{eq:vol_fracs}
\end{align}
Note that the volume fractions compensate for the additional GB volume and satisfy $\vfbulk + \vfgb = 1$.

In order to quantify the anisotropy in a microstructure, we further define a second-order structural tensor
\begin{align}
    \ull{S}_2 & {\color{rev} = \frac{1}{A} \intop_{\rm GB} \fn \otimes \fn \, {\rm d} A}= \sum\limits_{i=1}^{n_{\rm F}} \frac{A_i}{A} \ul{n}_i \otimes \ul{n}_i \in \mathrm{Sym}_{\geq 0}\lb\ffR^3\rb
    \label{eq:S2}
\end{align}
being the GB surface area. To give some intuition, the limits of this tensor for extreme but characteristic geometries are given: They are derived from the extension of the discrete sum of interface facets to a surface integral on a corresponding ellipsoid {\color{rev} defined by semi-axis  lengths $a_{x/y/z}$ (here $x/y/z$ stand for the principal axis of the ellipsoid defined by the columns of $\ull{Q}_{\rm S}$) which are gained from the diagonalization of the semi positive definite matrix $\ull{S}_2$ via}
{\color{rev}
\begin{align}
    \ull{S}_2 &= \ull{Q}_{\rm S} \ull{\Lambda}_{\rm S} \ull{Q}_{\rm S}^\mathsf{T}\, , &
    \ull{\Lambda}_{\rm S} & = \text{diag} \lb \frac{1}{a_x^2},\frac{1}{a_y^2},\frac{1}{a_z^2} \rb \, .
\end{align}}

\begin{figure}[h]
    \centering
    \begin{subfigure}[b]{0.25\textwidth}
        \centering
        \begin{tikzpicture}
        \node[anchor=south west, inner sep=0] (image) {\includegraphics[trim={3cm 1.5cm 3cm 1.5cm}, clip, width=\textwidth]{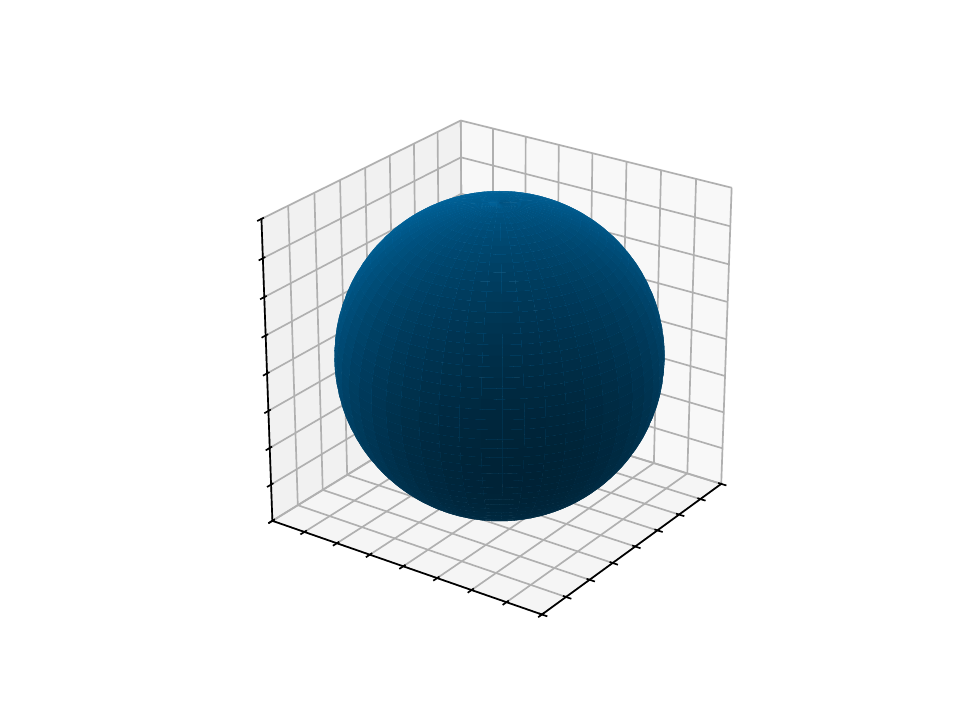}};
        \begin{scope}[xshift=1cm,x={(image.south east)},y={(image.north west)},scale=0.7]
            	\begin{scope}[z={(-0.866cm,-0.5cm)}, x={(0.866cm,-0.5cm)}, y={(0cm,1cm)}]
 		\draw[line width=1pt, arrows = {-Stealth[inset=0pt, length=10pt, angle'=40]}] (0,0,0) -- (1,0,0) node[below left] {$y$};
 		\draw[line width=1pt, arrows = {-Stealth[inset=0pt, length=10pt, angle'=40]}] (0,0,0) -- (0,1,0) node[below left, xshift=-3pt] {$z$};
 		\draw[line width=1pt, arrows = {-Stealth[inset=0pt, length=10pt, angle'=40]}] (0,0,0) -- (0,0,1) node[below right] {$x$};
        \end{scope}
        \node[xshift=2cm,yshift=-0.5cm,fill=white, rectangle,inner sep=3pt] {$a_x=a_y=a_z$};
        \node[xshift=1cm,yshift=-1.1cm,fill=white, rectangle,inner sep=3pt] {(a)};
 	\end{scope}
        \end{tikzpicture}
    \end{subfigure}
    \hfill
    \begin{subfigure}[b]{0.25\textwidth}
        \centering
                \begin{tikzpicture}
        \node[anchor=south west, inner sep=0] (image) {\includegraphics[trim={3cm 1.5cm 3cm 1.5cm}, clip, width=\textwidth]{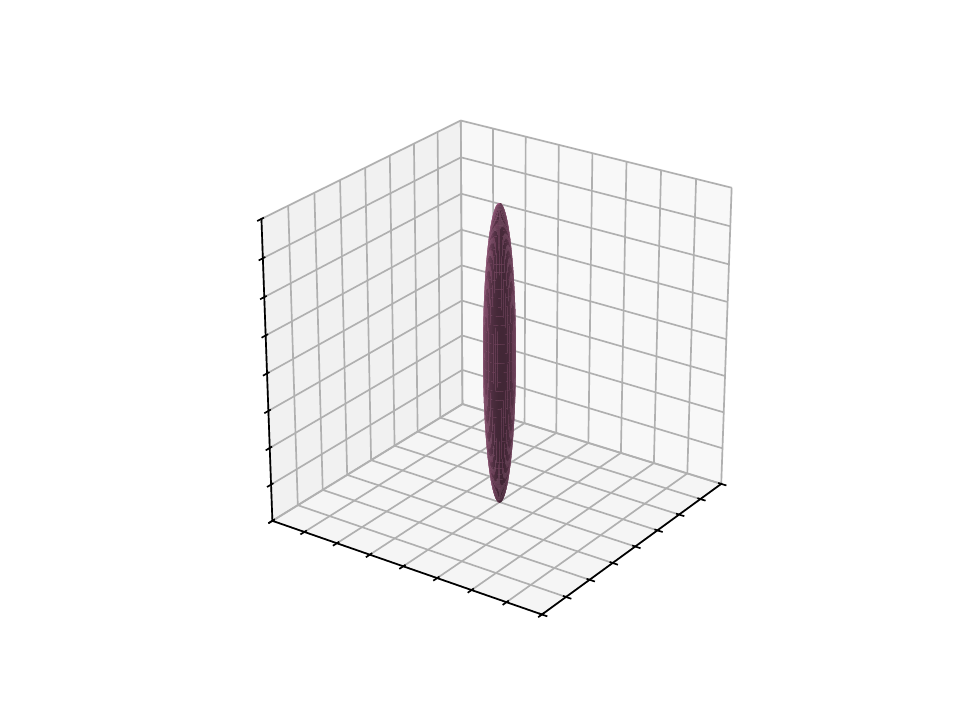}};
        \begin{scope}[xshift=1cm,x={(image.south east)},y={(image.north west)},scale=0.7]
            	\begin{scope}[z={(-0.866cm,-0.5cm)}, x={(0.866cm,-0.5cm)}, y={(0cm,1cm)}]
 		\draw[line width=1pt, arrows = {-Stealth[inset=0pt, length=10pt, angle'=40]}] (0,0,0) -- (1,0,0) node[below left] {$y$};
 		\draw[line width=1pt, arrows = {-Stealth[inset=0pt, length=10pt, angle'=40]}] (0,0,0) -- (0,1,0) node[below left, xshift=-3pt] {$z$};
 		\draw[line width=1pt, arrows = {-Stealth[inset=0pt, length=10pt, angle'=40]}] (0,0,0) -- (0,0,1) node[below right] {$x$};
        \node[xshift=1cm,yshift=-1.1cm,fill=white, rectangle,inner sep=3pt] {(b)};
        \end{scope}
        \node[xshift=2cm,yshift=-0.5cm,fill=white, rectangle,inner sep=3pt] {$a_x=a_y \ll a_z$};
 	\end{scope}
    \end{tikzpicture}
    \end{subfigure}
    \hfill
    \begin{subfigure}[b]{0.25\textwidth}
        \centering
                \begin{tikzpicture}
        \node[anchor=south west, inner sep=0] (image) {\includegraphics[trim={3cm 1.5cm 3cm 1.5cm}, clip, width=\textwidth]{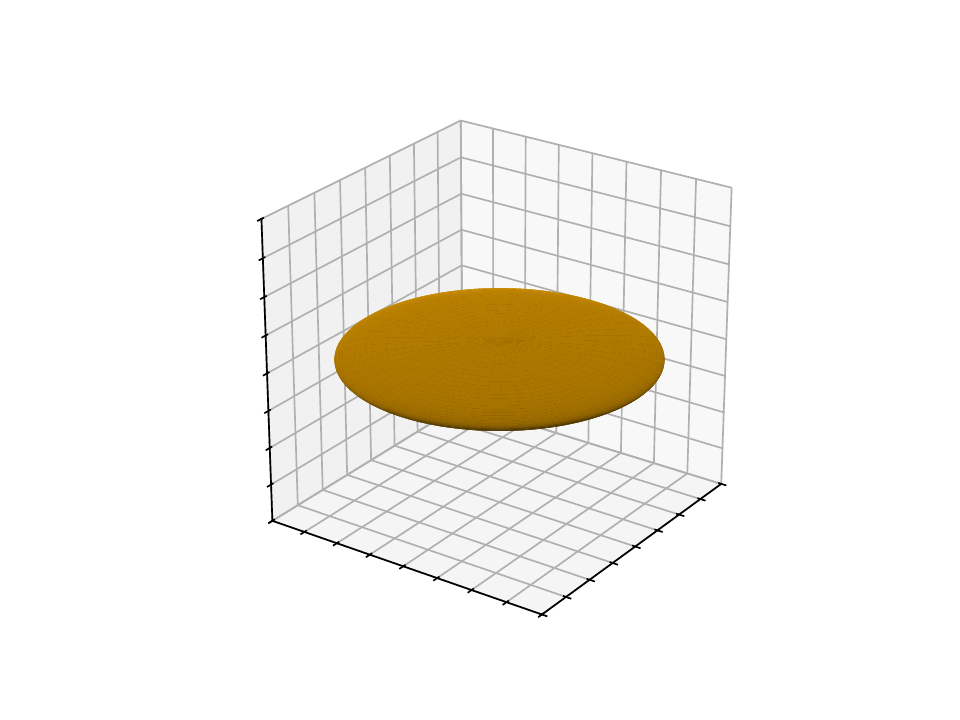}};
        \begin{scope}[xshift=1cm,x={(image.south east)},y={(image.north west)},scale=0.7]
            	\begin{scope}[z={(-0.866cm,-0.5cm)}, x={(0.866cm,-0.5cm)}, y={(0cm,1cm)}]
 		\draw[line width=1pt, arrows = {-Stealth[inset=0pt, length=10pt, angle'=40]}] (0,0,0) -- (1,0,0) node[below left] {$y$};
 		\draw[line width=1pt, arrows = {-Stealth[inset=0pt, length=10pt, angle'=40]}] (0,0,0) -- (0,1,0) node[xshift=-3pt,below left] {$z$};
 		\draw[line width=1pt, arrows = {-Stealth[inset=0pt, length=10pt, angle'=40]}] (0,0,0) -- (0,0,1) node[below right] {$x$};
        \node[xshift=1cm,yshift=-1.1cm,fill=white, rectangle,inner sep=3pt] {(c)};
        \end{scope}
        \node[xshift=2cm,yshift=-0.5cm,fill=white, rectangle,inner sep=3pt] {$a_x=a_y \gg a_z$};
 	\end{scope}
    \end{tikzpicture}
    \end{subfigure}
    \caption{Limit cases of a spheroid with semi-axis lengths $a_x, a_y, a_z$: (a) sphere, (b) prolate (needle-like), and (c) oblate (flake-like).}
    \label{fig:spheroids}
\end{figure}
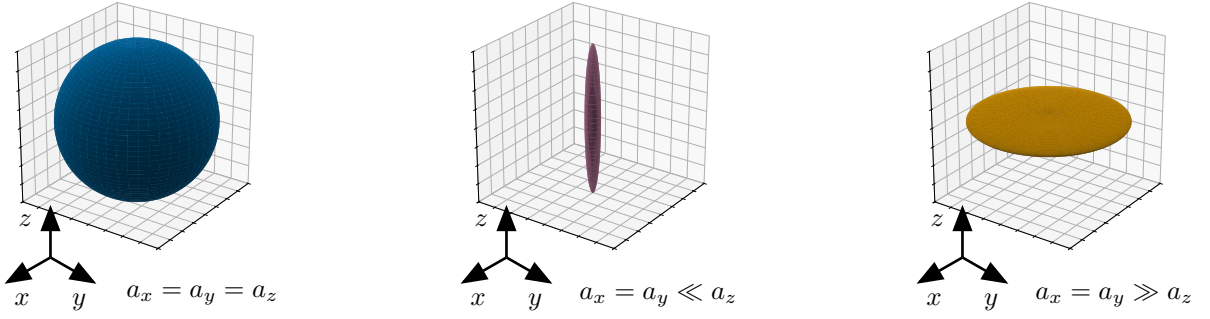

For reference, the results for perfectly isotropic geometries (as it appears in the limit case of $N\rightarrow\infty$), prolate ``needle''-like grains, and oblate ``flake-like'' grains are given:
\begin{align}
    \ull{S}_2^{\rm sphere} =
    \begin{bmatrix}
        1/3 & 0 & 0 \\
        0 & 1/3 & 0 \\
        0 & 0 & 1/3
    \end{bmatrix}\, , &&
    \ull{S}_2^{\rm prolate} \rightarrow
    \begin{bmatrix}
        1/2 & 0 & 0 \\
        0 & 1/2 & 0 \\
        0 & 0 & 0
    \end{bmatrix}\, , &&
    \ull{S}_2^{\rm oblate} \rightarrow 
    \begin{bmatrix}
        0 & 0 & 0 \\
        0 & 0 & 0 \\
        0 & 0 & 1
    \end{bmatrix}\, .
\end{align}
The corresponding spheroids are {\color{rev} defined as the isosurface
\begin{align}
\ul{x} \in \ffR^3 \text{ s. th. } \ul{x}^\mathsf{T} \ull{S}_2 \ul{x} &= 1 \, .
\end{align}
The isosurfaces for the three given examples are shown in \Cref{fig:spheroids}.} Note that the prolate and oblate limit cases correspond to the ratio of the semi-axes tending to infinity.

\subsubsection{Effective diffusion coefficients}
\label{Sec:eff}
{\color{rev}In the following, we derive accurate results for the effective diffusivity based on the obtained concentration fields. Following the framework of linear computational homogenization \cite{Ostoja-Starzewski2011}, we introduce the phase-wise volume averages of an arbitrary quantity $[\cdot]$ in bulk and GB building on the spatial average notation from \Cref{Sec:notation}:}
\begin{align}
    \langle[\cdot]\rangle_\Omega &\coloneqq \frac{1}{v_\Omega} \int\limits_\Omega [\cdot] \,\mathrm{d}V\, , \\
    \langle[\cdot]\rangle_\Vert &\coloneqq \frac{1}{v_{\rm GB}} \int\limits_{A}\int\limits_{-h}^h \lb\ull{I}-\ul{n}\otimes\ul{n}\rb[\cdot]\,\mathrm{d}\zeta_{\rm n}\,\mathrm{d}S\, , &&
    \langle[\cdot]\rangle_\perp \coloneqq \frac{1}{v_{\rm GB}} \int\limits_{A}\int\limits_{-h}^h \lb\ul{n}\otimes\ul{n}\rb[\cdot]\,\mathrm{d}\zeta_{\rm n}\,\mathrm{d}S\, .
\end{align}
For an arbitrary geometry and set of material parameters the global system in \cref{eq:FOM} is solved for different right-hand sides induced by the imposed unit gradients $\ol{\ul{g}}_k = \ul{e}_k$ for $k\in\{1,2,3\}$ to obtain the corresponding fluctuation fields. The phase-wise volume averages $\langle\ul{g}_k\rangle_{[\cdot]}$ have contributions from the imposed gradient field and the resulting fluctuation field:
\begin{align}
    \langle \ul{g}_k \rangle_\Omega = \ol{\ul{g}}_k + \langle \nabla \WT{c}_k \rangle_\Omega\, , 
    && \langle \ul{g}_k \rangle_\Vert = (\ull{I}-\ull{S}_2)\ol{\ul{g}}_k + \langle \nabla \WT{c}_k \rangle_\Vert\, ,
    && \langle \ul{g}_k \rangle_\perp = \langle \nabla \WT{c}_k \rangle_\perp\, .
    \label{eq:int_grad}
\end{align}
Note that $\langle\ul{g}_k\rangle_\perp$ is not affected by the imposed gradient since all nodes in the thickness direction of the interface share the same physical position due to the collapsed interface representation. To compute the averaged gradients of the fluctuation fields, we define phase-wise operators $\ffG_{[\cdot]}\in\ffR^{3\times d}$ to obtain the gradient field integrated over the domain. These will be of use later. This results in
\begin{align}
    \langle \nabla \ulWT{c}_k \rangle_\Omega &= \frac{1}{v_\Omega}\Gbulk \ulWT{c}_k & \text{with } \Gbulk &= \sum\limits_{i=1}^{n_{\rm el}^\Omega} \sum\limits_{j=1}^{n_{\rm GP}^{\rm T10}} \omega_j^{(i)} \ull{B}^{(i)}_\Omega(\ul{\zeta}_j) \ull{L}_\Omega^{(i)}\, , 
    \label{eq:G_bulk} \\
    \langle \nabla \ulWT{c}_k \rangle_\Vert &= \frac{1}{v_{\rm GB}}\Gpara \ulWT{c}_k & \text{with } \Gpara &= \frac{2h}{6} \sum\limits_{i=1}^{n_{\rm el}^{\rm GB}} \sum\limits_{j=1}^{n_{\rm GP}^{\rm C18}} \nu_j^{(i)} 
    \begin{bmatrix}
        \ull{B}_{\rm P2}^{(i)}(\ul{\zeta}_{\rm{t}(j)}) & 4\ull{B}^{(i)}_{\rm P2}(\ul{\zeta}_{\rm{t}(j)}) & \ull{B}^{(i)}_{\rm P2}(\ul{\zeta}_{\rm{t}(j)})
    \end{bmatrix} \ull{L}_{\rm GB}^{(i)}\, ,
    \label{eq:G_para} \\
    \langle \nabla \ulWT{c}_k \rangle_\perp &= \frac{1}{v_{\rm GB}}\Gperp \ulWT{c}_k & \text{with } \Gperp &= \sum\limits_{i=1}^{n_{\rm el}^{\rm GB}} \sum\limits_{j=1}^{n_{\rm GP}^{\rm C18}} \nu_j^{(i)}\ul{n}_i \begin{bmatrix}
        -\ul{N}_{\rm P2}^{(i)}(\ul{\zeta}_{\rm{t}(j)}) & \ul{0} & \ul{N}_{\rm P2}^{(i)}(\ul{\zeta}_{\rm{t}(j)})
    \end{bmatrix} \ull{L}_{\rm GB}^{(i)}\, .
    \label{eq:G_perp}
\end{align}
The overall volume average of the microscopic gradients and fluxes based on $\ol {\ul{g}}_k$ are then given as
\begin{align}
    \ul{G}_k &= \vfbulk\langle \ul{g}_k \rangle_\Omega + \vfgb \lb \langle \ul{g}_k \rangle_\Vert + \langle \ul{g}_k \rangle_\perp \rb\, , \\
    \ul{Q}_k &= - \lb \vfbulk D_\Omega \langle \ul{g}_k \rangle_\Omega +  \vfgb\lb D_\Vert \langle \ul{g}_k \rangle_\Vert + D_\perp \langle \ul{g}_k \rangle_\perp\rb\rb\, .
\end{align} 
Stacking $\ul{G}_k$ and $\ul{Q}_k$ for all three load cases ($k=1,2,3$),
\begin{align}
    \ull{G} = \begin{bmatrix}
        \ul{G}_1 & \ul{G}_2 & \ul{G}_3
    \end{bmatrix} \in\ffR^{3\times 3}\, , &&
    \ull{Q} = \begin{bmatrix}
        \ul{Q}_1 & \ul{Q}_2 & \ul{Q}_3
    \end{bmatrix} \in\ffR^{3\times 3}\, ,
    \label{eq:Qk}
\end{align}
yields the effective diffusion tensor on the mesoscale
\begin{align}
    \ol{\ull{D}} = -\ull{Q} \, \ull{G}^{-1} \in\ffR^{3\times 3}\, .
    \label{eq:Deff}
\end{align}
In the case where $D_\Omega=D_\Vert=D_\perp$ the definition of the volume fractions in \cref{eq:vol_fracs} compensates the additional GB volume and yields $\ull{Q}_k = -D_\Omega\ull{G}_k$ and therefore $\ol{\ull{D}}=D_\Omega \ull{I}$. Hence, the GB does not influence the response in this case.

\subsubsection{Voigt--Reuss bounds}
\label{Sec:Voigt-Reuss}
Analytical bounds for the effective material response, i.e., the coefficients of the effective diffusion tensor $\ull{D}$, exist. They are given by the weighted arithmetic and harmonic mean of the phase-specific material properties. In the following we refer to them as (upper) Voigt \cite{Voigt1889} and (lower) Reuss \cite{Reuss1929} bound $\ull{D}_{\rm V}$ and $\ull{D}_{\rm R}$ as commonly done in mechanics\footnote{Field-specific alternative terms for the same bounds exist, such as the Wiener bounds \cite{Wiener1912} in permittivity {\color{rev}or a diffusion-specific equivalent to the Voigt bound \cite{Hart1957}.}}. With the assumptions of an isotropic bulk (\Cref{Sec:bulk:diff}) and a transversely isotropic GB (Assumption \labelcref{A:D_GB}) as well as the geometry-specific characteristic quantities $\vfbulk, \vfgb, \ull{S}_2$ from \Cref{Sec:geo_quant} expressions for both bounds can be derived:
\begin{align}
    \ull{D}_{\rm V} = \vfbulk D_\Omega \ull{I} + \vfgb D_\Vert \lb \frac{D_\perp - D_\Vert}{D_\Vert} \ull{S}_2 + \ull{I} \rb\, , && 
    \ull{D}_{\rm R} = \lb \frac{\vfbulk}{D_\Omega}\ull{I} + \frac{\vfgb}{D_\Vert} \lb \frac{D_\Vert - D_\perp}{D_\perp} \ull{S}_2 + \ull{I} \rb\rb^{-1}\, .
    \label{eq:voigt}
\end{align}
{\color{rev} Note the presence of the structural tensor through $\ull{S}_2$ which captures the mean inclusion shape.}
Due to the transversely isotropic properties of the GB the bounds exhibit a dependence on the structural tensor $\ull{S}_2$ as defined in \cref{eq:S2}, i.e., the anisotropy of the geometry governs the anisotropy of the bounds. For $D_\Omega=D_\Vert=D_\perp$ the bounds yield a sharp estimate with $\ull{D}_V=\ull{D}_R$. In the limit case of an isotropic structure ($N\to\infty$) the bounds converge to
\begin{align}
    \ull{D}_{\rm V} \to \lb \vfbulk D_\Omega +  \vfgb\frac{D_\perp + 2 D_\Vert}{3} \rb \ull{I}\, , &&
    \ull{D}_{\rm R} \to \lb \frac{\vfbulk}{D_\Omega} + \vfgb \frac{D_\Vert + 2D_\perp}{3D_\perp D_\Vert} \rb^{-1} \ull{I}\, .
    \label{eq:reuss}
\end{align}

In order to derive a relative scaling of the effective diffusion tensor $\ol{\ull{D}}$ with respect to the Voigt--Reuss bounds, we follow the derivation introduced in the context of the Voigt--Reuss net~\cite{Keshav2025}. It is also roughly outlined in the following: The effective diffusion tensor $\ol{\ull{D}}$ for an arbitrary configuration needs to satisfy the derived Voigt and Reuss bounds in the following sense:
\begin{align}
    \ul{y}^\mathsf{T} \lb \ull{D}_{\rm V} - \ull{D}_{\rm R} \rb \ul{y} \geq \ul{y}^\mathsf{T} \lb \ol{\ull{D}} - \ull{D}_{\rm R} \rb \ul{y} \geq 0\, , && \forall \ul{y} \in\ffR^3\, .
\end{align}
Performing a Cholesky decomposition on the symmetric and positive-definite difference $\ull{D}_{\rm V} - \ull{D}_{\rm R}$ 
\begin{align}
    \ull{D}_{\rm V} - \ull{D}_{\rm R} = \ull{L}\,\ull{L}^\mathsf{T}\,  && \text{i.e., } \ull{L} = \lb \ull{D}_{\rm V} - \ull{D}_{\rm R} \rb^{1/2}\, ,
\end{align}
allows rewriting the bounding relations as
\begin{align}
    \ul{z}^\mathsf{T} \ul{z} \geq \ul{z}^\mathsf{T} \ull{L}^{-\mathsf{T}} \lb \ol{\ull{D}} - \ull{D}_{\rm R} \rb \ull{L}^{-1} \ul{z} \geq 0\, ,
\end{align}
if $\ul{y} = \ull{L}^{-1}\ul{z}$ with $\ul{z}\in\ffR^3$ is chosen. This can only be satisfied if the following eigendecomposition with restrictions exists:
\begin{align}
    \ull{L}^{-\mathsf{T}}\lb\ol{\ull{D}} - \ull{D}_{\rm R}\rb \ull{L}^{-1} = {\color{rev}\ull{U}^\mathsf{T}} \ull{\Lambda}\, {\color{rev}\ull{U}} && \text{with } \ull{\Lambda} = \rm{diag}\lb \lambda_i\rb\, , ~~ \lambda_i \in [0,1] ~~ (i\in\{1,2,3\})\, .
    \label{eq:voigt-reuss_relation}
\end{align}
The eigenvalues $\lambda_i$ might be used as a direct scalar measure to describe the relation of the material response to Voigt (close to 1) and Reuss (close to 0). This will be used for the study in \Cref{Sec:results:eff}. Note that the presented parametrization is inverse to the one in the Voigt--Reuss net \cite{Keshav2025}.

\subsection{Affine parameter dependence and dimensionless formulation}
\label{Sec:affine}
The system of equations stated in \cref{eq:FOM} exhibits a dependence on the diffusion coefficients $D_\Omega$, $D_\Vert$, and $D_\perp$ as well as the GB thickness parameter $h$. In the following, we further parameterize the geometry with respect to a reference geometry of edge length $l_0$ and introduce a uniform scaling by $l/l_0$. For a reference diffusion coefficient $D_0$ and the reference length $l_0$, the following set of five dimensionless scaling factors is obtained:
\begin{align}
    \pi_\Omega = \frac{D_\Omega}{D_0}\, , && \pi_\Vert = \frac{D_\Vert}{D_0}\, , &&  \pi_\perp = \frac{D_\perp}{D_0}\, , && \pi_h = \frac{h}{l_0}\, , && \pi_l = \frac{l}{l_0}\, .
    \label{eq:5d_param}
\end{align}

Interestingly, the Buckingham Pi Theorem \cite{Buckingham1914} yields only two independent variables for this five-dimensional parametrization{\color{rev}. By choosing} the two intrinsic parameters,
\begin{align}
    \Pi_\Vert &= \frac{\pi_\Vert}{\pi_\Omega}\frac{\pi_h}{\pi_l}\, , & \Pi_\perp &= \frac{\pi_\perp}{\pi_\Omega}\frac{\pi_l}{\pi_h}\, , \label{eq:Pi:intrinsic}
\end{align}
for each configuration $[\pi_\Omega, \pi_\Vert, \pi_\perp, \pi_h, \pi_l]$ the system of equations defined in \cref{eq:FOM} can be rewritten as
\begin{align}
    \left( \KbulkO+ \Pipara \KparaO + \Piperp \KperpO \right) \ulWT{c} = \ul{r}_\Omega^0 + \Pipara \ul{r}_\Vert^0 + \Piperp\ul{r}_\perp^0\, .
    \label{eq:affine_system}
\end{align}
Quantities from the reference geometry are indicated by the superscript $0$. A more detailed derivation can be found in \Cref{app:affine}.

{\color{rev}Hence, solving the system in \cref{eq:affine_system} for a given configuration $[\Pi_\Vert,\Pi_\perp]$ yields an entire 3D submanifold in the parametric space that shares the same discrete solution $\ulWT{c}$ for the fluctuation field. The submanifold is defined by \cref{eq:Pi:intrinsic}. In particular, this implies that the solution is not affected by:
\begin{itemize}
    \item the magnitude of $\pi_\Omega$, $\pi_\Vert$ and $\pi_\perp$ but rather their relation to each other,
    \item the magnitude of the geometrical features $\pi_h$ and $\pi_l$, i.e., a uniform scaling of the geometry length and GB thickness,
    \item a change in the GB volume fraction ($\pi_h/\pi_l$) if the effect is compensated by $\pi_\Omega$, $\pi_\Vert$ and $\pi_\perp$ (and vice versa).
\end{itemize}}

While the fluctuation field and the corresponding integrated gradients in \cref{eq:int_grad} have a lower intrinsic dimension, this does not hold for the flux field and the effective diffusivity tensor (see \Cref{app:affine:eff}). They scale directly with the absolute values of the phase-specific diffusion coefficients and volume fractions, which depend on the ratio of $\pi_h/\pi_l$. However, the computational effort for these postprocessing steps is linear and, therefore, insignificant as it reduces to scalar pre-factors, given the integrated gradients. The number of system configurations for which \cref{eq:affine_system} needs to be solved is considerably reduced, which limits the overall computational effort.

\begin{remark}
    The lengths $l_0$ and $l$ represent the edge length of the unit cell used. They directly relate to the grain size. By keeping $\pi_\Omega, \pi_\Vert, \pi_\perp$ and $\pi_h$ constant but varying $\pi_l$ a direct investigation of the grain size is possible.
\end{remark}


\section{A 2D example for validation}
\label{Sec:2d_validation}

\subsection{Setup}
\begin{figure}[h]
    \centering
    \includegraphics[width=0.95\textwidth]{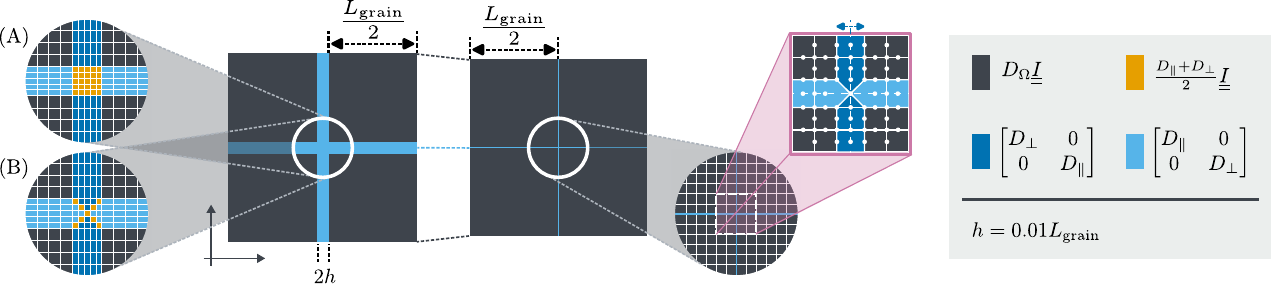}
    \caption{Sketch of the compared setups using a fully-resolved GB (left) and the proposed collapsed representation (right). Geometrical features and assigned material properties are given as well. For the fully resolved case, the two possible parametrizations of the junction domain (A) and (B) are shown. In this case, the GB width $2h$ is chosen to be two percent of the grain size $L_{\rm grain}$.}
    \label{fig:2d_sketch}
\end{figure}

To verify the choice of a second-order approximation for the concentration profile across the GB, a simplified 2D example is considered. In this case, a fully resolved representation of the GB domain is feasible and can be used as a reference solution. As depicted in \Cref{fig:2d_sketch}, the geometry consists of a square-shaped grain of sides with length $L_{\rm grain}$ in a periodic setting that is separated from itself by a GB layer of thickness $2h$.

In the following, we aim to compare the resulting fluctuation fields as well as the effective properties for the collapsed interface elements and for a fully resolved GB. Note that a fixed bulk volume yields different total geometry volumes (and, therefore, volume fractions) due to the implicit representation of the GB in the collapsed case:
\begin{align}
    v_{\rm bulk}^{\rm full} &= L_{\rm grain}^2\, , & v_{\rm GB}^{\rm full} &= 2\lb L_{\rm grain}\cdot 2h\rb + \lb 2h \rb^2\, , & v^{\rm full}_{\rm total} &= L_{\rm grain}^2 + 4L_{\rm grain} h + 4h^2\, , \\
    v_{\rm bulk}^{\rm coll} &= L_{\rm grain}^2\, , & v_{\rm GB}^{\rm coll} &= 2\lb L_{\rm grain} \cdot 2h\rb\, , & v_{\rm total}^{\rm coll} &= L_{\rm grain}^2 + 4L_{\rm grain}h\, .
    \label{eq:2d_vol}
\end{align}
{\color{rev} The volume of the GB and, thus, of the total domain shows a qualitative error in $\cO(h^2)$ due to potential GB gaps/overlaps at junctions. This theoretical mismatch is recovered in the above quantities.}

Further, the parametrization of the GB junction at the center is non-trivial in the fully resolved case. Hence, we investigate two reasonable variants as shown on the left of \Cref{fig:2d_sketch}:
\begin{enumerate}[label=(\Alph*)]
    \item isotropic diffusion behavior in the junction domain (i.e., arithmetic average of $D_\Vert$ and $D_\perp$),
    \item junction domain divided into triangles with properties inherited from neighboring GB segments.
\end{enumerate}

With the values given in \Cref{fig:2d_sketch}, the volume fractions for the fully-resolved and the collapsed setup are $f_{\rm GB}^{\rm full}\approx 3.88\%$ and $f_{\rm GB}^{\rm coll}\approx 3.85\%${\color{rev}, which confirms that the error on the order of $\cO(h^2)$ due to the collapsed instead of the fully resolved interface is negligible for realistic applications with $h\ll L_{\rm grain}$, justifying the validity of the chosen modeling approach.}

Following the model description above, the problem is complemented with periodic boundary conditions, and one node is fixed to an arbitrary concentration. We impose a macroscopic gradient $\ol{\ul{g}}$ as a driving force.

\subsection{Approximation quality of the proposed concentration profile within the interface}
\begin{figure}[h]
    \centering
    \includegraphics[width=\textwidth]{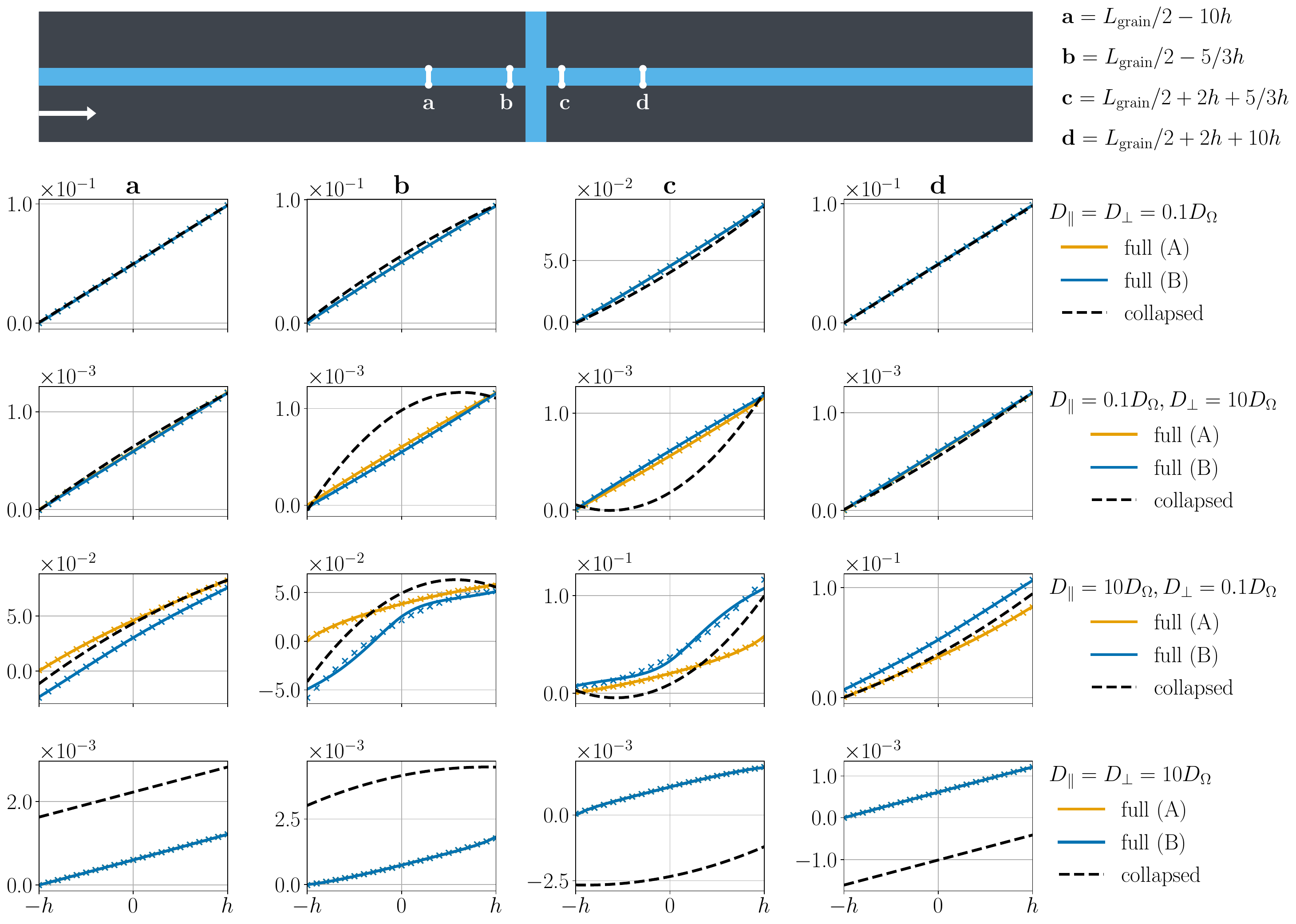}
    \caption{Concentration profiles across the GB in the positions \textbf{a}, \textbf{b}, \textbf{c} and \textbf{d} highlighted in the schematic at the top: Each row represents a different parameter configuration ($D_\Vert/D_\Omega$, $D_\perp/D_\Omega$). Least-squares second-order polynomial fits for the setups (A) and (B) (see \Cref{fig:2d_sketch}) are given by the $\times$-markers in the respective color. For the isotropic cases (top and bottom row), setups (A) and (B) coincide. The matching results for the proposed collapsed GB representation are shown for reference. For an easy comparison, the absolute concentration values were shifted such that the concentration value of setup (A) at position $-h$ resembles the zero value. Note that the scaling of the vertical concentration axes differs between the subplots. }
    \label{fig:2d_c_GB_dev_profile}
\end{figure}

In the following, we investigate whether the assumed second-order ansatz is sufficient to capture the relevant features of the concentration profile across the GB and how well the collapsed model approximates the fully resolved results. We choose the gradient $\ol{\ul{g}}=\begin{bmatrix} 4/5 & 3/5 \end{bmatrix}^\mathsf{T}$ to analyze the transversely isotropic GB parametrization. In \Cref{fig:2d_c_GB_dev_profile}, the concentration profiles for all three approaches at characteristic positions close and further away from the junction are shown. In addition, a least-squares fit of a second-order polynomial is added for the two fully resolved cases (A) and (B).

The choice of the parametrization for the junction clearly affects the resulting concentration profile in the anisotropic case. While the profiles resulting from approach (A) are perfectly captured by a second-order fit in all cases, setup (B) exhibits higher-order nonlinearity close to the junction. Hence, considering at least one term of even order is essential to properly capture the diffusion within the GB, while neglecting higher-order terms is a reasonable assumption.

The fully resolved concentration profiles are compared with the second-order reconstruction of the collapsed representation. Note that all subplots in \Cref{fig:2d_c_GB_dev_profile} exhibit a different scaling of the concentration axis. Further, the deviation from the concentration value of setup (A) at position $-h$ is displayed rather than the actual concentration values to allow for a better comparison. The jump in concentration captured by the proposed model aligns very well with the one of setup (B). However, deviations between both models can be observed at the junction close to the GB center line (cases \textbf{b}, \textbf{c}). For the direct vicinity of the junction, the collapsed interface elements cause inaccuracies by construction: The entire behavior within the junction is only modeled by a single shared middle plane node. In setups with large concentration jumps (e.g., caused by $D_\perp\ll D_\Omega$), this node will most likely represent an average of the surrounding top and bottom nodes of all affected interface elements. On average, this expected behavior is mitigated since an overestimation at position \textbf{b} will always cause an underestimation at position \textbf{c} due to the point symmetry of the concentration profiles.

It is noteworthy that Fisher-like transport models \cite{Fisher1951} with reduced representations cannot capture the large concentration jumps across the GB nor the detailed concentration profile. Further, the independent variation of $D_\Vert$ and $D_\perp$ clearly affects the resulting concentration field. Hence, the restriction to a single diffusion coefficient for the GB domain (either to account for jumps or for in-plane transport) is an oversimplifying assumption. The detailed effects will be studied in \Cref{Sec:results}.

\subsection{Approximation accuracy in effective properties}
To obtain the effective diffusion tensor $\ol{\ull{D}}$, we keep the geometrical setup from \Cref{fig:2d_sketch} fixed and solve the described diffusion problem for the two uniaxial load cases $\ol{\ul{g}}_k$ for $k\in\{1,2\}$. Due to the symmetric setup, we find $\ol{\ull{D}}=\ol D \ull{I}$. As before, the ratios of material parameters $D_\Vert/D_\Omega$ and $D_\perp/D_\Omega$ are varied. 

\begin{figure}[h]
    \centering
    \begin{subfigure}[b]{0.32\textwidth}
        \centering
        \includegraphics[width=\textwidth]{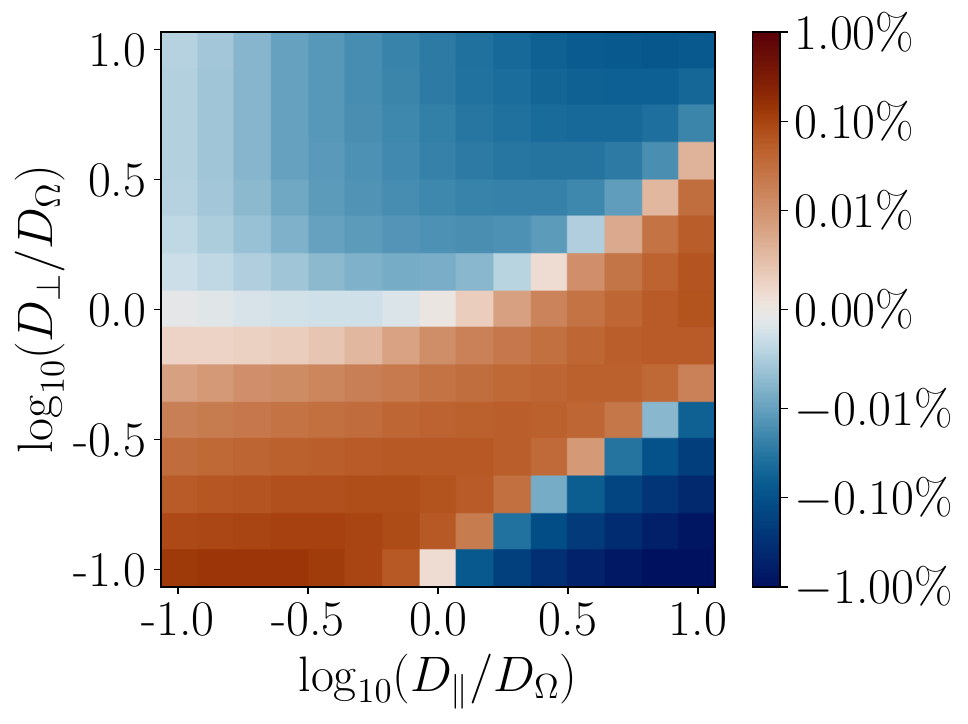}
        \caption{relative deviation from setup (A)}
        \label{fig:2d_D_eff_dev_avg}
    \end{subfigure}
    \hfill
    \begin{subfigure}[b]{0.32\textwidth}
        \centering
        \includegraphics[width=\textwidth]{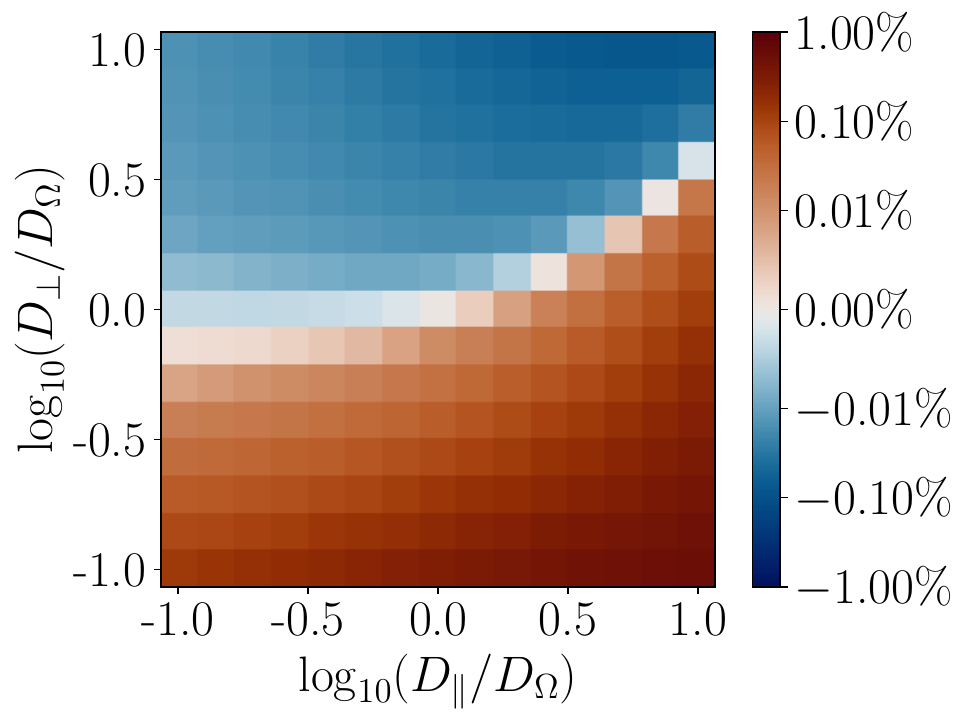}
        \caption{relative deviation from setup (B)}
        \label{fig:2d_D_eff_dev_tri}
    \end{subfigure}
    \hfill
    \begin{subfigure}[b]{0.3\textwidth}
        \centering
        \includegraphics[width=\textwidth]{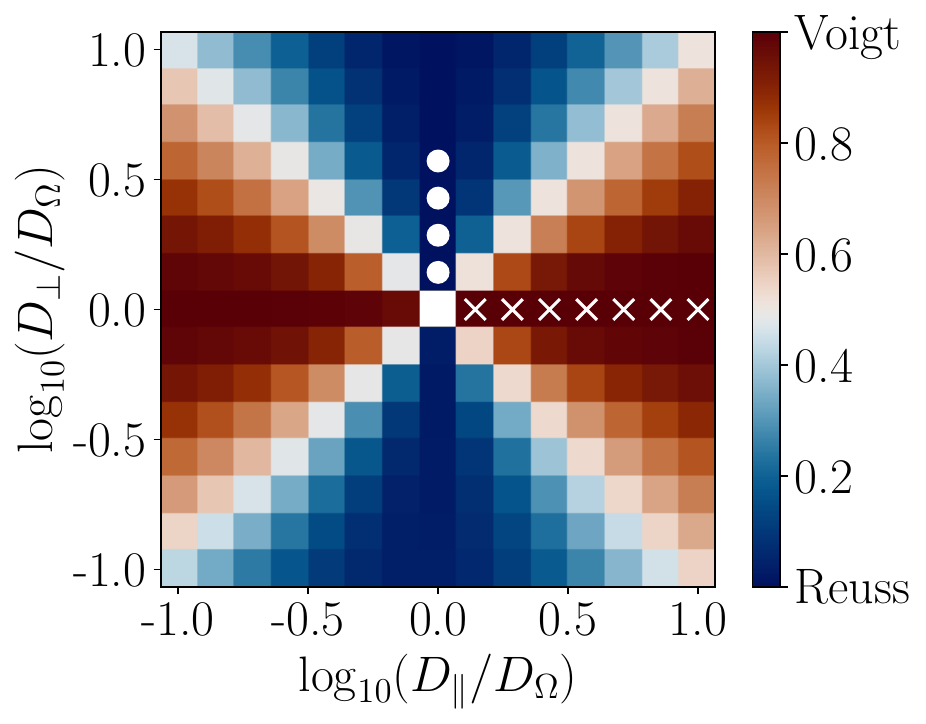}
        \caption{relative relation to bounds}
        \label{fig:2d_voigt_reuss_scaling}
    \end{subfigure}
    \caption{(a, b) Relative deviation in the effective diffusion coefficient $\ol{D}$ of the proposed model with respect to the fully resolved setups (A) and (B). Positive numbers indicate an overestimation, and negative numbers indicate an underestimation. For the trivial case of $D_\Vert=D_\perp=D_\Omega$ perfect agreement is found. (c) Relation of the effective diffusion coefficient from the proposed model to the Voigt--Reuss bounds. Violations of the Voigt and Reuss bounds are marked with $\times$ and $\circ$, respectively.}
\end{figure}

For the case of $D_\Omega=D_\Vert=D_\perp$, the effective diffusion coefficient $\ol{D}$ is found to perfectly recover a homogeneous material. For the sake of brevity, we omit a detailed interpretation of the dependence of $\ol{D}$ on the parameter space here and refer to \Cref{Sec:results:eff} instead. Only the deviation in the effective diffusion coefficient between the fully-resolved setups (A) and (B) and the collapsed GB representation is considered in \Cref{fig:2d_D_eff_dev_avg} and \ref{fig:2d_D_eff_dev_tri}, where a sound agreement was found (deviations less than 1\%). The deviations of both setups are similar except for the domain where $D_\Vert > D_\Omega$ and $D_\perp < D_\Omega$: \Cref{fig:2d_D_eff_dev_avg} mildly underestimates $\ol D$ while \Cref{fig:2d_D_eff_dev_tri} yields a slight overestimation. This implies that in this regime, both fully resolved setups differ, and the proposed model yields a meaningful in-between estimate. While the maximum deviation in (A) is larger than in (B), it exhibits better agreement on average.

For the given symmetric 2D setup, the relation to the Voigt--Reuss bounds as stated in \eqref{eq:voigt-reuss_relation} reduces to:
\begin{align}
    \lambda\lb\ol D\rb = \frac{\ol D - D_{\rm R}}{D_{\rm V} - D_{\rm R}}\, .
\end{align}
We recall that $\lambda=1$ corresponds to $\ol D=D_{\rm V}$ and $\lambda=0$ implies that $\ol D = D_{\rm R}$. Note that the Reuss bounds differ between the fully-resolved setups (A) and (B) by up to $0.5\%$, whereas the Voigt bounds are identical. For the scaling of the collapsed setup, as shown in \Cref{fig:2d_voigt_reuss_scaling}, the Reuss bound of setup (A) was used. For some configurations, slight (numerical) violations of the bounds are observed. This can be explained by the required approximations resulting from the proposed collapsed element and the corresponding artificially added volume, as well as the ambiguity of the ``correct'' Reuss bound {\color{rev} due to setup (A) vs. (B)}. However, these violations remain in a negligible range of less than $0.1$\%.

In summary, the validation example proves that the collapsed interface model resembles the fully resolved setups well in a global sense. Local deviations appear mainly close to the junction, where they are expected by construction. However, in these places, the results for the parametrizations (A) and (B) differ as well. For the computation of the effective diffusion coefficients, the volume averaging removes the majority of the point symmetric deviations and leads to a very good agreement with the fully resolved setups. {\color{rev} Most notably, the deviation of the collapsed interface model to either (A) or (B) is on the order of their mutual discrepancy in the Reuss bound, which is an effect due to the unknown behavior in the junction region. Furthermore, no need for quartic (or higher-order) modes in the through-thickness direction is identified in the current study.} In the following, the second-order ansatz is, therefore, used as a justified approximation choice for the GB profile.


\section{Diffusion analysis of polycrystalline microstructures}
\label{Sec:results}

\subsection{Objectives}
For the present study, an arbitrary geometry is chosen, which is constructed as a Voronoi tessellation with 100 grains. Geometry and bulk mesh were generated using the software \texttt{neper}~\cite{Quey2022, Quey2011}. The interface elements were added by extending standard approaches for inserting cohesive elements~\cite{Nguyen2014}. 
{\color{rev}
The structural tensor of the considered geometry, as defined in \cref{eq:S2}, is
\begin{align}
    \ull S_2 \approx \begin{bmatrix}
        0.3352 & 0.0017 & 0.0002 \\
        0.0017 & 0.3293 & 0.0033 \\
        0.0002 & 0.0033 & 0.3355
    \end{bmatrix}\, ,
\end{align}
and its eigenvalues $\lambda_i$ show the ratios
\begin{align}
    \frac{\lambda_1}{\lambda_0} \approx 0.9936\, , && \frac{\lambda_2}{\lambda_0}\approx 0.9713\, .
    \label{eq:S2_setup}
\end{align}
These justify the choice of the setup as a representative microstructure with a nearly isotropic structure.}
The considered discretization consists of 1,181,055 volume elements for the bulk and 78,074 cohesive elements for the GBs, resulting in 1,951,614 {\color{rev} nodes}.

Note that the volume fraction of the setup can be varied artificially for a fixed geometry by controlling the relation between the GB thickness parameter and geometry length scale $h/l$, which is one of the key advantages of the collapsed GB representation. To comply with the assumption of a thin GB that forms the foundation of the introduced thickness-collapsed GB model, we focus on systems with a small GB volume fraction of $\vfgb < 10\%$. In this study, we consider $D_\Vert\in[0.01~D_\Omega, 100~D_\Omega]$ and $D_\perp\in[0.01~D_\Omega, 100~D_\Omega]$ to form a reasonable range of GB diffusion coefficient ratios.

Given the proposed model, we now aim to explore the qualitative and quantitative characteristics of the resulting atomic diffusion and their dependence on the parameters $\ul{\pi}$ as introduced in \cref{eq:5d_param}. An in-depth understanding of these effects requires insight into the local diffusion mechanisms as well as the emerging global diffusion regimes. Through meaningful metrics and supporting visualizations, we underline the features that can be captured by the proposed model. Special attention is given to the impact of GBs. The study is based on and structured by the following guiding questions:
\begin{enumerate}
    \item \textbf{{\color{rev}Classification} of diffusion {\color{rev}behavior}}: How do GB diffusivities deviating from the bulk diffusivity affect the preferred paths of atomic transport?
    \item \textbf{Characterization of diffusion in GBs}: How do local transport mechanisms alter the diffusion characteristics of materials with polycrystalline microstructures? 
    \item \textbf{Quantification of GB impacts on effective diffusivity}: To what extent can the consideration of GB transport enhance or limit the overall diffusivity of the polycrystalline material?
\end{enumerate}
Relevant aspects and findings are discussed in the subsequent sections. The previously introduced reference geometry is used for all of the presented results.

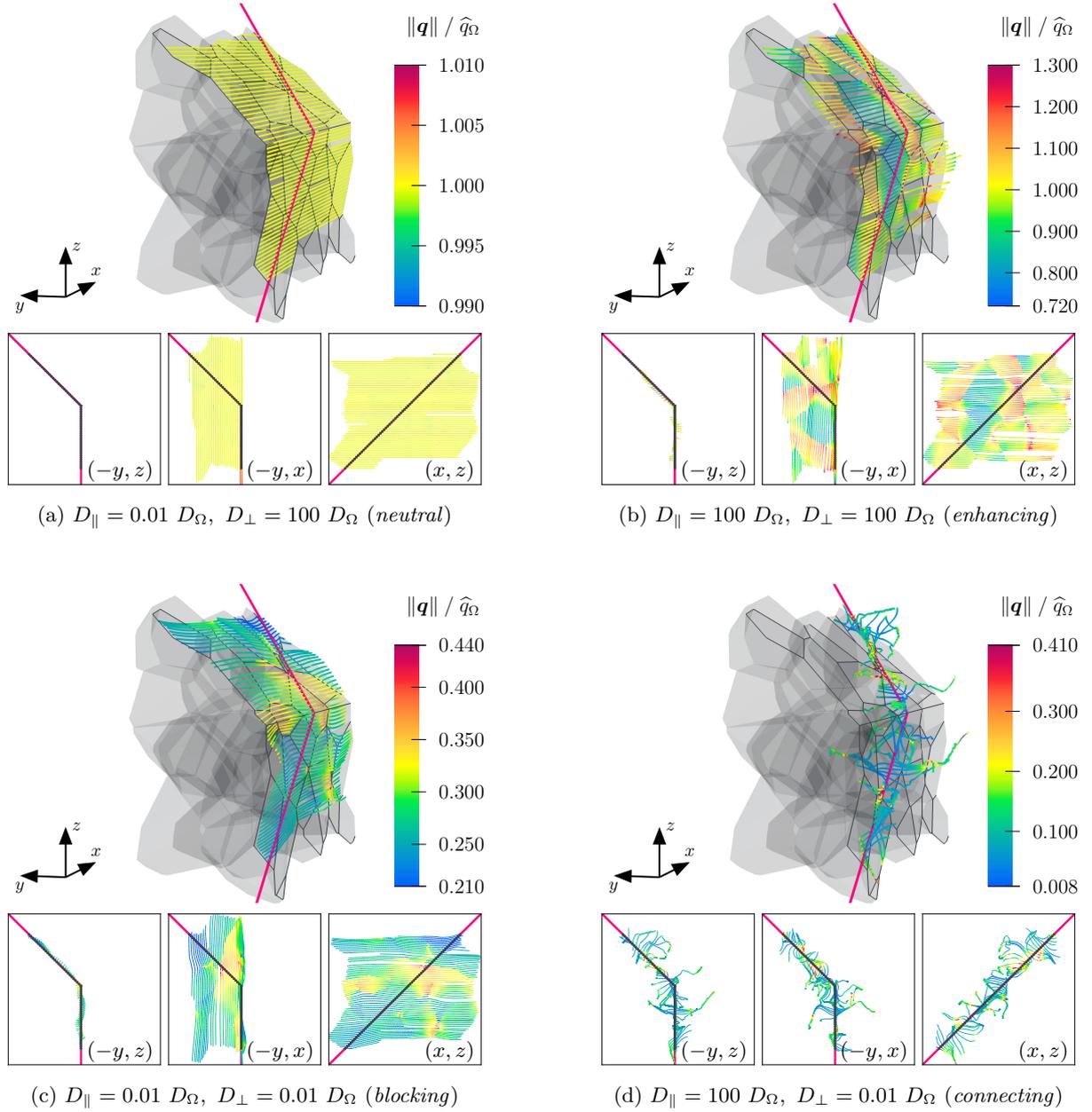
\begin{figure}[htbp!]
    \centering
    \begin{minipage}{0.45\textwidth}
        \centering
        \begin{subfigure}[b]{\textwidth}
            \begin{tikzpicture}
                \node[anchor=south west,inner sep=0] (image) at (0,0) {\includegraphics[height=0.7\textwidth,trim={22cm 7cm 18cm 7cm},clip]{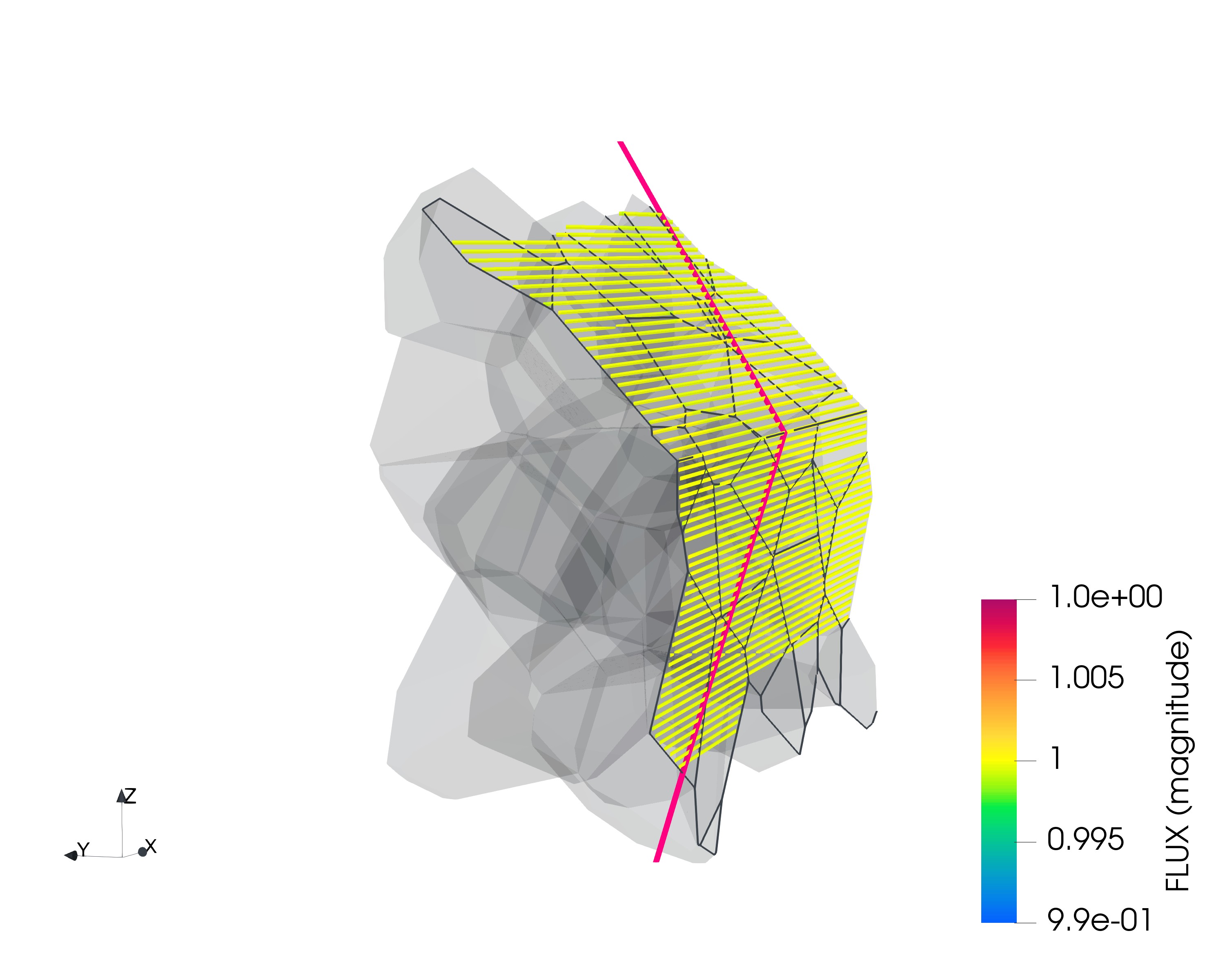}};
                \begin{scope}[x={(image.south east)},y={(image.north west)}]
                    \node[anchor=south east] at (0,0) {\includegraphics[width=0.2\textwidth]{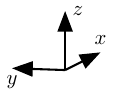}};
                    \node[anchor=south west] at (current bounding box.south east) {\includegraphics[width=0.25\textwidth]{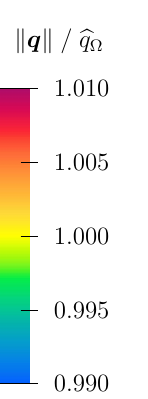}};
                \end{scope}
            \end{tikzpicture}
        \end{subfigure}
        \vspace{0.5cm}
        \begin{subfigure}[b]{\textwidth}
            \begin{tikzpicture}
                \node[anchor=south west,inner sep=0] (image) at (0,0) {\includegraphics[width=\textwidth]{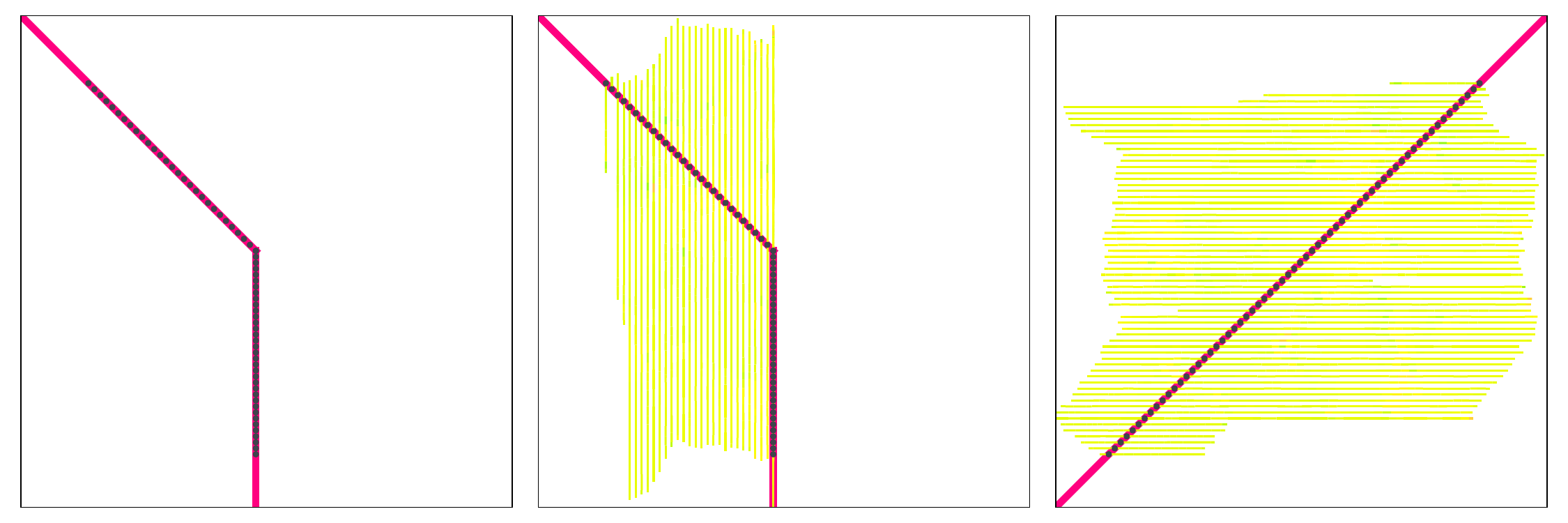}};
                \begin{scope}[x={(image.south east)},y={(image.north west)}]
                    \node[anchor=south east] at (0.33,0.) {\small$(-y,z)$};
                \end{scope}
                \begin{scope}[x={(image.south east)},y={(image.north west)}]
                    \node[anchor=south east] at (0.66,0.) {\small$(-y,x)$};
                \end{scope}
                \begin{scope}[x={(image.south east)},y={(image.north west)}]
                    \node[anchor=south east] at (0.99,0.) {\small$(x,z)$};
                \end{scope}
            \end{tikzpicture}
            \caption{$D_\Vert=0.01~D_\Omega,~D_\perp=100~D_\Omega$ (\textit{neutral})}
            \label{fig:geo_streamlines_a}
        \end{subfigure}
    \end{minipage}
    \hfill
    \begin{minipage}{0.45\textwidth}
        \centering
        \begin{subfigure}[b]{\textwidth}
            \begin{tikzpicture}
                \node[anchor=south west,inner sep=0] (image) at (0,0) {\includegraphics[height=0.7\textwidth,trim={22cm 7cm 18cm 7cm},clip]{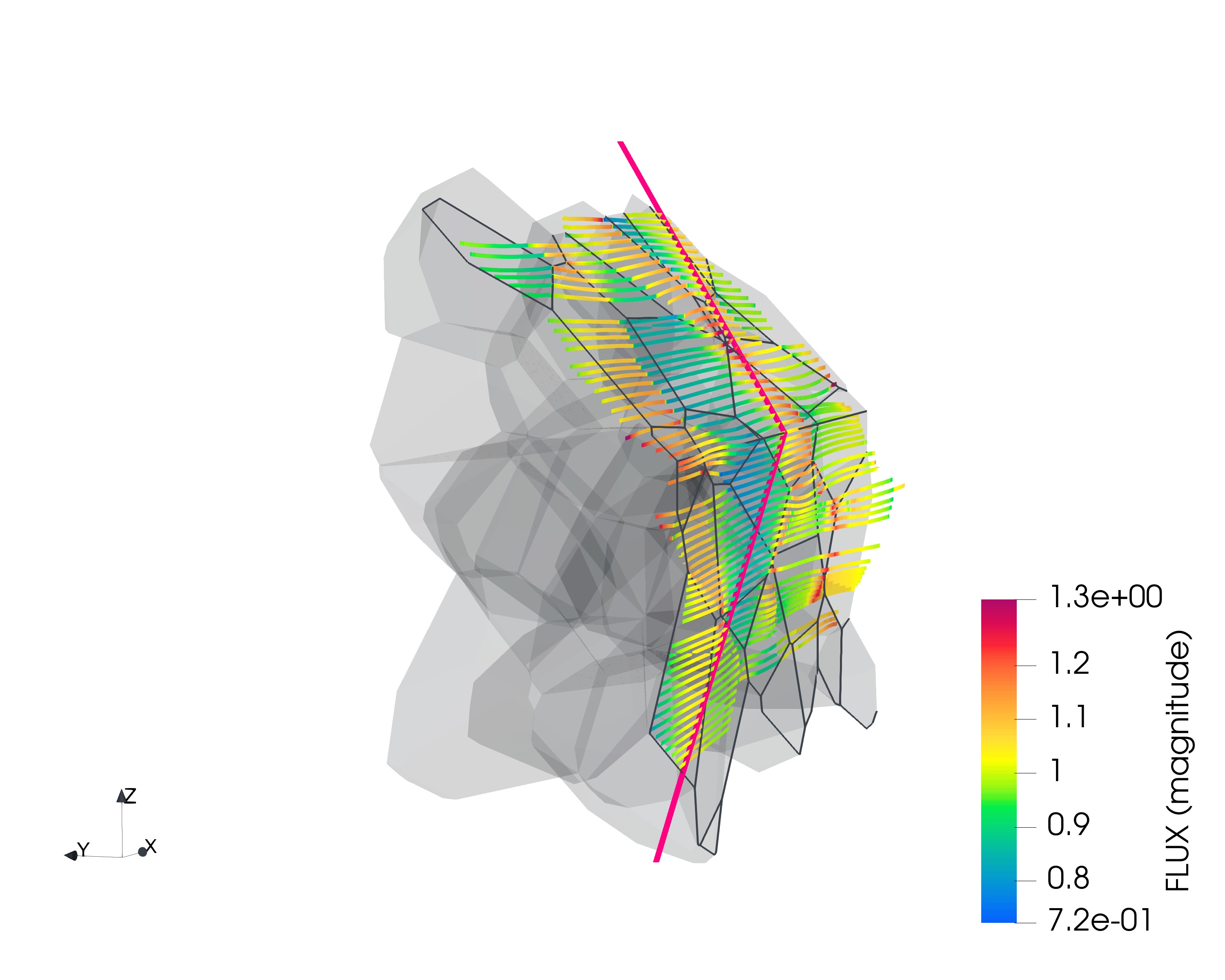}};
                \begin{scope}[x={(image.south east)},y={(image.north west)}]
                    \node[anchor=south east] at (0,0) {\includegraphics[width=0.2\textwidth]{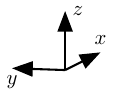}};
                    \node[anchor=south west] at (current bounding box.south east) {\includegraphics[width=0.25\textwidth]{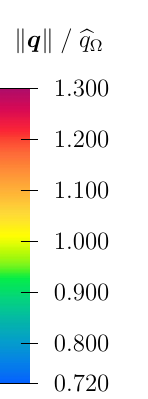}};
                \end{scope}
            \end{tikzpicture}
        \end{subfigure}
        \vspace{0.5cm}
        \begin{subfigure}[b]{\textwidth}
            \begin{tikzpicture}
                \node[anchor=south west,inner sep=0] (image) at (0,0) {\includegraphics[width=\textwidth]{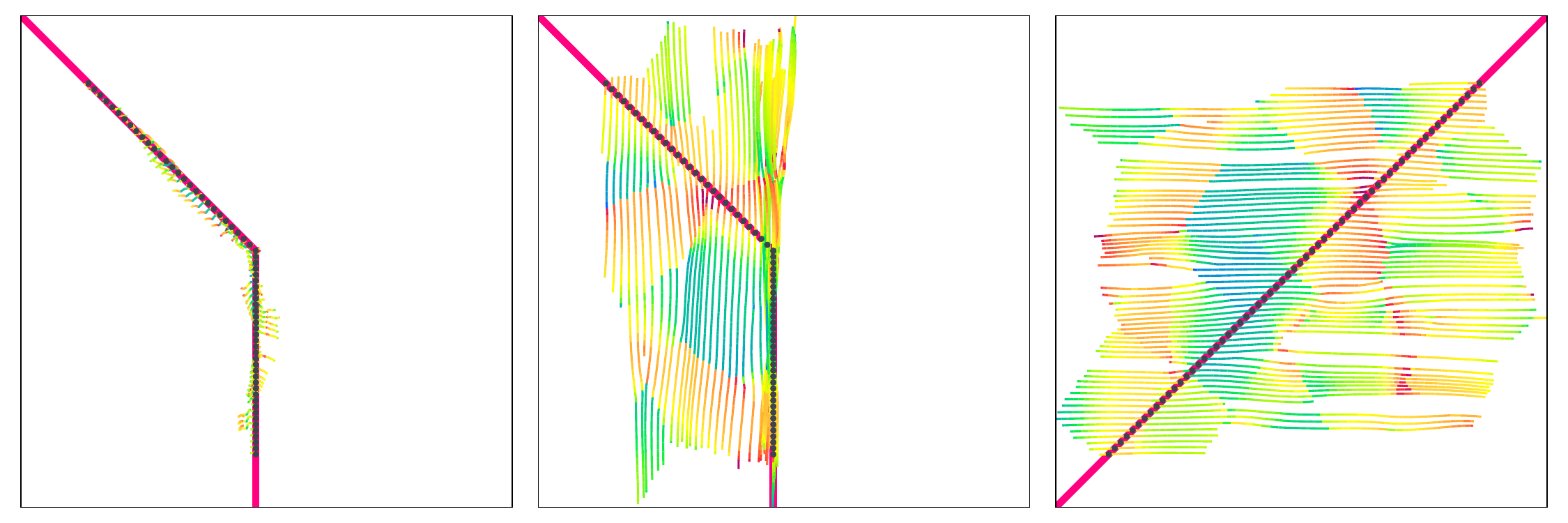}};
                \begin{scope}[x={(image.south east)},y={(image.north west)}]
                    \node[anchor=south east] at (0.33,0.) {\small$(-y,z)$};
                \end{scope}
                \begin{scope}[x={(image.south east)},y={(image.north west)}]
                    \node[anchor=south east] at (0.66,0.) {\small$(-y,x)$};
                \end{scope}
                \begin{scope}[x={(image.south east)},y={(image.north west)}]
                    \node[anchor=south east] at (0.99,0.) {\small$(x,z)$};
                \end{scope}
            \end{tikzpicture}
            \caption{$D_\Vert=100~D_\Omega,~D_\perp=100~D_\Omega$ (\textit{enhancing})}
            \label{fig:geo_streamlines_b}
        \end{subfigure}
    \end{minipage}
    \begin{minipage}{0.45\textwidth}
        \centering
        \begin{subfigure}[b]{\textwidth}
            \begin{tikzpicture}
                \node[anchor=south west,inner sep=0] (image) at (0,0) {\includegraphics[height=0.7\textwidth,trim={22cm 7cm 18cm 7cm},clip]{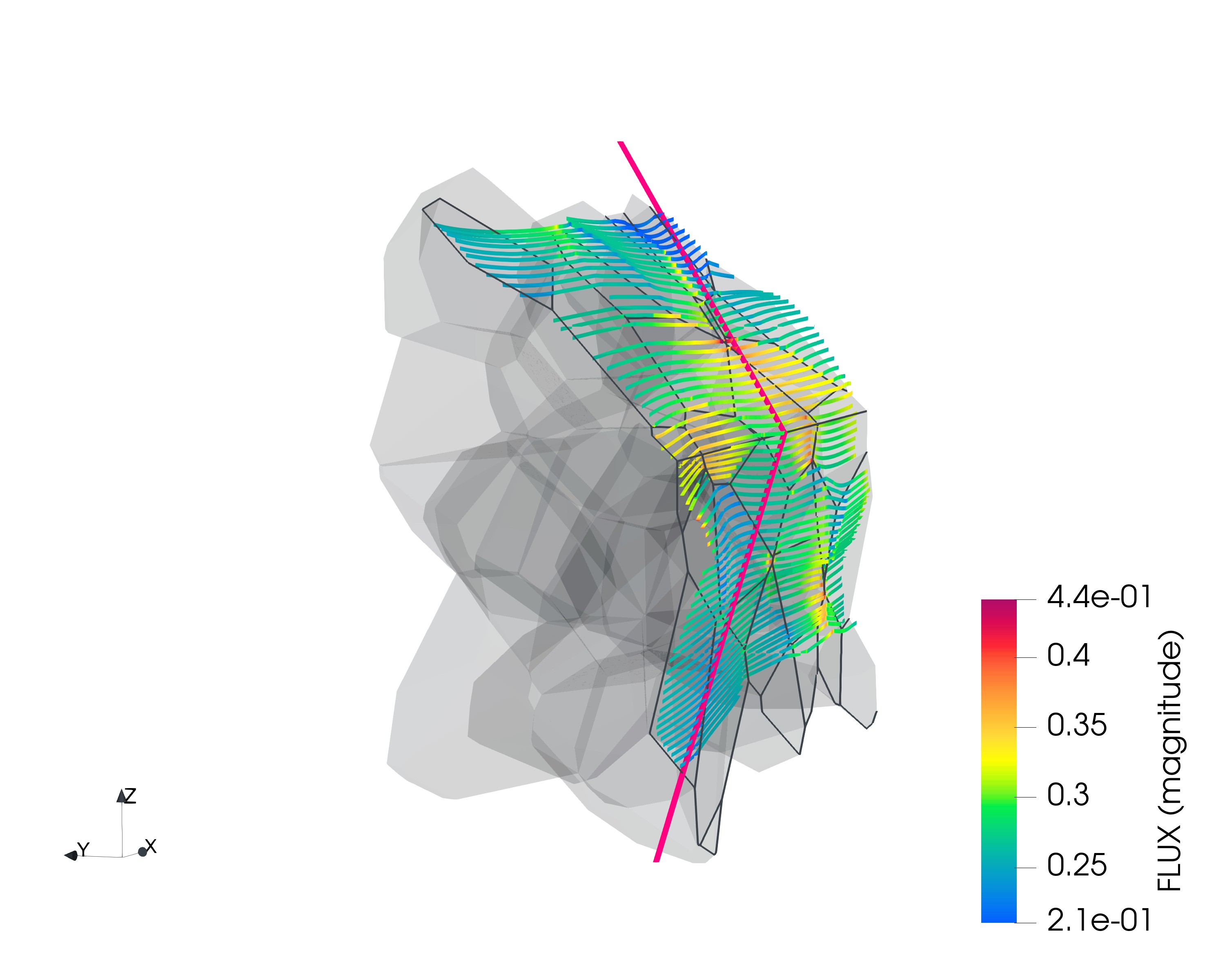}};
                \begin{scope}[x={(image.south east)},y={(image.north west)}]
                    \node[anchor=south east] at (0,0) {\includegraphics[width=0.2\textwidth]{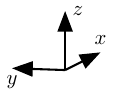}};
                    \node[anchor=south west] at (current bounding box.south east) {\includegraphics[width=0.25\textwidth]{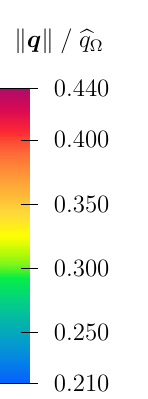}};
                \end{scope}
            \end{tikzpicture}
        \end{subfigure}
        \vspace{0.5cm}
        \begin{subfigure}[b]{\textwidth}
            \begin{tikzpicture}
                \node[anchor=south west,inner sep=0] (image) at (0,0) {\includegraphics[width=\textwidth]{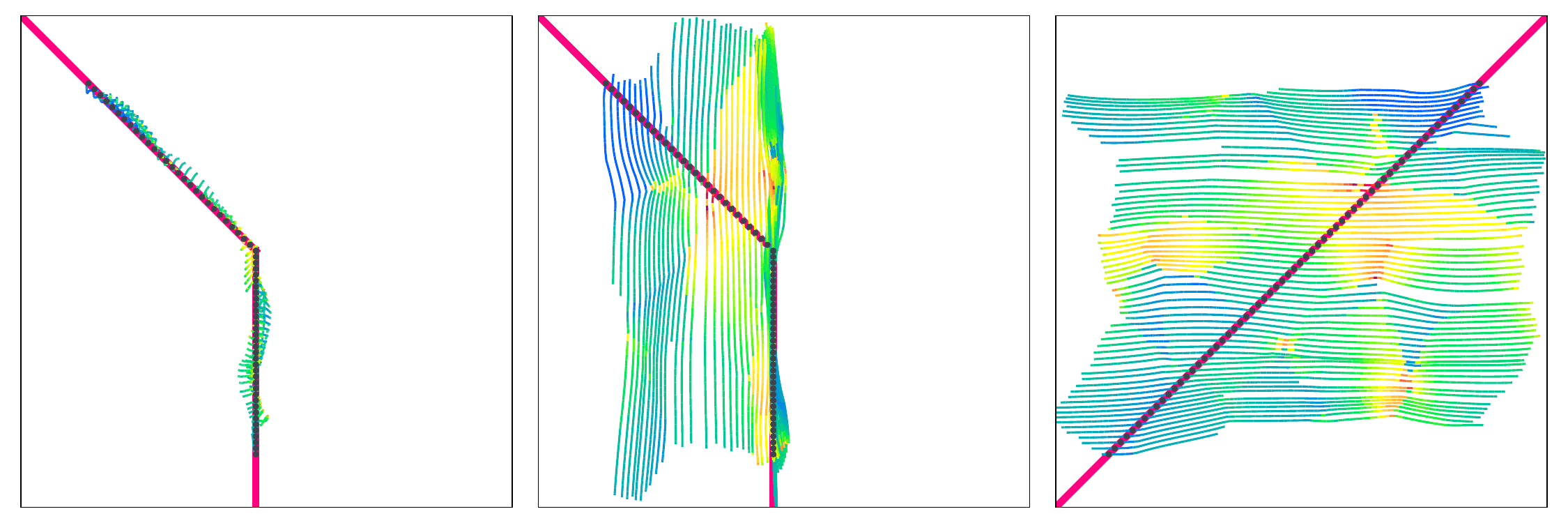}};
                \begin{scope}[x={(image.south east)},y={(image.north west)}]
                    \node[anchor=south east] at (0.33,0.) {\small$(-y,z)$};
                \end{scope}
                \begin{scope}[x={(image.south east)},y={(image.north west)}]
                    \node[anchor=south east] at (0.66,0.) {\small$(-y,x)$};
                \end{scope}
                \begin{scope}[x={(image.south east)},y={(image.north west)}]
                    \node[anchor=south east] at (0.99,0.) {\small$(x,z)$};
                \end{scope}
            \end{tikzpicture}
            \caption{$D_\Vert=0.01~D_\Omega,~D_\perp=0.01~D_\Omega$ (\textit{blocking})}
        \label{fig:geo_streamlines_c}
        \end{subfigure}
    \end{minipage}
    \hfill
    \begin{minipage}{0.45\textwidth}
        \centering
        \begin{subfigure}[b]{\textwidth}
            \begin{tikzpicture}
                \node[anchor=south west,inner sep=0] (image) at (0,0) {\includegraphics[height=0.7\textwidth,trim={22cm 7cm 18cm 7cm},clip]{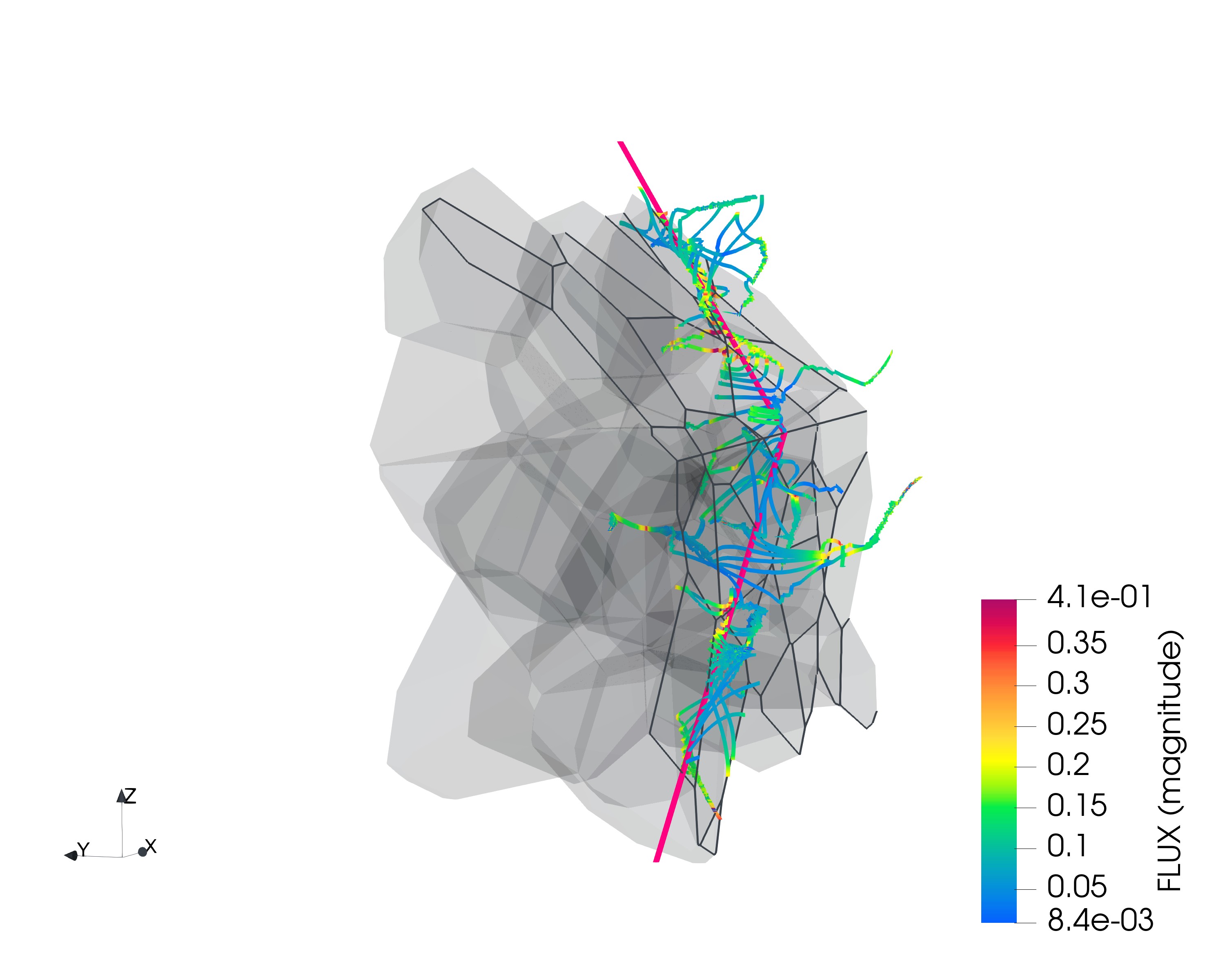}};
                \begin{scope}[x={(image.south east)},y={(image.north west)}]
                    \node[anchor=south east] at (0,0) {\includegraphics[width=0.2\textwidth]{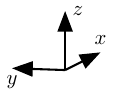}};
                    \node[anchor=south west] at (current bounding box.south east) {\includegraphics[width=0.25\textwidth]{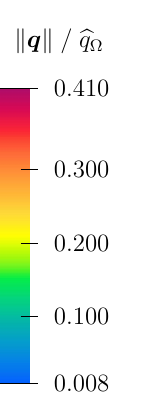}};
                \end{scope}
            \end{tikzpicture}
        \end{subfigure}
        \vspace{0.5cm}
        \begin{subfigure}[b]{\textwidth}
            \begin{tikzpicture}
                \node[anchor=south west,inner sep=0] (image) at (0,0) {\includegraphics[width=\textwidth]{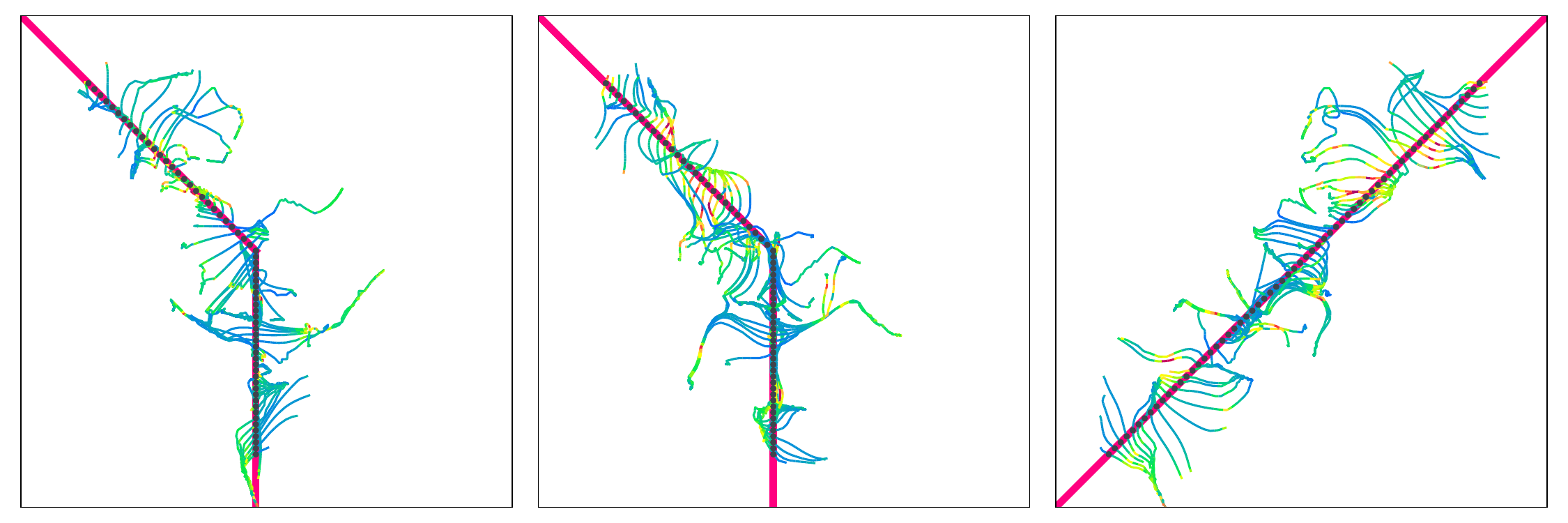}};
                \begin{scope}[x={(image.south east)},y={(image.north west)}]
                    \node[anchor=south east] at (0.33,0.) {\small$(-y,z)$};
                \end{scope}
                \begin{scope}[x={(image.south east)},y={(image.north west)}]
                    \node[anchor=south east] at (0.66,0.) {\small$(-y,x)$};
                \end{scope}
                \begin{scope}[x={(image.south east)},y={(image.north west)}]
                    \node[anchor=south east] at (0.99,0.) {\small$(x,z)$};
                \end{scope}
            \end{tikzpicture}
            \caption{$D_\Vert=100~D_\Omega,~D_\perp=0.01~D_\Omega$ (\textit{connecting})}
            \label{fig:geo_streamlines_d}
        \end{subfigure}
    \end{minipage}
    \caption{Selected streamlines stemming from the dedimensionalized bulk flux field are shown for the case of an imposed gradient $\ol{\ul{g}}$ along the $x$-axis. The flux is dedimensionalized with $\widehat{q}_\Omega=\|-D_\Omega \ol{\ul{g}}\|$. The seeds of the streamlines were equidistantly spaced along the pink lines. For a better comparison, the respective projections (of a selection of streamlines) onto the planes of the coordinate system are given as well. The relative GB thickness parameter was fixed to $h=0.0032l$ for all configurations, resulting in a volume fraction of $\vfgb\approx 7.80\%$. The material parameters for the GB are given for the respective subplots together with the transport regimes introduced in \Cref{tab:regimes}. The color map was chosen to highlight the local differences in the flux magnitude.}
    \label{fig:geo_streamlines}
\end{figure}

\subsection{{\color{rev}Classification} of diffusion {\color{rev}behavior}}
\label{Sec:results:localizing}
\begin{figure}[h!]
    \centering
    \includegraphics[width=0.7\linewidth]{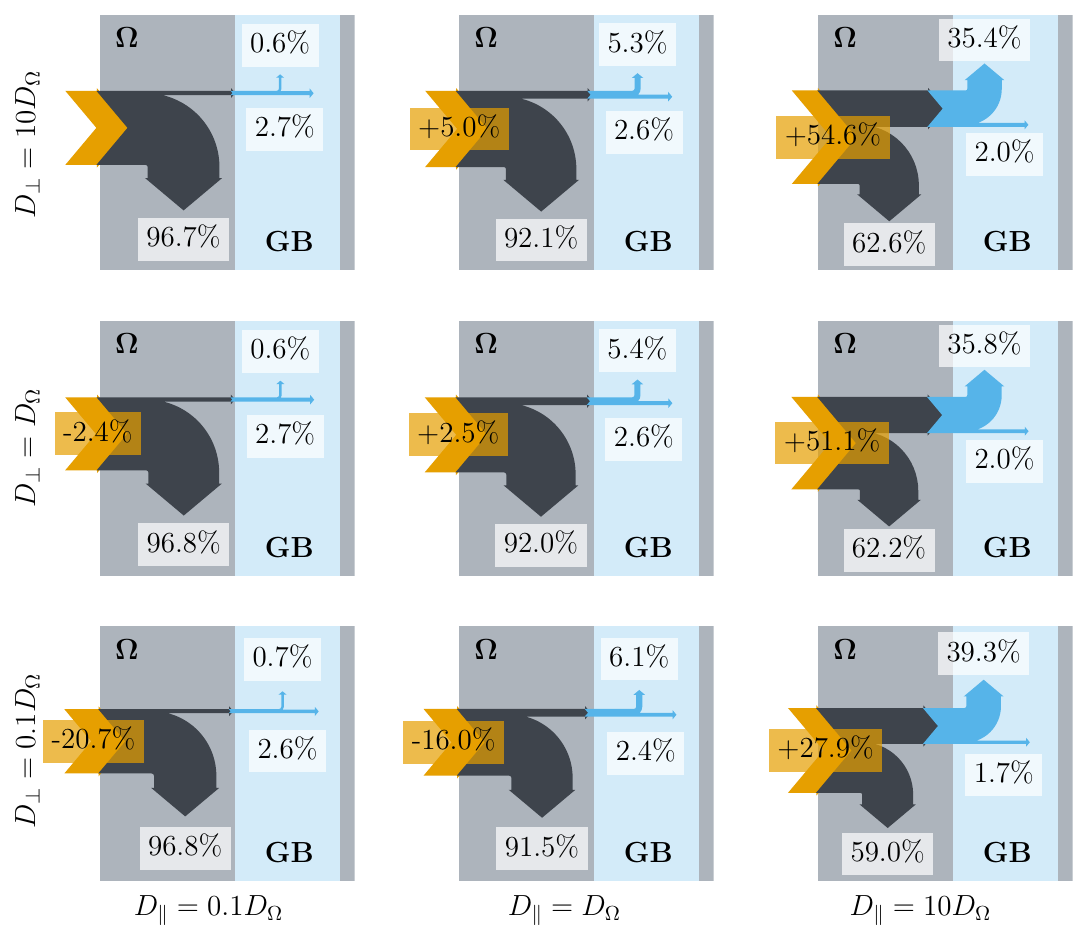}
    \caption{Split of the overall magnitude of integrated flux in bulk domain (gray area $\Omega$) as well as in-plane and normal direction in the GB (blue area) for varying material parameters ($D_\Vert/D_\Omega$, $D_\perp/D_\Omega$) and an imposed gradient in global $x$-direction. We fixed $h=0.0032l$ for all configurations (i.e., $\vfgb\approx 7.80\%$). The yellow boxes represent the relative change in total flux with respect to the configuration on the top left. The split of the total flux in bulk and the two GB contributions is given in the white boxes.}
    \label{fig:relation_fluxes}
\end{figure}

As stated in \Cref{Sec:surface:diffusion}, the material response tends to that of a homogeneous bulk material with idealized interfaces not only in the case of a vanishing GB thickness ($h\rightarrow 0$) but also if $D_\perp \gg D_\Omega$ and, simultaneously, $D_\Vert \ll D_\Omega$. Streamlines for the flux field as depicted in \Cref{fig:geo_streamlines_a} are parallel to the ones of the imposed gradient field inside the bulk. The flux magnitude remains uniform. The partitioning of the accumulated flux between bulk and GB in \Cref{fig:relation_fluxes} underlines that the overall transport is clearly bulk-dominated in this case. In general, however, the flux distribution exhibits a strong dependence on the GB parameters: The diffusion coefficient in normal direction $D_\perp$ has a limiting character, i.e., for small $D_\perp$ (i.e., $D_\perp \ll D_\Omega$) atomic diffusion in the bulk is severely hindered compared to the case {\color{rev}of idealized interfaces}. Opposed to that, large $D_\Vert$ (i.e., $D_\Vert\gg D_\Omega$) leads to a significant enhancement of the in-plane flux and a slight increase of the absolute normal transport (although the relative share is reduced). The accumulated flux in the bulk, however, remains unaffected. Hence, increasing $D_\Vert$ introduces additional diffusion paths on the connected GB network.

When taking a closer look at the local flux field in the bulk, the streamlines in \Cref{fig:geo_streamlines} show how the diffusion paths starting in the same seed points in the bulk vary solely depending on the material properties of the GBs. 
In contrast to the bulk-like case, locally varying flux fields and diffusion paths are observed for all other limit cases: If we allow for additional diffusion along the GBs (i.e., configuration in \Cref{fig:geo_streamlines_b}) or introduce GBs with a hindering and limiting effect (i.e., configuration in \Cref{fig:geo_streamlines_c}), the magnitude of the flux clearly reflects the distribution and shape of the grains and of the GBs. We obtain discontinuities in the magnitude of the flux across the GBs and slightly deviating diffusion paths.

The case in \Cref{fig:geo_streamlines_d} with enhanced transport along the GB but limited ability to cross the GB deserves special attention. The diffusion paths within the bulk domain are completely governed by the local structure and deviate heavily from the direction of the imposed gradient. Some of the streamlines exhibit an oscillating behavior along certain GB facets. Note that these streamlines are only based on the flux field of the bulk domain and do, therefore, not reflect the transport behavior within the GBs for now. This, however, underlines the need to thoroughly investigate the detailed transport behavior within the GBs as we will do in \Cref{Sec:results:character}.

\begin{figure}[h]
    \centering
    \begin{subfigure}[b]{0.23\textwidth}
        \begin{tikzpicture}
            \node[anchor=south west,inner sep=0] (image) at (0,0) {\includegraphics[width=0.9\textwidth,trim={0 0 10cm 15cm},clip]{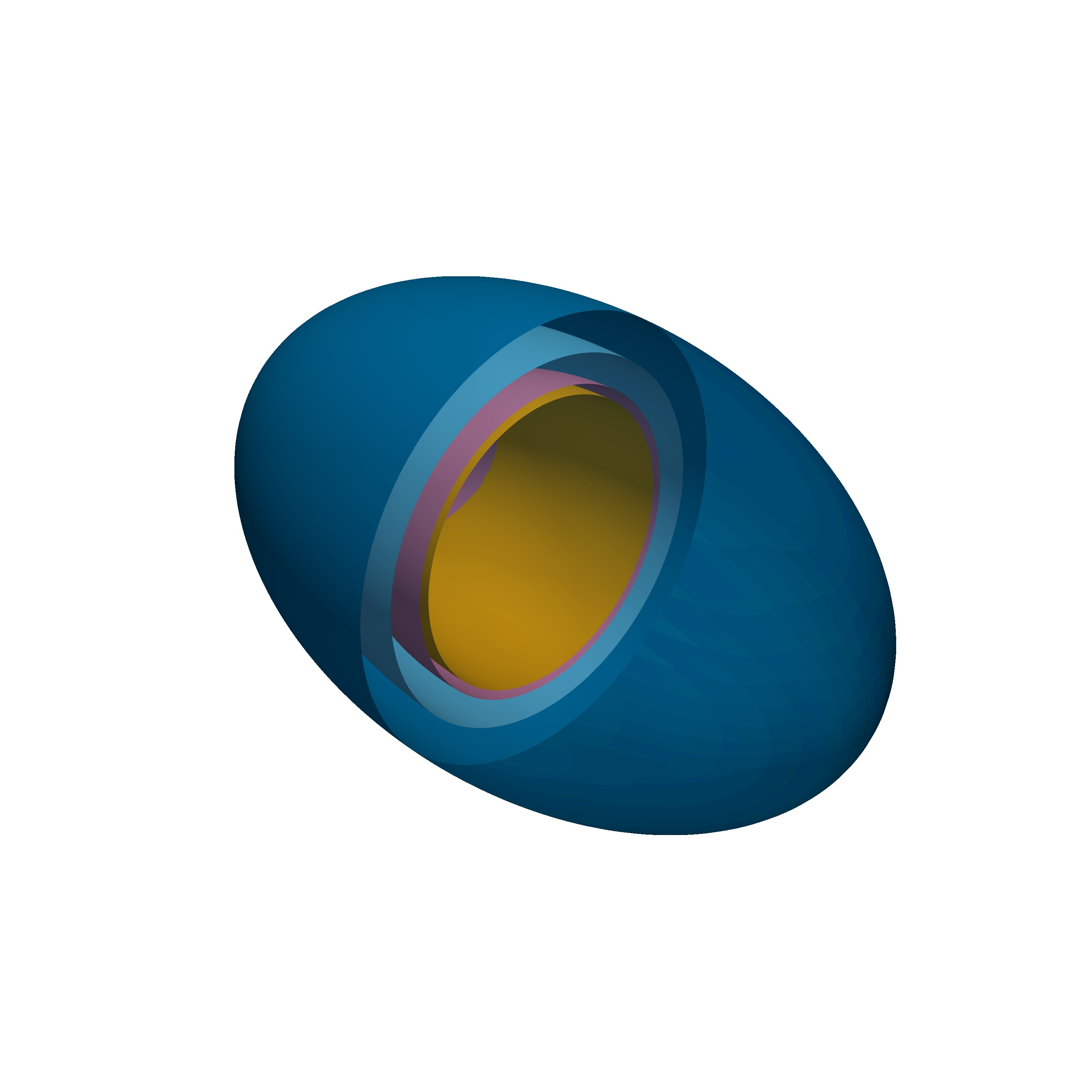}};
            \begin{scope}[x={(image.south east)},y={(image.north west)}]
                \node[anchor=south west] at (0,0) {\includegraphics[width=0.3\textwidth]{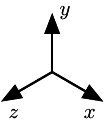}};
            \end{scope}
        \end{tikzpicture}
    \end{subfigure}
    \hfill
    \begin{subfigure}[b]{0.23\textwidth}
        \begin{tikzpicture}
            \node[anchor=south west,inner sep=0] (image) at (0,0) {\includegraphics[width=0.9\textwidth, trim={0 0 10cm 13cm},clip]{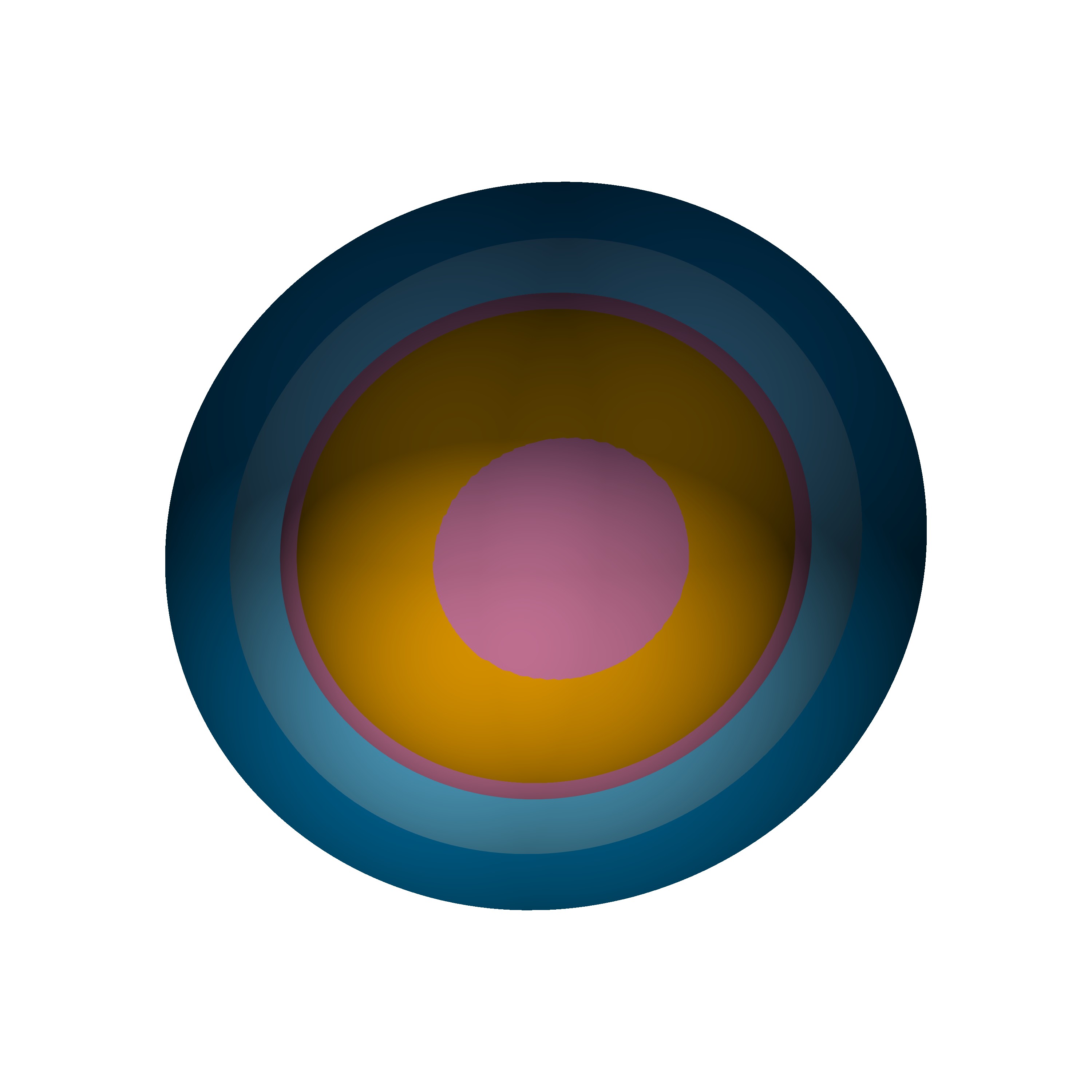}};
            \begin{scope}[x={(image.south east)},y={(image.north west)}]
                \node[anchor=south west] at (0,0) {\includegraphics[width=0.3\textwidth]{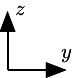}};
            \end{scope}
        \end{tikzpicture}
    \end{subfigure}
    \hfill
    \begin{subfigure}[b]{0.23\textwidth}
        \begin{tikzpicture}
            \node[anchor=south west,inner sep=0] (image) at (0,0) {\includegraphics[width=0.9\textwidth, trim={0 0 6cm 15cm},clip]{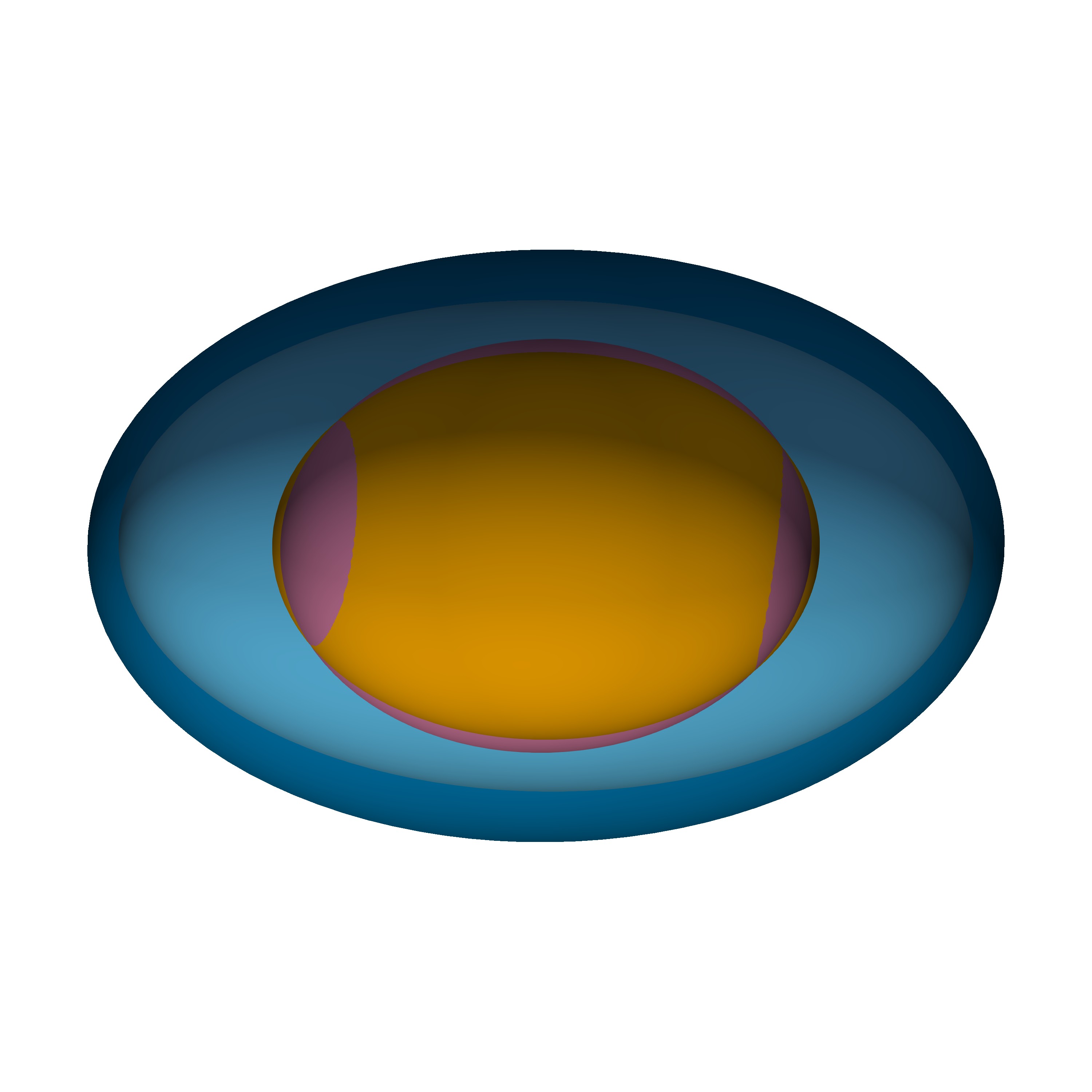}};
            \begin{scope}[x={(image.south east)},y={(image.north west)}]
                \node[anchor=south west] at (0,0) {\includegraphics[width=0.3\textwidth]{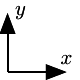}};
            \end{scope}
        \end{tikzpicture}
    \end{subfigure}
    \hfill
    \begin{subfigure}[b]{0.17\textwidth}
        \centering
        \includegraphics[width=\textwidth]{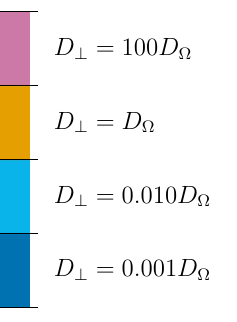}
    \end{subfigure}
    \caption{Ellipsoids representing the relative second moment of the dedimensionalized flux-field in the bulk domain as defined in \cref{eq:moment} for an imposed gradient in $x$-direction $\ol{\ul g} = \begin{bmatrix} 1 & 0 & 0\end{bmatrix}$ and different parameter configurations. The ellipsoids are constructed based on the eigenvectors of $\ull{R}$ scaled with the square roots of the corresponding eigenvalues. We fixed $D_\Vert = 10 ~D_\Omega$ for all cases and varied $D_\perp/D_\Omega$ as shown in the color bar. The shown cuts and perspectives highlight the intersections of the latter two ellipsoids.}
    \label{fig:ellipsoid}
\end{figure}

While the streamlines help gain a feel for different regimes, they are of limited use when it comes to comparing potentially very similar parameter configurations. The relevant features enhanced by the streamlines comprise local variations in flux magnitude as well as the orientation of the flux. We, therefore, define the second moment of the flux field in the bulk domain $\Omega$ as 
\begin{align}
    \ull{R} = \frac{1}{|\Omega|}\int\limits_{\Omega} \ul{q} \otimes \ul{q} \, {\rm{d}}V - \langle \ul{q} \rangle_\Omega \otimes \langle \ul{q} \rangle_\Omega
    \label{eq:moment}
\end{align}
accounting for the volume averaged flux $\langle \ul{q} \rangle_\Omega$. The second moment corresponds to the covariance matrix of the 3D atomic flux field, where the diagonal entries directly correspond to the variance in the flux along the respective coordinate axes. Hence, we propose it as a suitable metric for directly comparing configurations. Note that the computation of the metric is performed using numerical integration in this context. In \Cref{fig:ellipsoid}, the ellipsoids corresponding to $\ull{R}$ for different configurations are shown in order to visually enhance the directions of the largest variation (half axis radii of the ellipsoid correspond to square roots of the eigenvalues of $\ull{R}$). Note that only the ratio $D_\perp/D_\Omega$ was varied for these visualizations, i.e., $D_\Vert=10 D_\Omega$ was held constant. The larger the components of $\ull{R}$ are, the more local variations appear, corresponding to an increased average curvature/curlyness of the streamlines. In the case of an (almost) perfectly uniform flux field as displayed in \Cref{fig:geo_streamlines_a} for the bulk-like regime, all components of $\ull{R}$ should tend to zero, i.e., streamlines are straight.

In all displayed cases, the eigenvector with the dominating eigenvalue is close to the direction of loading, i.e., the $x$-axis in this case, as suggested by \Cref{fig:geo_streamlines}. Therefore, any local deviation in magnitude and/or direction leads to a larger absolute variance along this direction. The general shape of the ellipsoids is close to a spheroid, as can be seen in the middle perspective, due to the close-to-isotropic property of the structural tensor in \cref{eq:S2_setup}.

For $D_\perp \geq D_\Omega$, the eigenvalues show only minor variations. Therefore, the two innermost ellipsoids are hard to distinguish. The most relevant change is that the ellipsoid becomes more similar to a sphere, which implies that the flux variance depends less on the loading direction. This also causes an intersection of those two ellipsoids. For $D_\perp < D_\Omega$, the magnitude of the eigenvalues grows rapidly, underlining how the limited ability to enter and leave the GB forces mobile species to take alternative routes and detours within the bulk as well as the GB. This also causes an increased variance in the non-loading directions.

Based on the observations presented in this section, we define four regimes of transport based on the impact of GBs as shown in \Cref{tab:regimes}.
In the following, we use these terms when referring to the four limit cases in order to keep the descriptions concise.

\begin{table}[h]
    \centering
    \begin{tabular}{c||c|c}
        & $D_\Vert\ll D_\Omega$ & $D_\Vert \gg D_\Omega$ \\
        \hline
        \hline
        \multirow{ 2}{*}{$D_\perp\gg D_\Omega$} & \textbf{NEUTRAL}:  & \textbf{ENHANCING}:  \\
        & idealized interfaces & additional transport in GBs \\
        \hline
        \multirow{ 2}{*}{$D_\perp\ll D_\Omega$} & \textbf{BLOCKING}: & \textbf{CONNECTING}: \\
        & hindered transport & GB segments as channels
    \end{tabular}
    \caption{Diffusion regimes to characterize the impact of GB transport on the overall system response.}
    \label{tab:regimes}
\end{table}

\subsection{Characterization of diffusion in GBs}
\label{Sec:results:character}
Approaches that account for concentration jumps across GBs \cite{Peng2024,Bai2020} can recover the local flux of mobile species across the GB. For a Fisher-like model, the in- and ouflux of the GB were related to the in-plane gradient~\cite{Han2013}. However, beyond that, the condensed representation of the GB in the Fisher-like model generally remains a severe limitation when investigating the local transport of mobile species within the GB. In this context, the three-node-collapsed analytical design of the proposed GB diffusion model from \cref{eq:c_ansatz} and \cref{eq:tensor_prod} is of particular importance and interest. It allows us to analyze the relevant modes of GB transport and enables the full reconstruction of the 3D GB volume and corresponding concentration profiles as discussed in the following.

\begin{figure}[h]
    \centering
    \begin{subfigure}[t]{0.2\textwidth}
        \centering
        \includegraphics[width=\textwidth]{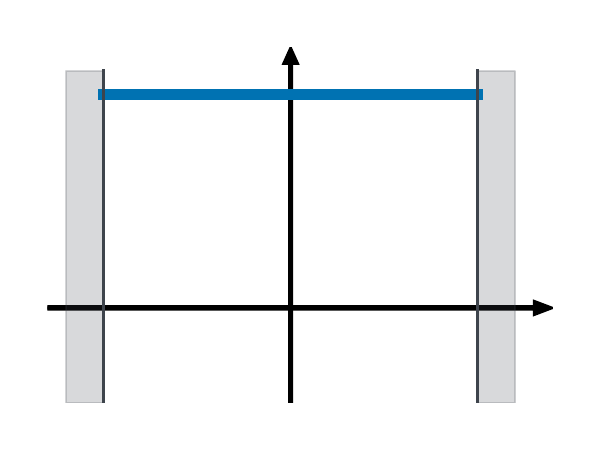}
        \caption{\color{rev}$\alpha_0$}
    \end{subfigure}
    \hfill
    \begin{subfigure}[t]{0.2\textwidth}
        \centering
        \includegraphics[width=\textwidth]{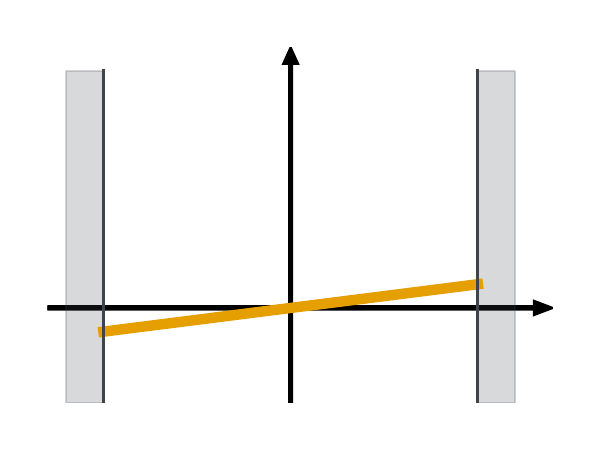}
        \caption{\color{rev}$\alpha_1 \zeta_{\rm n}/h$}
    \end{subfigure}
    \hfill
    \begin{subfigure}[t]{0.2\textwidth}
        \centering
        \includegraphics[width=\textwidth]{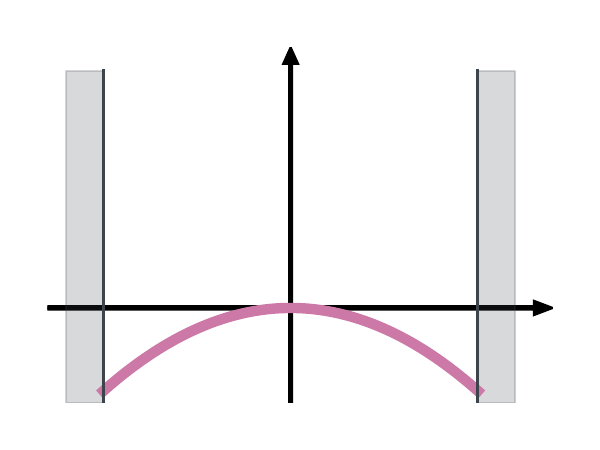}
        \caption{\color{rev}$\alpha_2 \zeta_{\rm n}^2/h^2$}
    \end{subfigure}
    \hfill
    \begin{subfigure}[t]{0.2\textwidth}
        \centering
        \includegraphics[width=\textwidth]{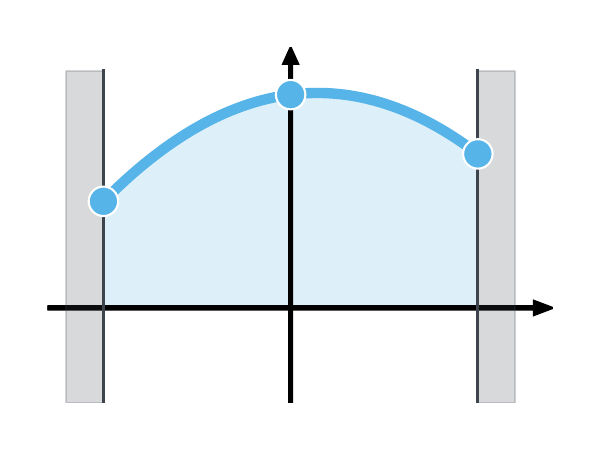}
        \caption{\color{rev}$\sum_{j=1}^2\alpha_j \zeta_{\rm n}^j/h^j$}
    \end{subfigure}
    \caption{\color{rev} Constant, linear and quadratic concentration modes with coefficients $\alpha_0$, $\alpha_1$ and $\alpha_2$ together form the analytical concentration profile as defined in \cref{eq:c_ansatz}.}
    \label{fig:modes_example}
\end{figure}

\begin{figure}[h]
    \centering
    \includegraphics[width=0.32\textwidth]{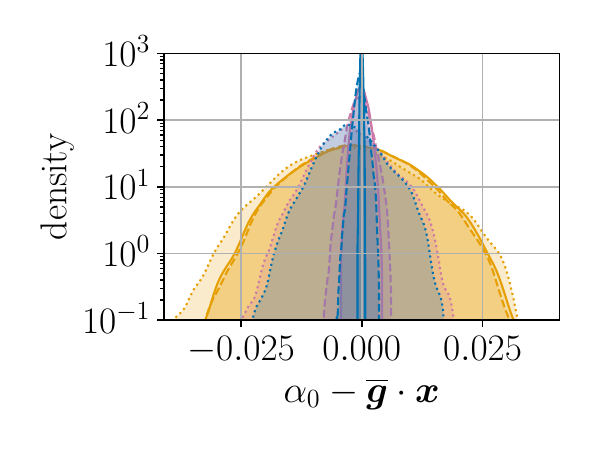}
    \hfill
    \includegraphics[width=0.32\textwidth]{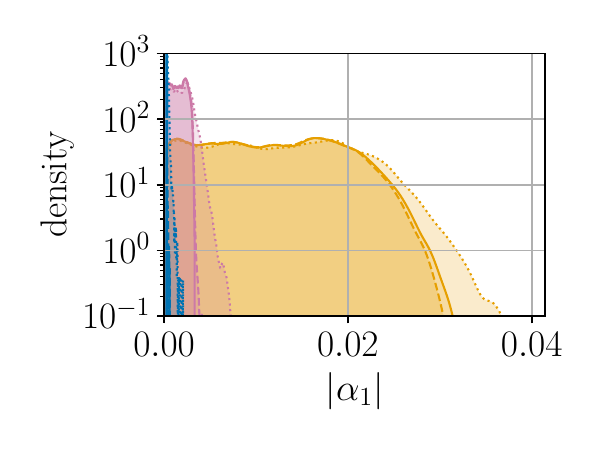}
    \hfill
    \includegraphics[width=0.32\textwidth]{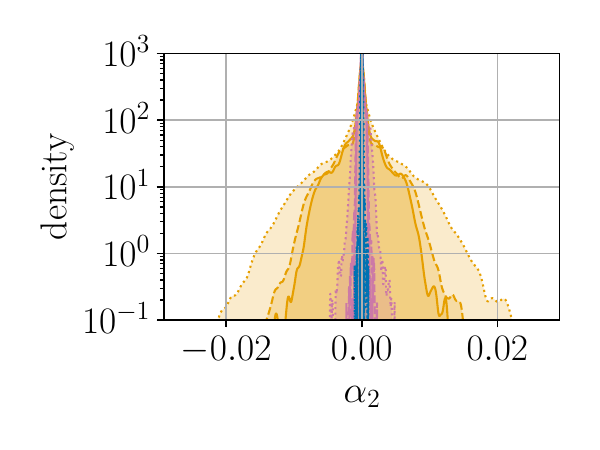}
    \\
    \includegraphics[width=\textwidth, trim={0 2cm 0 4cm}, clip]{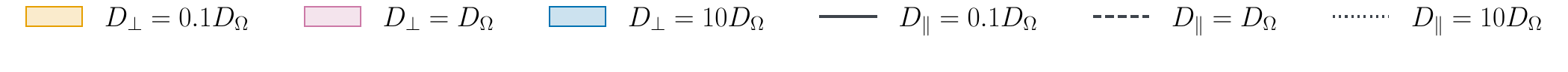}
    \caption{Geometry-wide distribution of element-averaged $\alpha_0-\overline{\fg}\cdot\fx$ (fluctuation-based offset, cf. \cref{eq:homogenization}), $\alpha_1$ (jump) and $\alpha_2$ (channeling) in \cref{eq:c_ansatz} for different parameter configurations: The choice of $D_\perp/D_\Omega$ is indicated by the color whereas $D_\Vert/D_\Omega$ is determined by the linestyle. The GB thickness was fixed to be $h=0.0032 l$ ($\vfgb\approx 7.80\%$) to match the results in \Cref{fig:grain_concentration}). The density profiles were computed using kernel density estimation on the element averages (independent of the element volumes). Note that this might introduce a slight bias depending on the variance in the element volumes, e.g., close to grain edges or vertices.}
    \label{fig:alpha_stat}
\end{figure}

The coefficients of the analytical GB profile in \cref{eq:c_ansatz} {\color{rev}(visualization of the underlying modes in \Cref{fig:modes_example})} are: 
\begin{align}
    \alpha_0(\fx_\Vert) &= c(\ul{\zeta}_{\rm t}, \zeta_{\rm n}=0)\, , \\
    \alpha_1(\fx_\Vert) &= \frac{1}{2} \lb c(\ul{\zeta}_{\rm t}, \zeta_{\rm n}=h) - c(\ul{\zeta}_{\rm t}, \zeta_{\rm n}=-h) \rb\, , \\ 
    \alpha_2(\fx_\Vert) &= \frac{1}{2} \lb c(\ul{\zeta}_{\rm t}, \zeta_{\rm n}=h) - 2c(\ul{\zeta}_{\rm t}, \zeta_{\rm n}=0) + c(\ul{\zeta}_{\rm t}, \zeta_{\rm n}=-h) \rb\, .
\end{align}
They determine the magnitude of the constant ($\alpha_0$), linear ($\alpha_1$) and quadratic ($\alpha_2$) modes in the model. Hence, their distribution over the entire GB network can be used to characterize GB transport depending on the parameter configuration as shown in \Cref{fig:alpha_stat}. While the coefficient $\alpha_0$ holds responsible for the offset concentration of the GB, the first-order coefficient $\alpha_1(\fx_\Vert)$ is directly related to the concentration jump $\llbracket c \rrbracket$ via $2|\alpha_1 (\fx_\Vert)|= \llbracket c \rrbracket = |c(\ul{\zeta}_{\rm t}, \zeta_{\rm n}=h) - c(\ul{\zeta}_{\rm t}, \zeta_{\rm n}=-h)|$. The second-order coefficient $\alpha_2(\fx_\Vert)$ appears in the net influx into the GB, which we define as 
\begin{align}
    q_{\rm in}(\fx_\Vert) &= D_\perp\lb\frac{\partial c(\ul{\zeta}_{\rm t},\zeta_{\rm n})}{\partial \zeta_{\rm n}}\bigg|_{\zeta_{\rm n} = h} - \frac{\partial c(\ul{\zeta}_{\rm t},\zeta_{\rm n})}{\partial \zeta_{\rm n}}\bigg|_{\zeta_{\rm n} = -h} \rb
    =  D_\perp \frac{4}{h}\alpha_2(\fx_\Vert),
    & \text{with } q_{\rm in}
    \begin{cases}
        > 0: \quad \text{influx} \\
        < 0: \quad \text{outflux}
    \end{cases}
    \label{eq:net_influx}
\end{align}
based on the normal fluxes on both sides of the GB. 

\begin{figure}[h]
    \centering
    \begin{subfigure}[b]{0.32\textwidth}
        \centering
        \begin{tikzpicture}
            \node[anchor=south west,inner sep=0] (image) at (0,0) {\includegraphics[width=0.7\textwidth, trim={30cm 0 30cm 0}, clip]{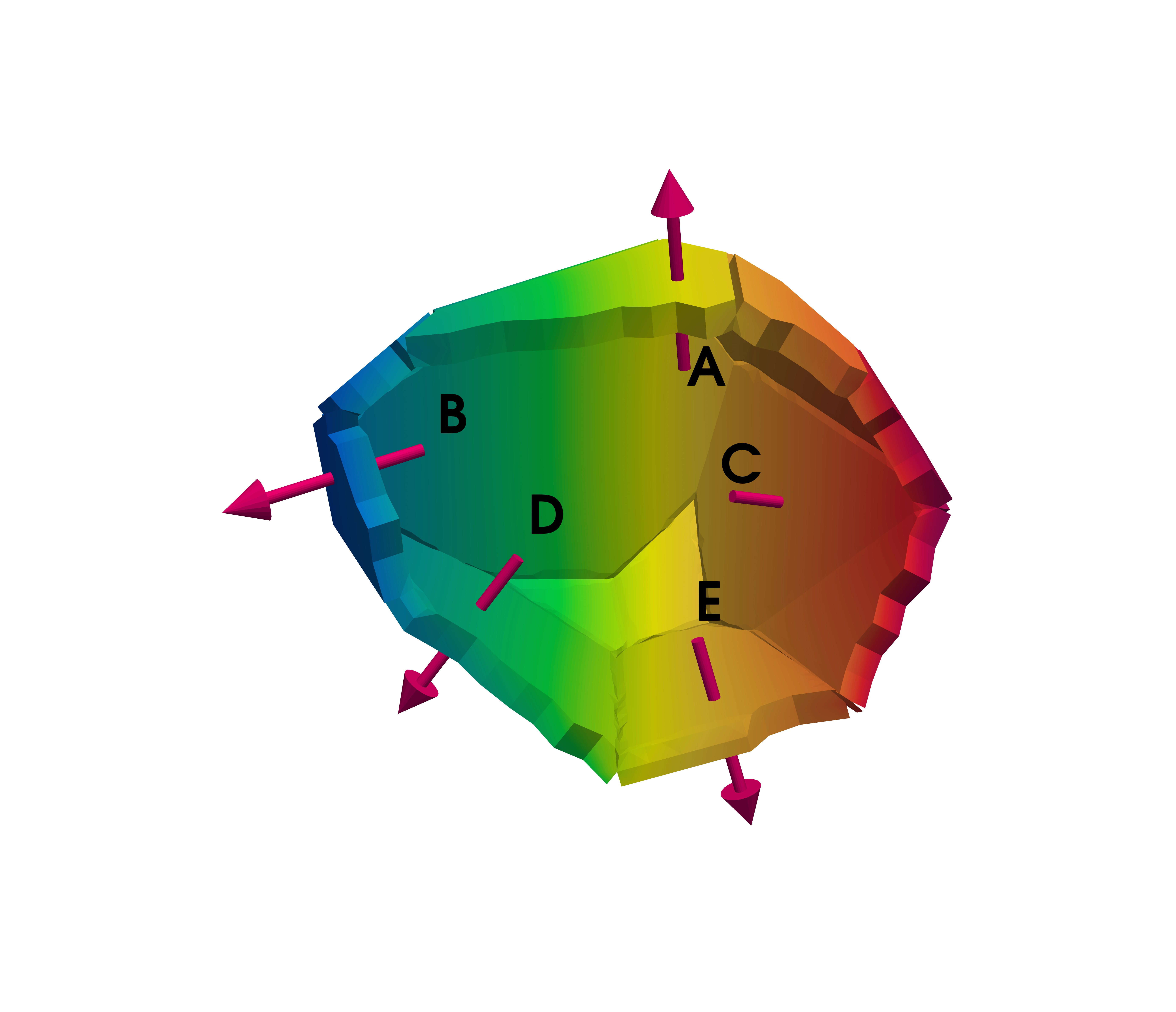}};
            \node[xshift=0.2cm,yshift=4cm,fill=white, rectangle,inner sep=3pt] {(a)};
            \begin{scope}[x={(image.south east)},y={(image.north west)}]
                \node[anchor=south west] at (0,0) {\includegraphics[scale=0.6]{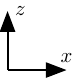}};
                \node[anchor=south west] at (current bounding box.south east) {\includegraphics[width=0.3\textwidth]{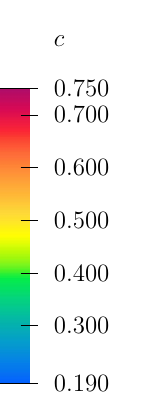}};
            \end{scope}
        \end{tikzpicture}
    \end{subfigure}
        \hfill
    \begin{subfigure}[b]{0.32\textwidth}
        \centering
        \begin{tikzpicture}
            \node[anchor=south west,inner sep=0] (image) at (0,0) {\includegraphics[width=0.7\textwidth, trim={30cm 0 30cm 0}, clip]{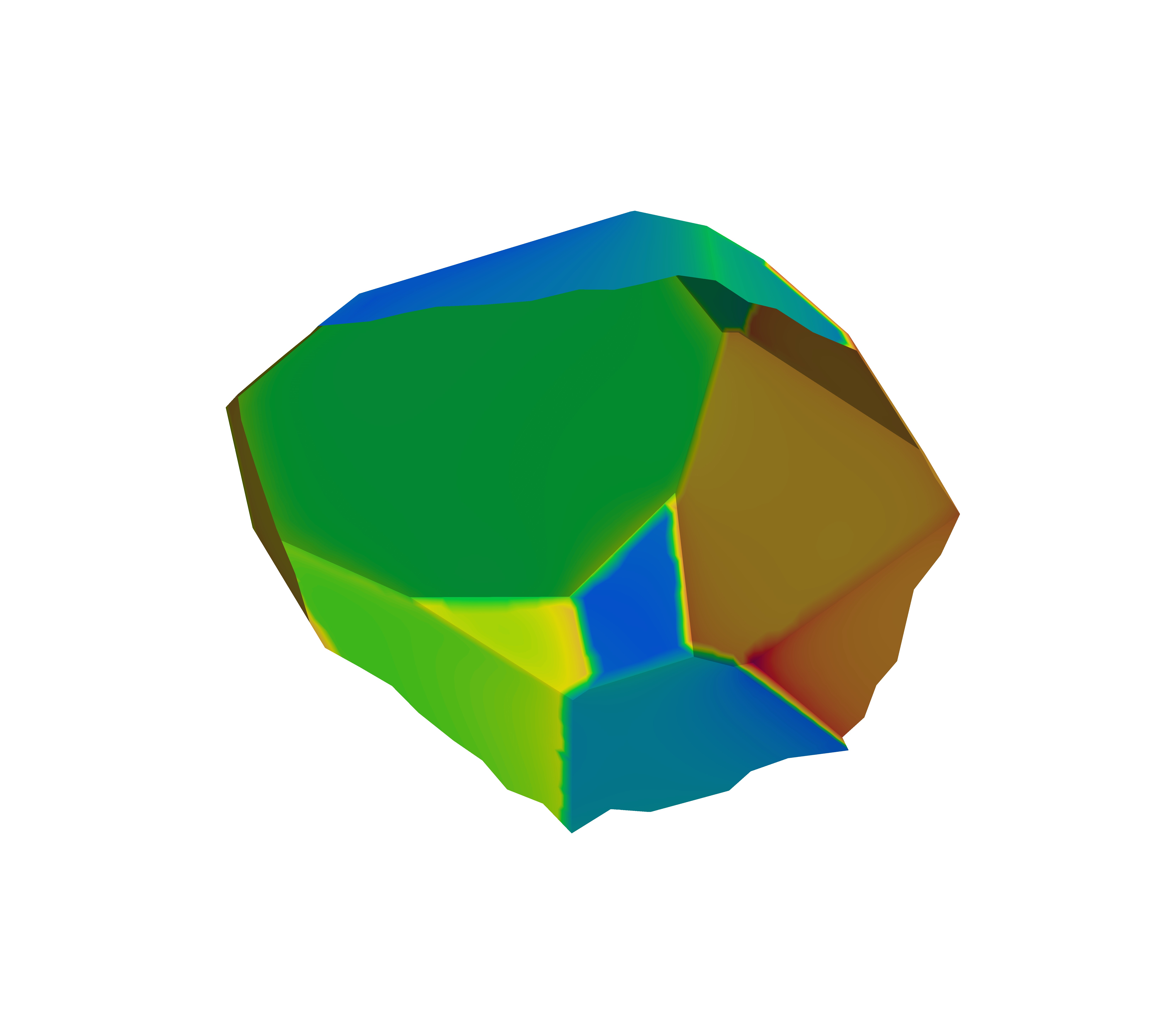}};
            \node[xshift=0.2cm,yshift=4cm,fill=white, rectangle,inner sep=3pt] {(b)};
            \begin{scope}[x={(image.south east)},y={(image.north west)}]
                \node[anchor=south west] at (0,0) {\includegraphics[scale=0.6]{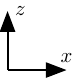}};
                \node[anchor=south west] at (current bounding box.south east) {\includegraphics[width=0.3\textwidth]{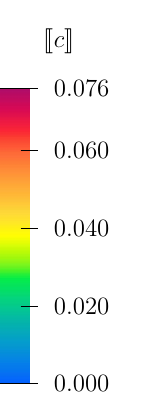}};
            \end{scope}
        \end{tikzpicture}
    \end{subfigure}
        \hfill
    \begin{subfigure}[b]{0.32\textwidth}
        \centering
        \begin{tikzpicture}
            \node[anchor=south west,inner sep=0] (image) at (0,0) {\includegraphics[width=0.7\textwidth, trim={30cm 0 30cm 0}, clip]{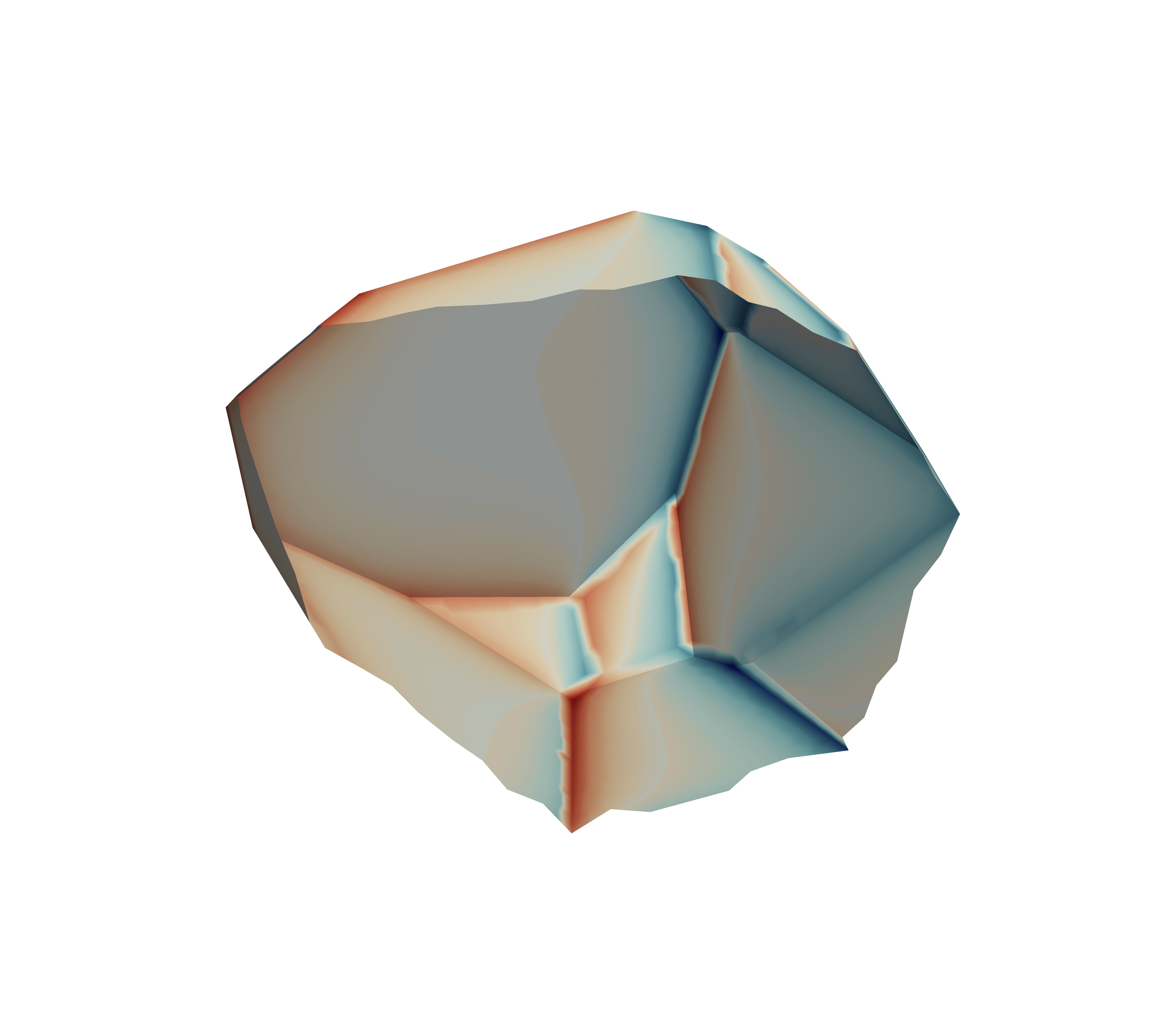}};
            \node[xshift=0.2cm,yshift=4cm,fill=white, rectangle,inner sep=3pt] {(c)};
            \begin{scope}[x={(image.south east)},y={(image.north west)}]
                \node[anchor=south west] at (0,0) {\includegraphics[scale=0.6]{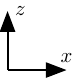}};
                \node[anchor=south west] at (current bounding box.south east) {\includegraphics[width=0.3\textwidth]{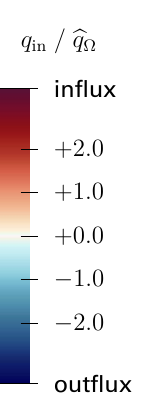}};
            \end{scope}
        \end{tikzpicture}
    \end{subfigure}
        \\ [0.5cm]
    \begin{subfigure}[b]{0.2\textwidth}
        \centering
        \begin{tikzpicture}
            \node[anchor=south west,inner sep=0] (image) at (0,0) {\includegraphics[width=\textwidth, trim={0 0 20cm 20cm}, clip]{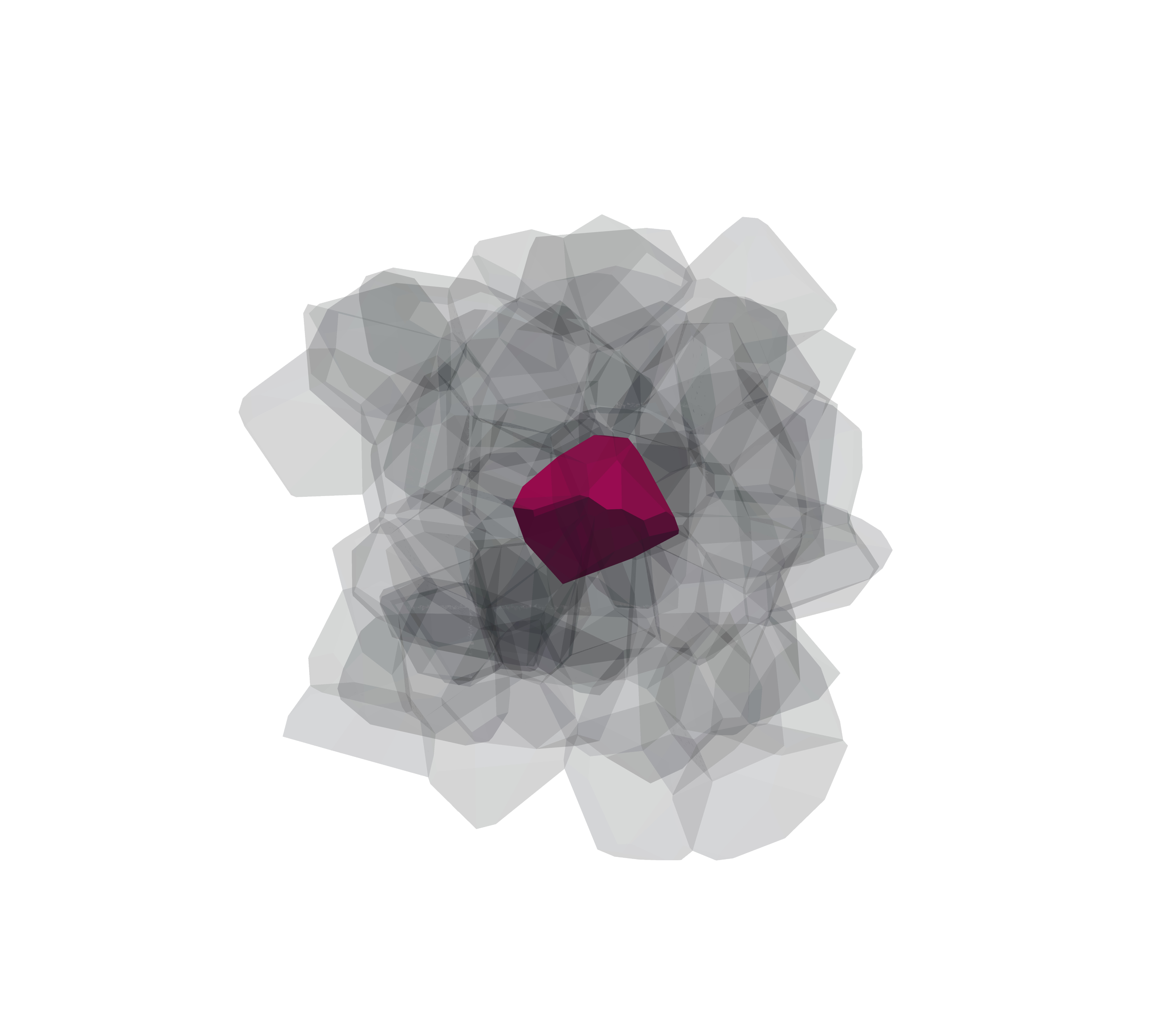}};
            \node[xshift=0.2cm,yshift=2.5cm,fill=white, rectangle,inner sep=3pt] {(d)};
            \begin{scope}[x={(image.south east)},y={(image.north west)}]
                \node[anchor=south west] at (0,0) {\includegraphics[scale=0.6]{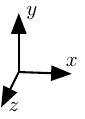}};
            \end{scope}
        \end{tikzpicture}
    \end{subfigure}
        \hfill
    \begin{subfigure}[b]{0.15\textwidth}
        \centering
        \includegraphics[width=\textwidth]{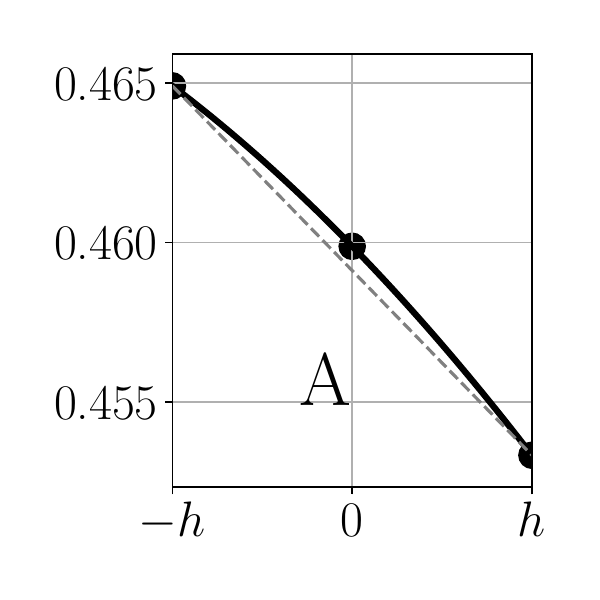}
    \end{subfigure}
        \hfill
    \begin{subfigure}[b]{0.15\textwidth}
        \centering
        \includegraphics[width=\textwidth]{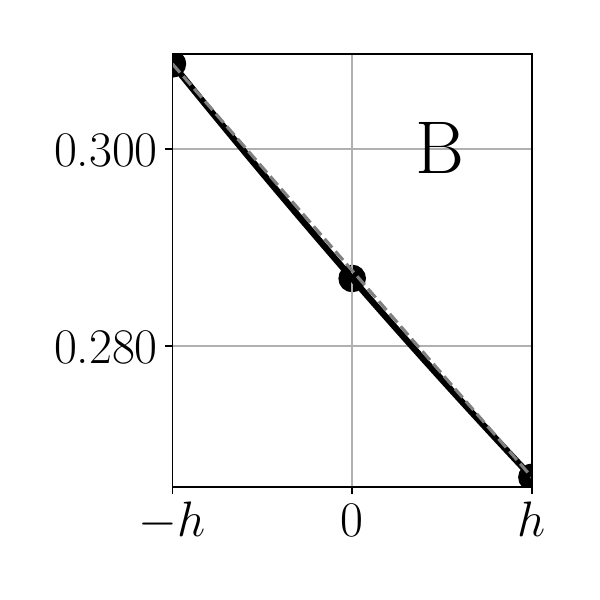}
    \end{subfigure}
        \hfill
    \begin{subfigure}[b]{0.15\textwidth}
        \centering
        \includegraphics[width=\textwidth]{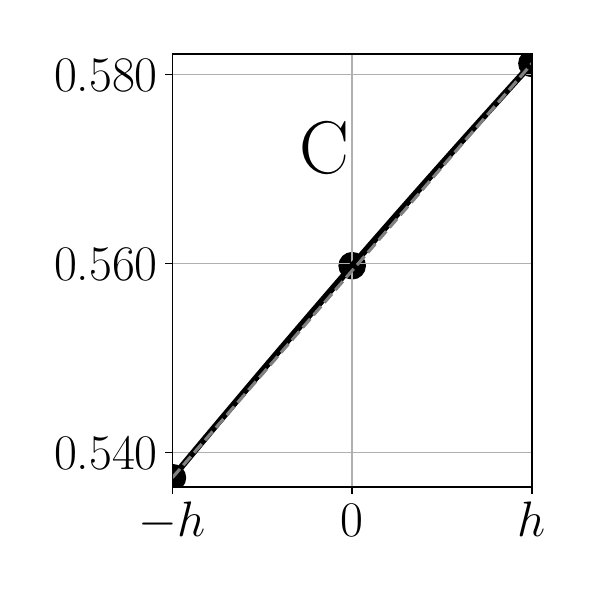}
    \end{subfigure}
        \hfill
    \begin{subfigure}[b]{0.15\textwidth}
        \centering
        \includegraphics[width=\textwidth]{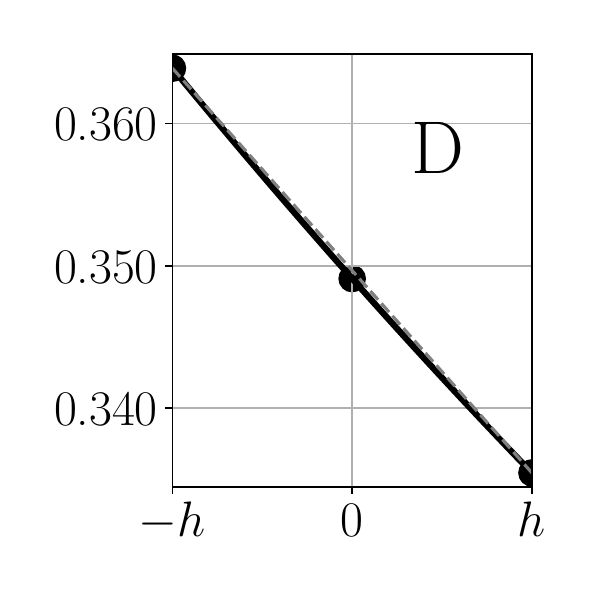}
    \end{subfigure}
        \hfill
    \begin{subfigure}[b]{0.15\textwidth}
        \centering
        \includegraphics[width=\textwidth]{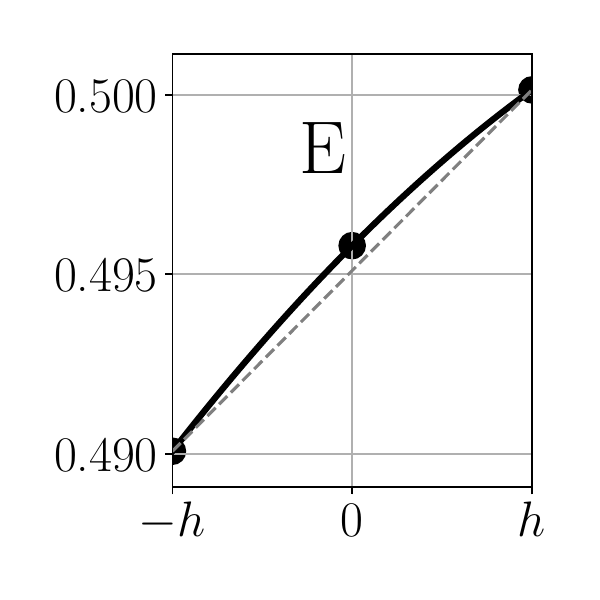}
    \end{subfigure}
    \caption{In-depth analysis of the GB domain for a selected grain (position in the geometry displayed at the bottom left) which is cut in half and a fixed parameter configuration ($D_\Vert=10 D_\Omega$, $D_\perp=0.1D_\Omega$, $h=0.0032l$, $\vfgb\approx 7.80\%$). (a) Reconstruction of the concentration profile. Note that the GB plane was extruded by a factor of $0.01l>h$ in this case to enhance the visibility of the discussed effects. For selected points, the concentration profile across the GB (in the direction of the respective arrow) is shown in (d). The dashed lines are added to underline the effect of the second-order mode. (b) Absolute concentration jump $\llbracket c\rrbracket$ across the GB. (c) Net influx into the GB as defined in \cref{eq:net_influx}. The flux is dedimensionalized with $\widehat{q}_\Omega=\|-D_\Omega \ol{\ul{g}}\|$. The sign of $q_{\rm in}$ determines the flow direction as indicated.}
    \label{fig:grain_concentration}
\end{figure}

The reconstruction of the 3D GB volume is achieved by shifting the top and bottom node layer of a given collapsed interface outwards and inwards, respectively, by the GB thickness parameter $h$. In the following, an arbitrary grain within the geometry (highlighted in \Cref{fig:grain_concentration}) is used for an in-depth analysis. In \Cref{fig:grain_concentration}, the concentration field within the extruded GB volume is shown for the selected grain to underline the capabilities of this reconstruction approach.

In the following the interplay of \Cref{fig:alpha_stat} and \Cref{fig:grain_concentration} allows for detailed insights into the global relevance of the different modes as well as their dependence on local features such as interface orientation.

Starting from the limit case of the \textit{neutral} regime, the concentration jump across the GB, represented by $\alpha_1(\fx_\Vert)$, grows significantly for $D_\perp\ll D_\Omega$ as shown in \Cref{fig:alpha_stat}. Increasing $D_\perp/D_\Omega$ (that is, lowering the resistance) leads to negligible concentration jumps. This implies a strong dependence of the jump magnitude on the orientation of the interface with respect to the direction of the imposed gradient which can be verified in \Cref{fig:grain_concentration} for the \textit{connecting} regime\footnote{In fact the largest concentration jumps and second order coefficients can be observed within the \textit{connecting} regime (see \Cref{fig:alpha_stat}) which was the motivation for the choice of this parameter configuration in \Cref{fig:grain_concentration}.}: Larger jumps are observed the more the normal vector aligns with the direction of the imposed gradient (i.e., the $x$-direction in the case of \Cref{fig:grain_concentration}). Due to the nearly isotropic structure of the geometry, the distribution of $\alpha_1(\fx_\Vert)$ in \Cref{fig:alpha_stat} is relatively uniform for $D_\Vert\ll D_\Omega$. However, for $D_\Vert\gg D_\Omega$ the enhanced transport along the GB leads to a stronger impact of the local geometry arrangement and smears out the distribution profile, leading to even larger jumps and increased spatial variability of the concentration profile.

The second-order term $\alpha_2(\fx_\Vert)$ accounts for the channeling effect of atomic transport along GBs. The distribution in \Cref{fig:alpha_stat} shows a similar dependence on $D_\perp$ as described for the case of $\alpha_1$: A larger resistance in normal direction ($D_\perp\ll D_\Omega$) enables the formation of a second-order profile. The range of $\alpha_2(\fx_\Vert)$ is additionally enhanced by increasing $D_\Vert/D_\Omega$. Hence, the \textit{connecting} regime exhibits the strongest channeling effects that we explicitly wanted to account for by Assumption \labelcref{A:quadratic}. In alignment with the concentration jump, the second-order mode also depends on the orientation of the interface: In the displayed concentration profiles of \Cref{fig:grain_concentration}, a growing second-order mode can be observed as the GB plane aligns with the direction of the imposed gradient. This can be better understood when considering the $\alpha_2$-related net influx for the selected grain in the top right of \Cref{fig:grain_concentration}: It shows a net influx of mobile species on the left of the facets and an outflux on the right of the facets. A net outflux ($q_{\rm in}<0$) corresponds to an accumulation of mobile species at the middle plane, whereas a net influx ($q_{\rm in}>0$) is the result of a local minimum of mobile species at the GB center. This induces enhanced transport along the GB middle plane from local maximum to local minimum, which is consistent with the symmetric density profiles that were found for $\alpha_2$ in \Cref{fig:alpha_stat}. The role of GB junctions with respect to the second-order modes should also be mentioned briefly: At any junction where different GB segments meet, the central node on the middle layer is shared among all segments. This approximation necessarily introduces artifacts along grain edges and corners that might locally alter the reconstructed profiles and modes.
{\color{rev} However, the diffusion behavior at multi-junctions (corners, edges) has not been studied sufficiently to our knowledge. Given this uncertainty in the behavior, the results from the 2D study already highlight that the proposed GB model is able to closely match the results for different setups at the junctions. We just want to highlight that the local behavior in these regions should be interpreted with care.
This holds, e.g., for extreme second-order modes close to the junctions in \Cref{fig:grain_concentration} and can lead to the rapid decay of the $\alpha_2$-density profiles in \Cref{fig:alpha_stat} in the \textit{connecting} regime.}

\begin{figure}[h]
    \centering
    \begin{subfigure}[b]{0.45\textwidth}
        \begin{tikzpicture}
            \node[anchor=south west,inner sep=0] (image) at (0,0) {\includegraphics[width=0.75\textwidth,trim={30cm 10cm 30cm 20cm},clip]{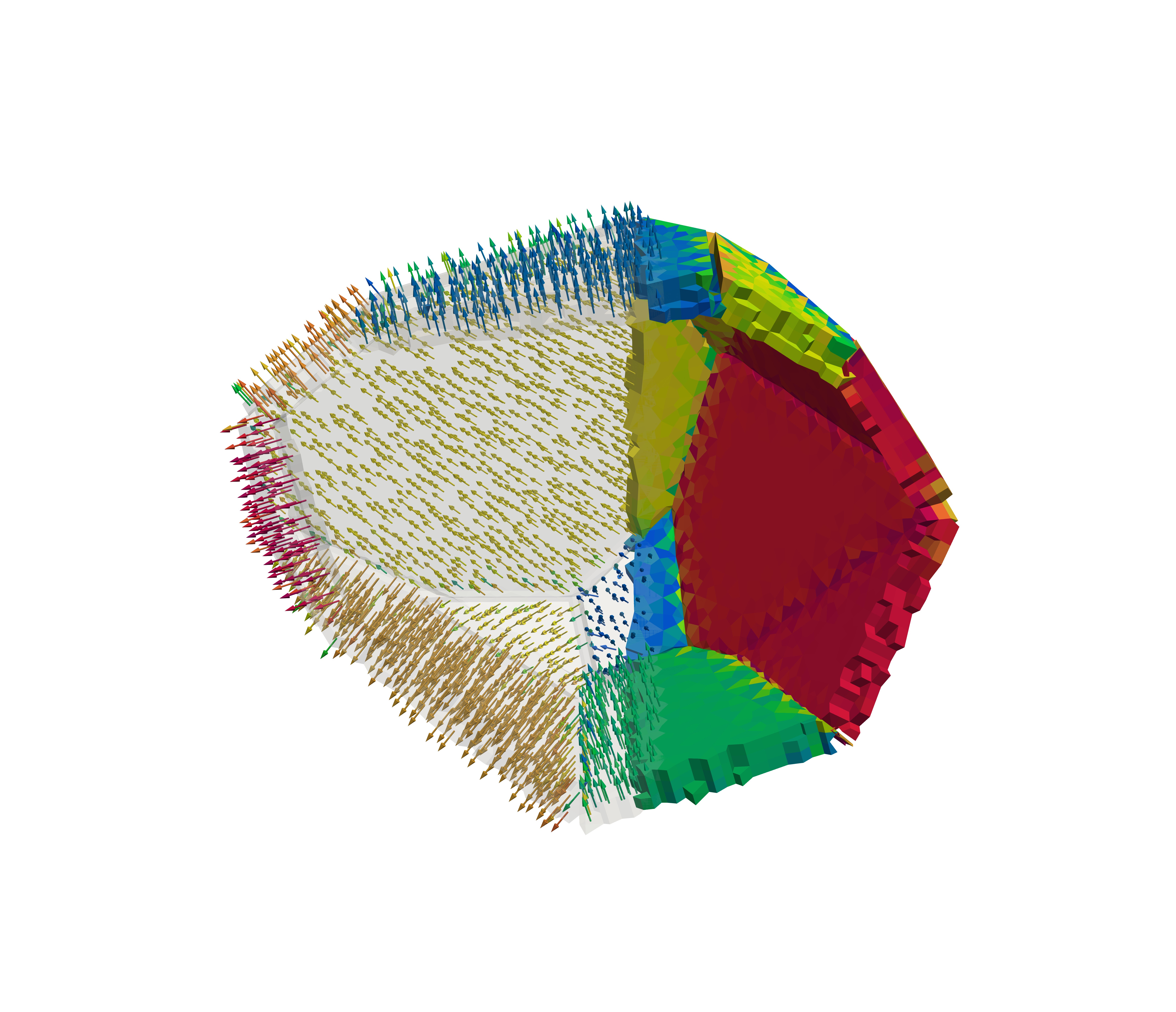}};
            \begin{scope}[x={(image.south east)},y={(image.north west)}]
                \node[anchor=south west] at (0,0) {\includegraphics[width=0.2\textwidth]{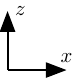}};
                \node[anchor=south west] at (current bounding box.south east) {\includegraphics[width=0.25\textwidth]{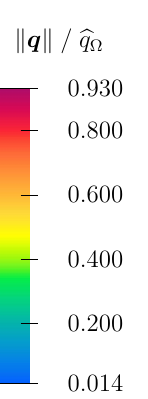}};
            \end{scope}
        \end{tikzpicture}
        \caption{$D_\Vert=0.01D_\Omega,~D_\perp=100D_\Omega$ (\textit{neutral})}
        \label{fig:grain_flux_a}
    \end{subfigure}
        \hfill
    \begin{subfigure}[b]{0.45\textwidth}
        \begin{tikzpicture}
            \node[anchor=south west,inner sep=0] (image) at (0,0) 
            {\includegraphics[width=0.75\textwidth,trim={30cm 10cm 30cm 20cm},clip]{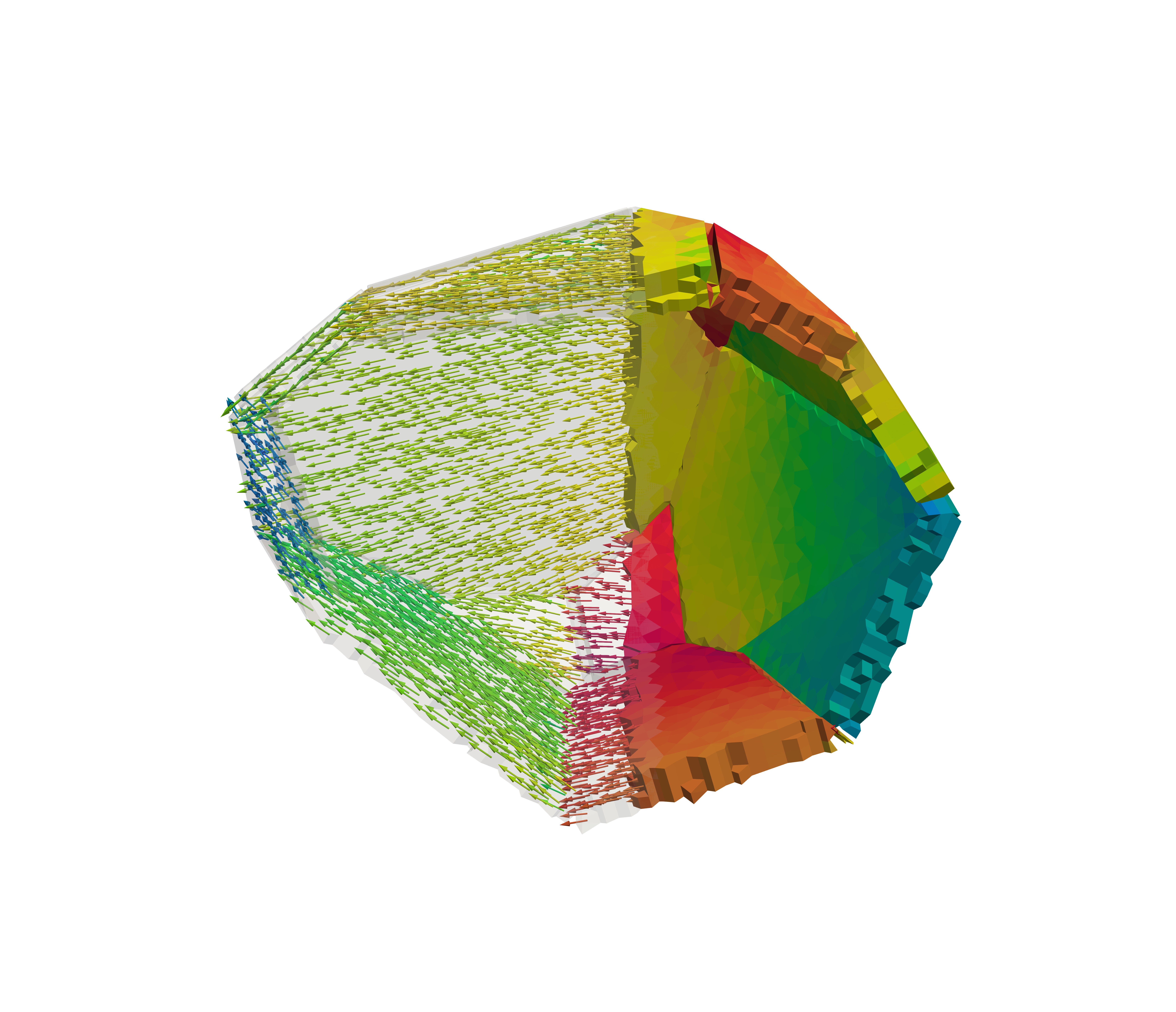}};
            \begin{scope}[x={(image.south east)},y={(image.north west)}]
                \node[anchor=south west] at (0,0) {\includegraphics[width=0.2\textwidth]{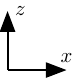}};
                \node[anchor=south west] at (current bounding box.south east) {\includegraphics[width=0.25\textwidth]{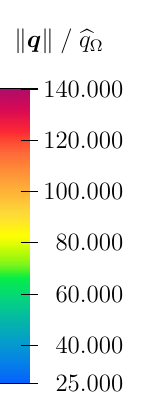}};
            \end{scope}
        \end{tikzpicture}
        \caption{$D_\Vert=100D_\Omega,~D_\perp=100D_\Omega$ (\textit{enhancing})}
        \label{fig:grain_flux_b}
    \end{subfigure}
        \\
    \begin{subfigure}[b]{0.45\textwidth}
        \begin{tikzpicture}
            \node[anchor=south west,inner sep=0] (image) at (0,0) 
            {\includegraphics[width=0.75\textwidth,trim={30cm 10cm 30cm 20cm},clip]{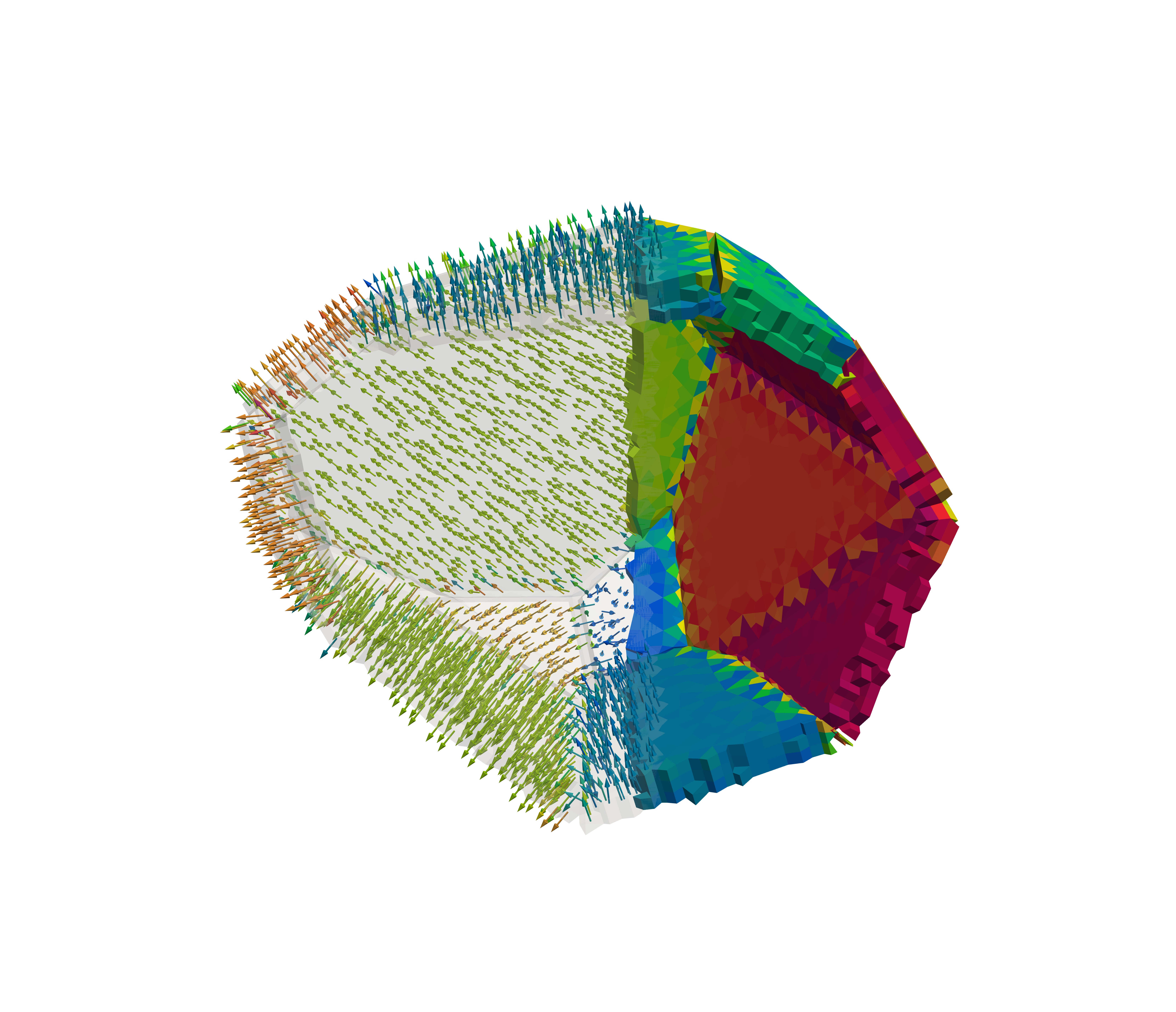}};
            \begin{scope}[x={(image.south east)},y={(image.north west)}]
                \node[anchor=south west] at (0,0) {\includegraphics[width=0.2\textwidth]{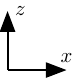}};
                \node[anchor=south west] at (current bounding box.south east) {\includegraphics[width=0.25\textwidth]{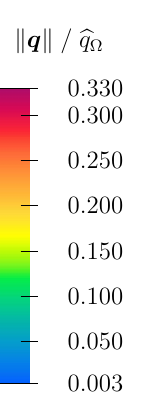}};
            \end{scope}
        \end{tikzpicture}
        \caption{$D_\Vert=0.01D_\Omega,~D_\perp=0.01D_\Omega$ (\textit{blocking})}
        \label{fig:grain_flux_c}
    \end{subfigure}
        \hfill
    \begin{subfigure}[b]{0.45\textwidth}
        \begin{tikzpicture}
            \node[anchor=south west,inner sep=0] (image) at (0,0) 
            {\includegraphics[width=0.75\textwidth,trim={30cm 10cm 30cm 20cm},clip]{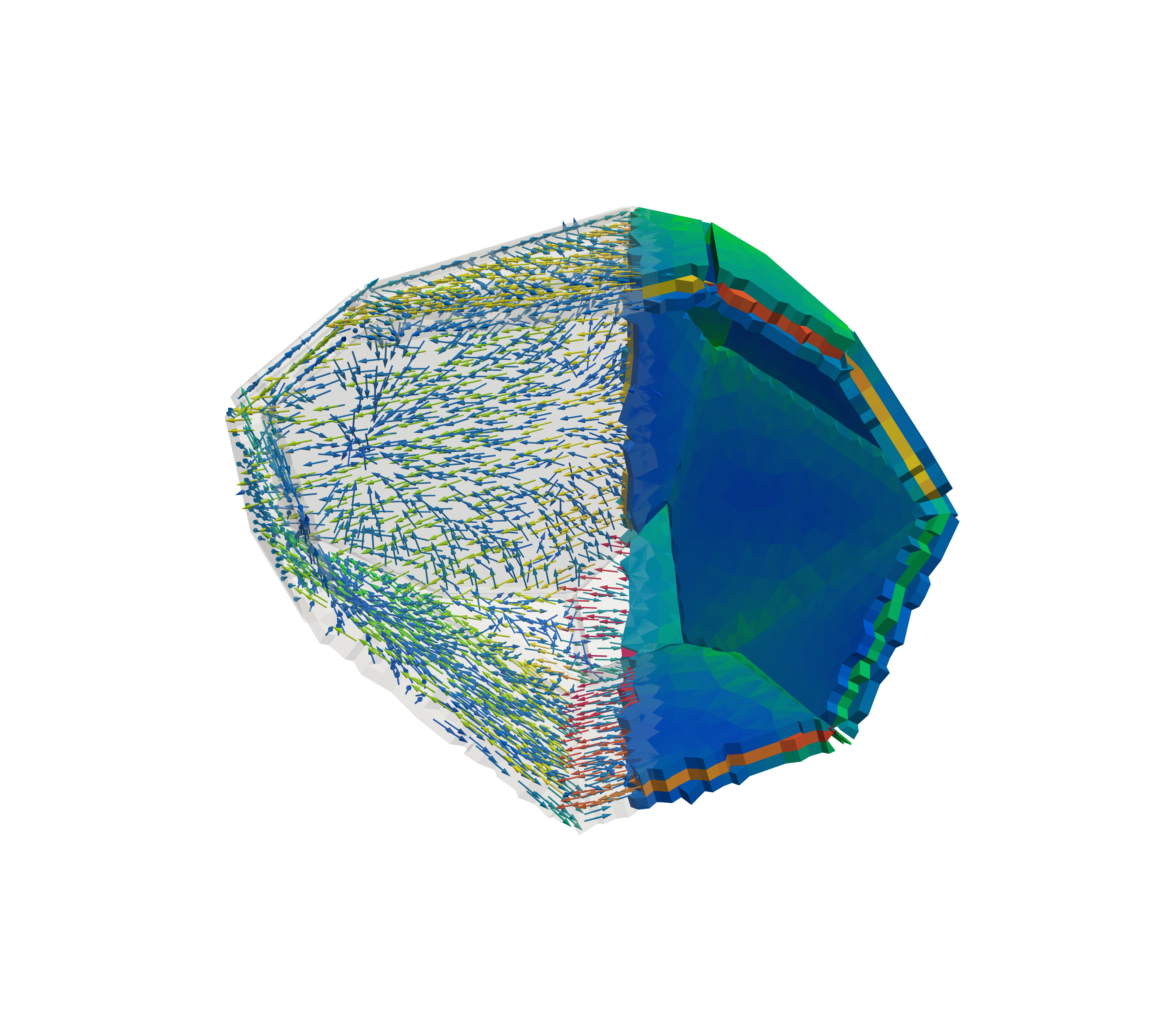}};
            \begin{scope}[x={(image.south east)},y={(image.north west)}]
                \node[anchor=south west] at (0,0) {\includegraphics[width=0.2\textwidth]{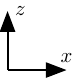}};
                \node[anchor=south west] at (current bounding box.south east) {\includegraphics[width=0.25\textwidth]{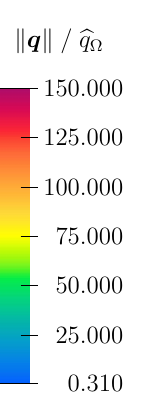}};
            \end{scope}
        \end{tikzpicture}
        \caption{$D_\Vert=100D_\Omega,~D_\perp=0.01D_\Omega$ (\textit{connecting})}
        \label{fig:grain_flux_d}
    \end{subfigure}
    \caption{Dedimensionalized flux within the extruded GB of the selected grain introduced in \Cref{fig:grain_concentration} for an imposed gradient $\ol{\ul{g}}$ in $x$-direction: The flux is dedimensionalized with $\widehat{q}_\Omega=\|-D_\Omega \ol{\ul{g}}\|$. Magnitude as well as direction of the flux field are displayed for varying combinations of $D_\Vert/D_\Omega$ and $D_\perp/D_\Omega$. In accordance with \Cref{fig:grain_concentration} we set $h=0.0032l$ such that $\vfgb\approx 7.80\%$. However, the GB was extruded by $0.01l$ to show an enhanced GB effect.}
    \label{fig:grain_flux}
\end{figure}

To shed light on the interplay of the discussed GB transport mechanisms, the GB flux fields of the four limiting cases of the different regimes are compared. In \Cref{fig:grain_flux} 3D fields are reconstructed in the same way as the concentration profiles in \Cref{fig:grain_concentration}. This enables direct insight into where mobile species enter, diffuse along, and exit the GB. 

For the \textit{neutral} (\Cref{fig:grain_flux_a}) and the \textit{blocking} regime (\Cref{fig:grain_flux_c}), the flux aligns with the interface normal. Its magnitude exhibits a significant dependence on the respective facet orientation, as observed for the first mode in \Cref{fig:grain_concentration}, where the flux increases the more aligned the facet normal and direction of the imposed gradient are. For the \textit{blocking} regime, however, the magnitude of the flux is further altered by the local geometry arrangement. 

Opposed to that, the increase in $D_\Vert/D_\Omega$ leads to more in-plane dominated transport in case of the \textit{enhancing} (\Cref{fig:grain_flux_b}) and \textit{connecting} regime (\Cref{fig:grain_flux_d}). Despite that, the characteristics of the two regimes still differ significantly. For the \textit{enhancing} regime, the flux magnitude increases if interfaces align with the direction of the imposed gradient, indicating that mobile species do not deviate significantly from the shortest diffusion paths in a global sense. This aligns with the properties of the second-order mode in \Cref{fig:grain_concentration}. At the same time the local geometry causes variance in magnitude within the individual facets.

Lastly, the channeling effect in the \textit{connecting} regime as observed in the concentration profiles of \Cref{fig:grain_concentration} is clearly visible in \Cref{fig:grain_flux_d}: An amplified in-plane flux within the middle layer of the GB can be observed that is impacted by the orientation of the GB segments. In the top and bottom layers of the GB, which are directly coupled with the bulk domain, flux happens in the inverse direction. The channeling effect causes a suction effect that alters the transport behavior in the surrounding bulk domain. Hence, the transport within the grains themselves becomes less relevant, and the GB channels connect the individual grains, which only show little fluctuation from constant concentration levels. The significant local variations in the orientation of the in-plane flux hint at the strong impact of the local geometry arrangement. 

\subsection{Quantification of GB impacts on effective diffusivity}
\label{Sec:results:eff}
\begin{figure}[h]
    \centering
    \begin{subfigure}[t]{\textwidth}
        \centering
        \begin{subfigure}[b]{0.32\textwidth}
            \centering
            \includegraphics[width=\textwidth]{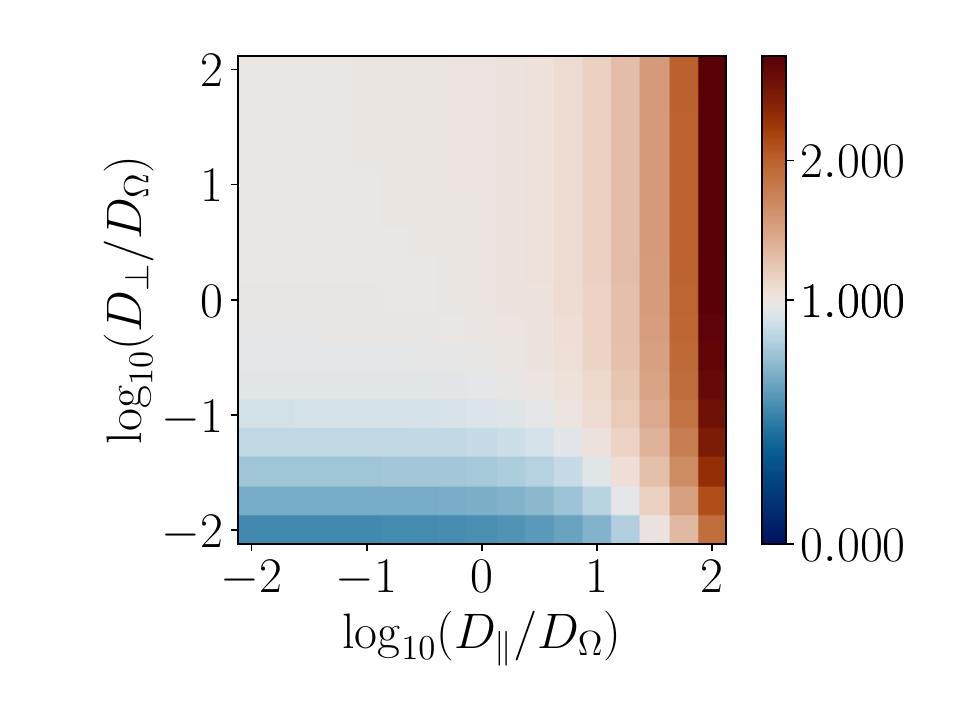}
            \caption{$\lambda_0$}
            \label{fig:D_eff_-3.0_eigval_1_largest}
        \end{subfigure}
        \hfill
        \begin{subfigure}[b]{0.32\textwidth}
            \centering
            \includegraphics[width=\textwidth]{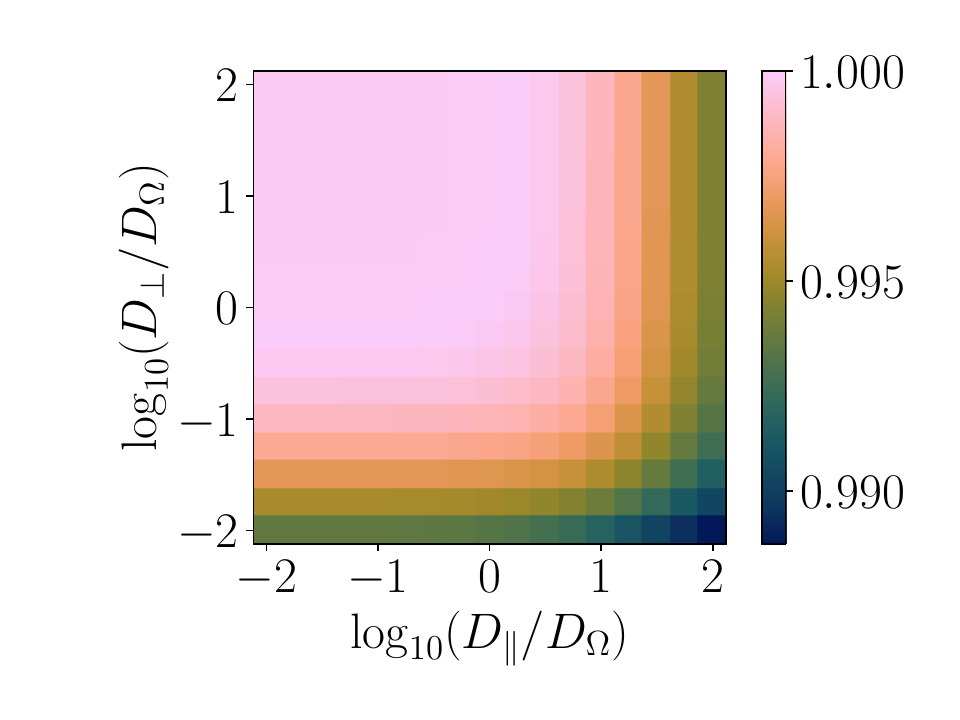}
            \caption{$\lambda_1/\lambda_0$}
            \label{fig:D_eff_h_-3.0_eigval_2_rel}
        \end{subfigure}
        \hfill
        \begin{subfigure}[b]{0.32\textwidth}
            \centering
            \includegraphics[width=\textwidth]{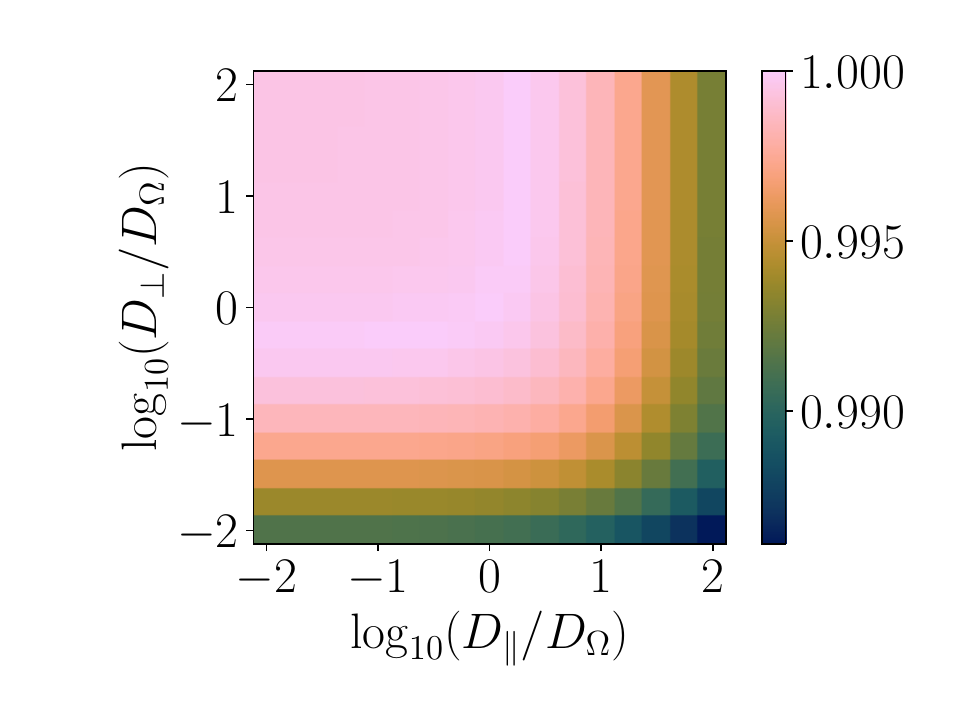}
            \caption{$\lambda_2/\lambda_0$}
            \label{fig:D_eff_h_-3.0_eigval_3_rel}
        \end{subfigure}
        \captionsetup{labelformat=empty}
        \caption*{Relation $h=0.0010l$ ($\vfgb\approx 2.67\%$) was fixed for all of the above parameter variations.}
    \end{subfigure}

    \begin{subfigure}[t]{\textwidth}
        \centering
        \begin{subfigure}[b]{0.32\textwidth}
            \centering
            \includegraphics[width=\textwidth]{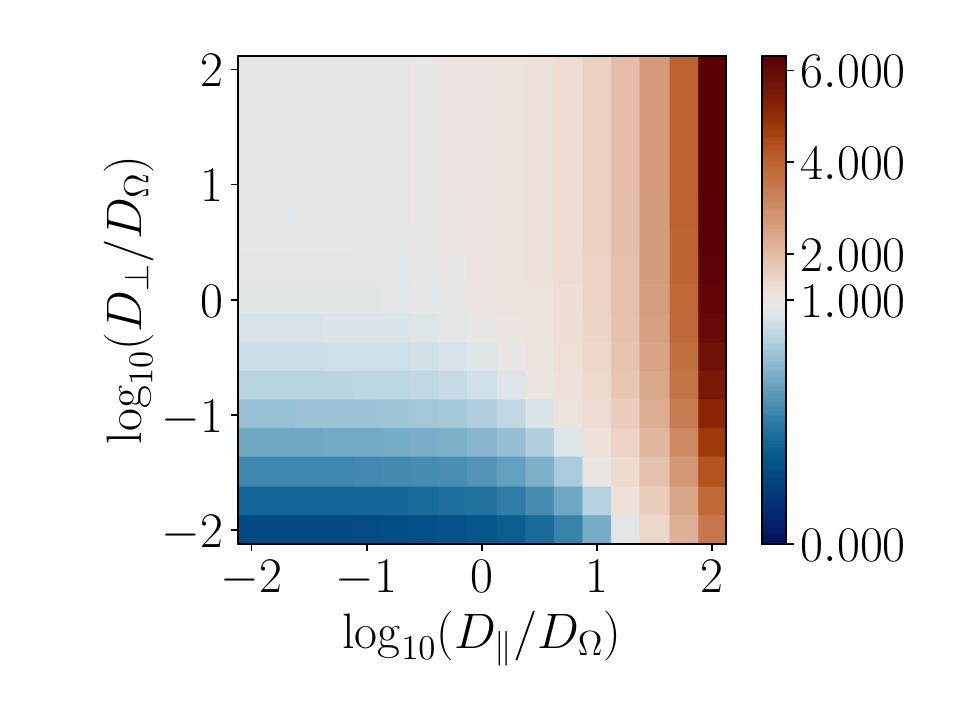}
            \caption{$\lambda_0$}
            \label{fig:D_eff_-2.5_eigval_1_largest}
        \end{subfigure}
        \hfill
        \begin{subfigure}[b]{0.32\textwidth}
            \centering
            \includegraphics[width=\textwidth]{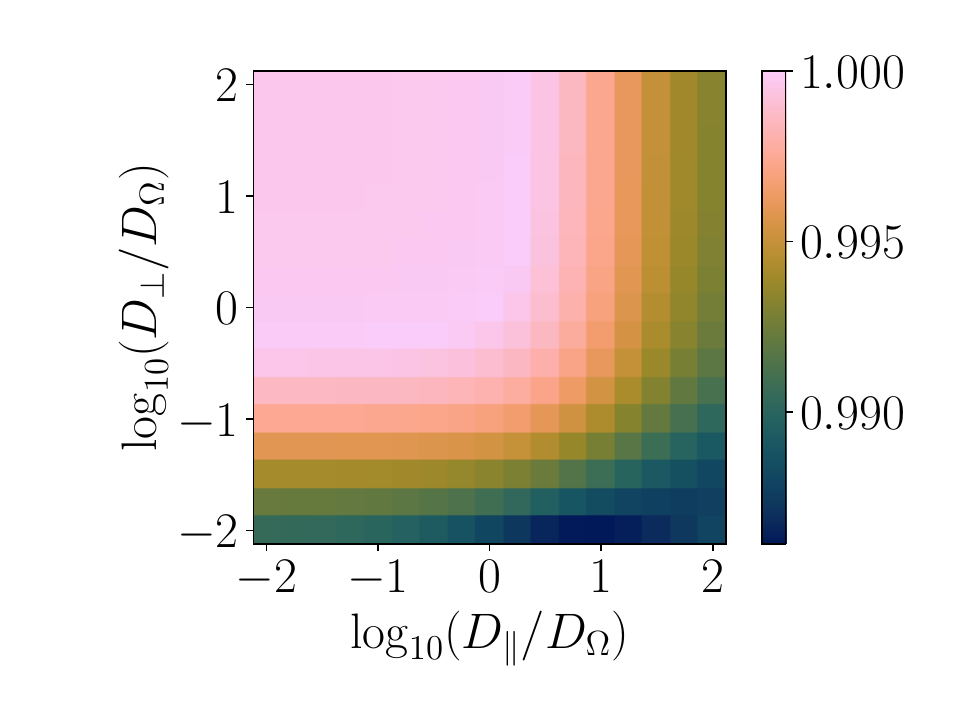}
            \caption{$\lambda_1/\lambda_0$}
            \label{fig:D_eff_h_-2.5_eigval_2_rel}
        \end{subfigure}
        \hfill
        \begin{subfigure}[b]{0.32\textwidth}
            \centering
            \includegraphics[width=\textwidth]{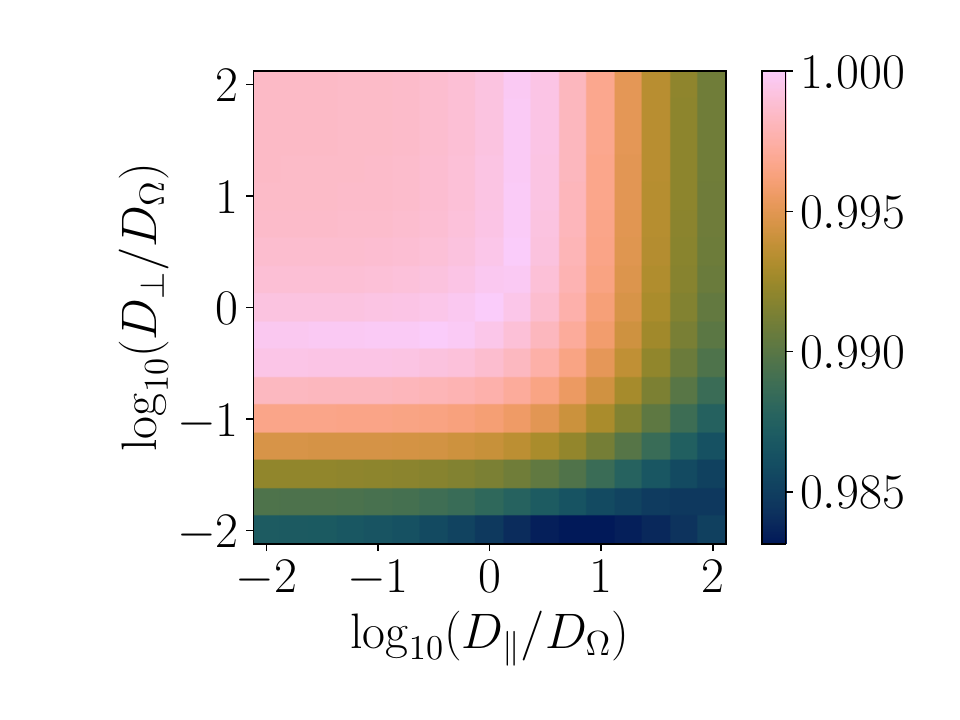}
            \caption{$\lambda_2/\lambda_0$}
            \label{fig:temp_D_eff_h_-2.5_eigval_3_rel}
        \end{subfigure}
        \captionsetup{labelformat=empty}
        \caption*{Relation $h=0.0032l$ ($\vfgb\approx 7.80\%$) was fixed for all of the above parameter variations.}
    \end{subfigure}
    \caption{Eigenvalues $\lambda_i$ of the dedimensionalized effective diffusivity tensor $\ol{\ull{D}} / D_\Omega$ for varying diffusion parameters in the GB. For comparability, the largest eigenvalue $\lambda_0$ and the ratios of $\lambda_1$ and $\lambda_2$ are given, respectively. }
    \label{fig:D_eff}
\end{figure}

In many applications, it is the average behavior or effective diffusion tensor on the mesoscale resulting from a certain microscale geometry and material data that is of primary interest. Hence, the effects of parameter variations on the effective behavior are further investigated. The eigenvalues of the diffusion tensor $\ol{\ull{D}}$ for different slices of the parameter space are shown in \Cref{fig:D_eff}.

The largest eigenvalue of the diffusion tensor is a suitable indicator for the behavior of the entire tensor due to the close-to-isotropic structure of the considered geometry (cf. \cref{eq:S2_setup}). For the \textit{neutral} regime, the effective eigenvalue tends towards that of the bulk. Opposed to that, the effective diffusion coefficients are maximized in the case of the \textit{enhancing} regime due to the additional GB transport. It is further increased for a larger thickness parameter $h$ of the GB and, therefore, larger volume fractions of the more conductive GB. This effect is clearly governed by $D_\Vert \gg D_\Omega$ since the \textit{connecting} regime also exhibits a larger eigenvalue, albeit to a lesser extent. Considering that small $D_\perp$ (or equivalently large~$h$) limits the ability to cross the GB, the effective eigenvalue is found to be decreased below $D_\Omega$ for the \textit{blocking} regime. In summary, the effective diffusion behavior is mainly altered for $D_\perp\ll D_\Omega$ and/or $D_\Vert\gg D_\Omega$. This again stresses the importance of accounting for an anisotropic GB, i.e., for $D_\Vert$ and $D_\perp$ separately.

In general, the underlying anisotropy of the polycrystalline geometry affects the effective diffusion tensor structurally and quantitatively. However, \Cref{fig:D_eff} shows that the geometrical anisotropies have a varying impact within the different parametric regimes. For the bulk-like behavior in the \textit{neutral} regime, the specific arrangement and orientation of the GB segments do not really play a role. In this case we find $\lambda_0 \approx \lambda_1 \approx \lambda_2${\color{rev}, which is also hinted at by the near perfectly isotropic structural tensor $\ull{S}_2$, see also \cref{Sec:geo_quant}. The authors would like to emphasize that the collapsed GB model operates independent of the microstructural anisotropy, which will be studied explicitly in future studies.}

As pointed out in \Cref{Sec:results:localizing}, the diffusion behavior becomes less dependent on the GB segment orientation and more dependent on the local geometry arrangement for $D_\perp\ll D_\Omega$. This observation is underlined by the effective diffusion tensor showing the most pronounced anisotropy in the \textit{blocking} and \textit{connecting} regime. A similar but weaker effect can be seen for the \textit{enhancing} regime. The use of GBs as preferred diffusion paths also stresses the dependence on geometry and anisotropic features.

\begin{figure}[h]
    \begin{subfigure}[b]{0.32\textwidth}
        \centering
        \includegraphics[width=\textwidth]{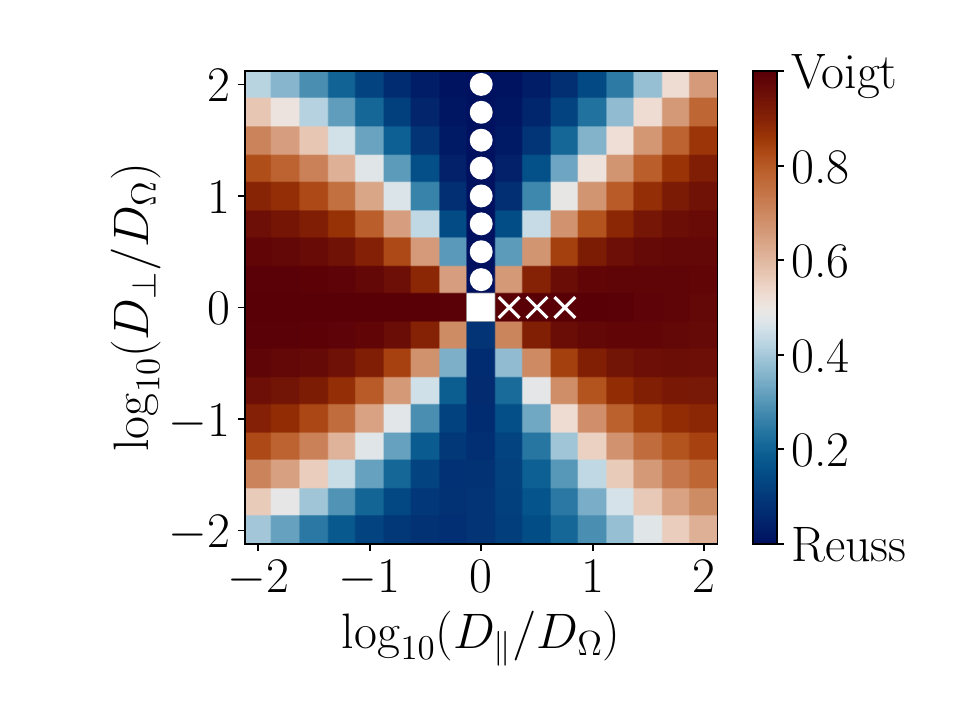}
        \caption{$h=0.0010l~(\vfgb\approx 2.67\%)$}
        \label{fig:D_eff_rel_Voigt_Reuss_h_-3.0}
    \end{subfigure}
    \hfill
    \begin{subfigure}[b]{0.32\textwidth}
        \centering
        \includegraphics[width=\textwidth]{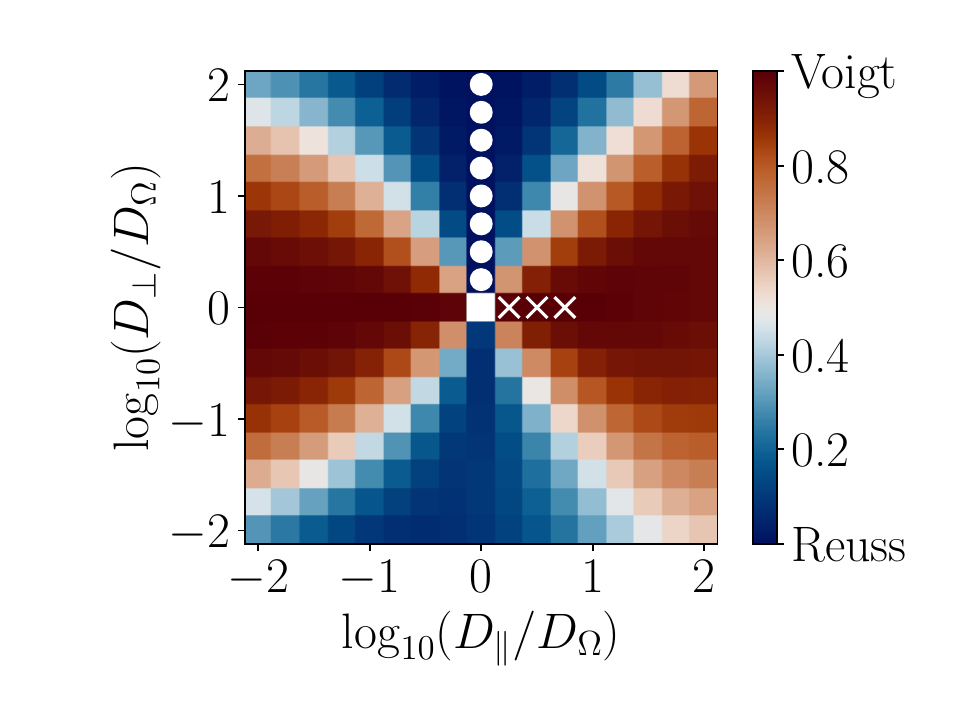}
        \caption{$h=0.0018l~(\vfgb\approx 4.65\%)$}
        \label{fig:D_eff_rel_Voigt_Reuss_h_-2.75}
    \end{subfigure}
    \hfill
    \begin{subfigure}[b]{0.32\textwidth}
        \centering
        \includegraphics[width=\textwidth]{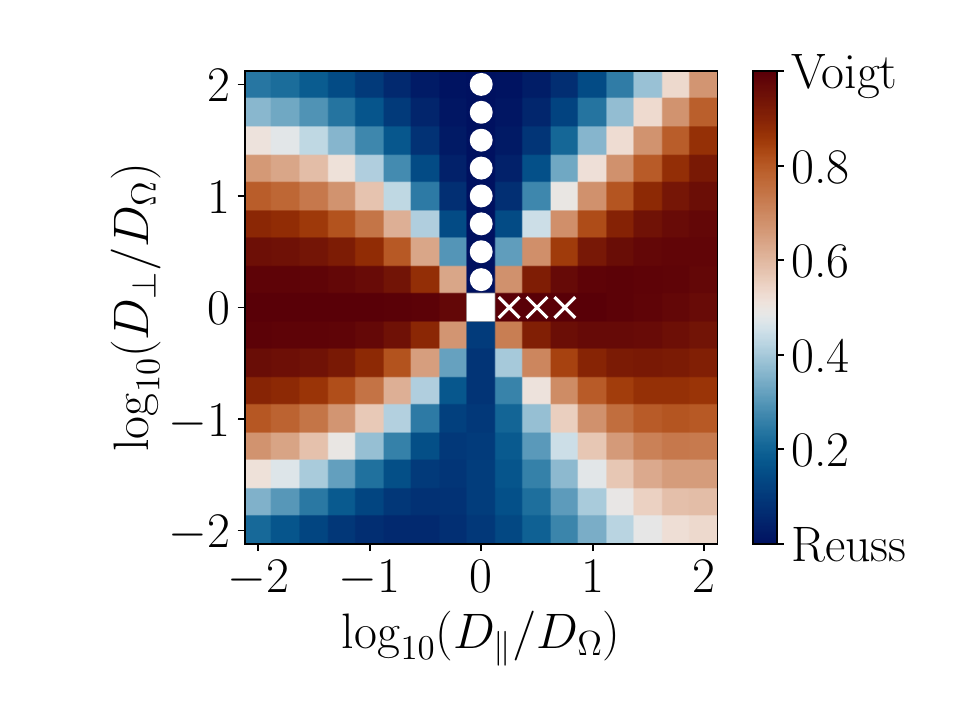}
        \caption{$h=0.0032l~(\vfgb\approx 7.80\%)$}
        \label{fig:D_eff_rel_Voigt_Reuss_h_-2.5}
    \end{subfigure}
    \caption{\protect Relation of the effective diffusivity tensor to the analytical Voigt bound (value 1) and the Reuss bound (value 0) for varying GB thickness parameter $h$. The computations follow the remarks in \Cref{Sec:Voigt-Reuss}. Numerical violations of the bounds are marked with $\circ$ for the Reuss bound and $\boldsymbol{\times}$ for the Voigt bound.}
    \label{fig:Voigt-Reuss}
\end{figure}

The relation of $\ol{\ull{D}}$ with respect to the analytical Voigt--Reuss bounds as derived in \cref{eq:voigt} and \cref{eq:reuss} is shown in \Cref{fig:Voigt-Reuss}. The Voigt--Reuss relations for the given 3D polycrystal example show a great similarity with the 2D example (cf. \Cref{fig:2d_voigt_reuss_scaling}): For extreme (that is very large of very small) $D_\Vert$-configurations the solution is close to the Voigt bound while for extreme $D_\perp$-configurations it is close to the Reuss bound. This behavior stems from the observation mentioned above that $D_\perp$, in general, has a rather hindering impact on the overall diffusion behavior and, therefore, pushes the effective diffusion tensor closer to the lower bound. Opposed to that, the additional transport that is enabled through in-plane diffusion along the GBs yields effective responses close to the upper bound. For an isotropic GB ($D_\Vert=D_\perp$), the effective diffusion coefficient roughly corresponds to Hill's estimate \cite{Hill1952}, which is the arithmetic mean of the upper (Voigt) and lower (Reuss) bounds. Interestingly, a similar effect can be observed for cases when $D_\Vert D_\perp = 1$. This implies that $D_\Vert$ and $D_\perp$ can partly balance out each other's effects. 

As discussed for the 2D validation example, slight numerical violations of the bounds are sometimes obtained in a narrow parameter range, as highlighted. However, their magnitude is less than 0.1\%, meaning it is in a similar range compared to the 2D example. Further, the structure of the relation in the parameter space gradually loses its symmetry for increased volume fractions, i.e., larger values for $h/l$. One possible interpretation is the growing inapplicability of the proposed model and the compensation (see \Cref{Sec:eff}) for too-thick GBs. Note that the determination of a distinct limit value for the validity of the model, however, is non-trivial.

{\color{rev}
\section{Application to the solid-state electrolyte \texorpdfstring{Li\textsubscript{6}PS\textsubscript{5}Cl}{Li6PS5Cl}}
\label{Sec:case_study}

The argyrodite \Argyro{} is considered a promising candidate for solid-state electrolyte applications, see, e.g., Ref.~\cite{Deiseroth2008Jan}. In a recent study~\cite{Ou2024}, diffusion parameters in the form of the Arrhenius relation (see \cref{eq:arrhenius}) were obtained from molecular dynamics simulations for the bulk and three types of GBs in \Argyro{} based on \textit{ab initio}-based machine-learning interatomic potentials. A complex GB impact on atomic diffusivity was revealed, leading to temperature-dependent, anisotropic diffusion behavior in the GB regions. Depending on the adjacent bulk structures, GBs may enhance or hinder atomic diffusion, exhibiting diffusion coefficients that differ by up to approximately one order of magnitude from the bulk at room temperature. For FE simulations of \Argyro{} with polycrystalline microstructures, both the Fisher model~\cite{Gibbs1966, Fisher1951} and the classical cohesive element approach~\cite{Peng2024} are inadequate for describing essential GB effects due to mismatches in the applicable diffusion regimes. In contrast, the GB diffusion model presented in this study is well-suited for application to \Argyro{}. For brevity, we consider the isotropic anion-ordered bulk together with the GB type $\Sigma 5(001)[001]$ here since it shows the most significant anisotropy among the GB types studied in Ref.~\cite{Ou2024}. Parameters and further remarks on the simulation setup are given in \Cref{app:realistic_setup}. 
}

\begin{figure}[h]
    \centering
    \includegraphics[width=0.8\textwidth]{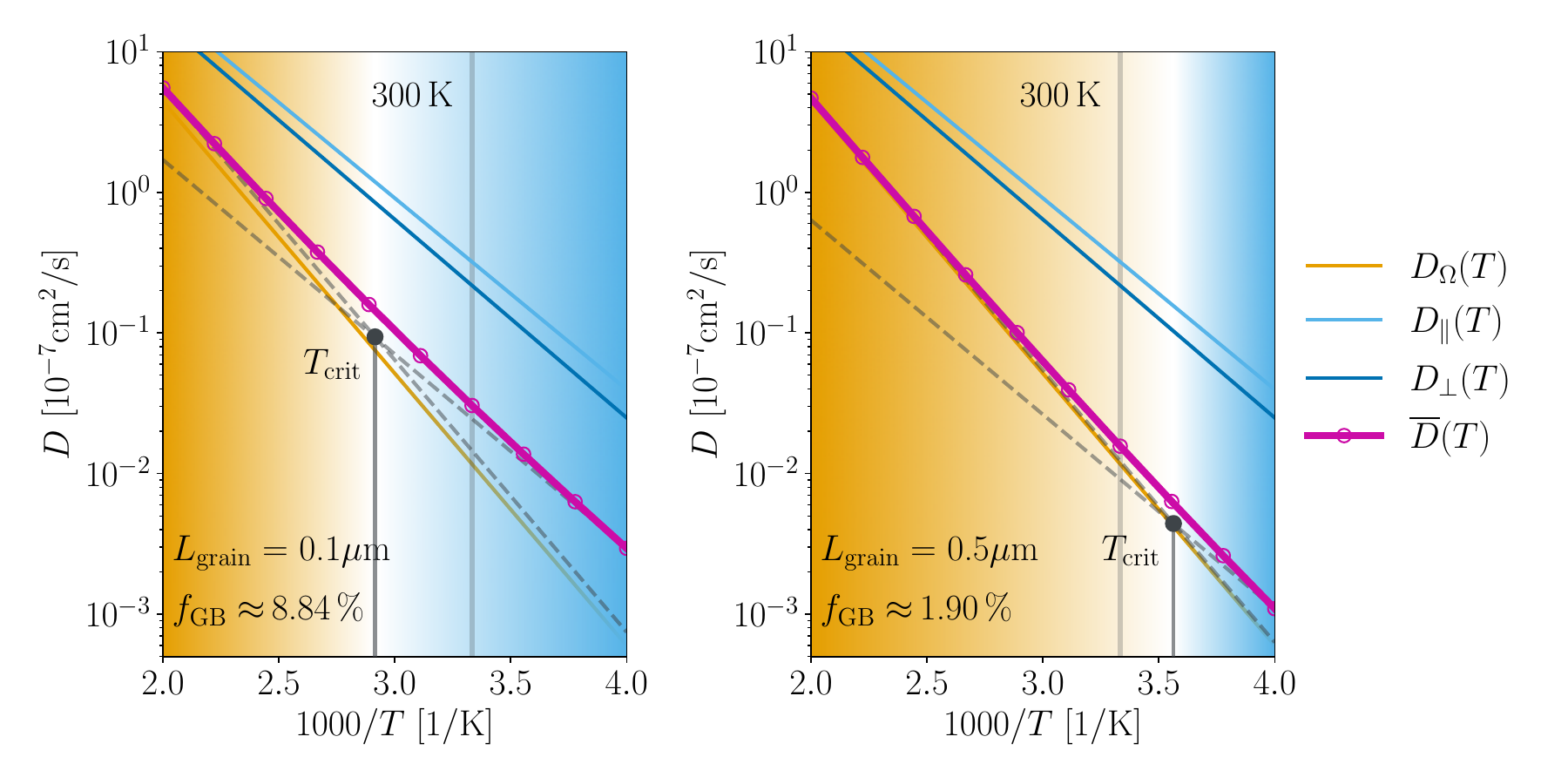}
    \caption{Temperature dependence of the largest eigenvalue of the effective diffusion tensor for the solid-state electrolyte \Argyro{} with a polycrystalline microstructure. Two exemplar average grain sizes $L_{\rm grain}$ with resulting volume fractions $f_{\rm GB}$ are shown. The structure-specific parameters $D_\Omega$, $D_\Vert$ and $D_\perp$, which follow the Arrhenius relation from~\cref{eq:arrhenius}, are also shown. To underline the non-Arrhenius response of the polycrystal, the critical temperature $T_{\rm crit}$~\cite{Heo2021} (see \Cref{app:realistic_setup}) is constructed (dashed lines). {\color{rev}It splits the temperature domain into a bulk-dominated regime (yellow) and a GB-dominated regime (blue).} For reference, room temperature ($T=300$ K) is marked by the gray vertical line.}
    \label{fig:T_Sigma5}
\end{figure}

The effective diffusion coefficient and the non-Arrhenius behavior for the considered polycrystalline geometry as a function of temperature is shown in \Cref{fig:T_Sigma5}. Two representative average grain sizes $L_{\rm grain}$ have been chosen to show the size effect. Note that both choices satisfy $f_{\rm GB}<10\%$, i.e., the assumption of a clear length separation between grain size and GB thickness is reasonable. For large temperatures, the effective response is strongly bulk-dominated. For lower temperatures, the contrast between the bulk and GB diffusivities becomes more significant, leading to a larger GB impact on the effective diffusivities. To underline the transition from the bulk- to the GB-dominated regime, we construct the critical temperature $T_{\rm crit}$ following Ref.~\cite{Heo2021} (see \Cref{app:realistic_setup}), and highlight it in \Cref{fig:T_Sigma5}. Results show that the critical temperature $T_{\rm crit}$ and, therefore, the regimes of bulk- and GB-dominated response are clearly dependent on the GB volume fraction. At higher GB volume fractions, the diffusivity of polycrystalline \Argyro{} is more strongly influenced by GB effects. 

{\color{rev}
In experiments, the exact grain size and grain morphology of \Argyro{} remain hard to control and determine~\cite{Wang2024}, and experimentally observed diffusion behaviors are subject to many sources of uncertainty. A direct comparison of simulation results and experimental observations for \Argyro{} is challenging. The affine parametric formulation of the model and mesh-independent scaling of grain size and GB thickness in the collapsed representation allow for computationally efficient sensitivity studies. This facilitates a comprehensive understanding of diffusion behaviors for \Argyro{} and contributes to its inverse optimization.} Further, the assumption of identical properties in all GB segments does not hold for \Argyro{}~\cite{Ou2024}. Similar effects have been partially studied in Ref.~\cite{Peng2024} by splitting the set of GB segments into subsets with uniform properties, which can lead to percolation paths. {\color{rev}Such a reparametrization can be implemented as a straightforward extension of the presented model to simulate \Argyro{} with a more realistic polycrystalline microstructure. 
}


{\color{rev}
\section{Summary and outlook}
\label{Sec:conclusion}
We have proposed and validated a novel collapsed-interface model for simulating transversely isotropic atomic diffusion properties of grain boundaries (GBs). Finite element simulations based on this model provide a more accurate and realistic, yet computationally efficient alternative to the overly simplified averaging approaches typically used to upscale atomistic diffusion data for predicting the effective diffusivity of polycrystalline materials, e.g., strict upper and lower bounds \cite{Reuss1929,Wiener1912,Voigt1889} being overly broad.

Through theoretical analysis of a simple channel flow problem and validation using a two-dimensional simulation, we have shown the necessity of employing modes that are even, i.e., at least a second-order approximation when modeling the concentration field inside GBs. The resulting quadratic prismatic diffusion element captures concentration offsets, jumps, and channeling effects. Furthermore, it can be collapsed along its thickness direction, yielding a zero-thickness interface element. These elements have three nodes seemingly at the same location: two linked to the different solid neighbors and one virtual node at half the thickness of the GB. Thereby, an accurate representation of GB effects is enabled without requiring a full 3D meshing of the thin interface layers, ruling out mesh related issues and simplifying the discretization process considerably. Furthermore, our approach facilitates the study of both grain size and interface thickness effects on the same mesh, offering significant advantages in both preprocessing cost as well as regarding computational efficiency. The latter is realized by exploiting the affine parameter dependence of the problem. Dimensional analysis further helps to drastically reduce the number of required simulations by moving from an initial five-dimensional parameter space (three diffusivities and two length parameters) to two intrinsic parameters called $\Pi_\Vert, \Pi_\perp$. This implies that each discrete solution solves a full 3D space of parameter variations, reducing the number of simulations to a manageable amount. This reduction could also be further exploited by a surrogate model as done for the effective response of simpler GB models (e.g., Ref.~\cite{Peng2024}) to replace overly simplified averages (e.g., Refs.~\cite{Yoon2023,Dawson2018}).

With a postprocessing step, the full GB domain and concentration profiles can be reconstructed for in-depth analysis by leveraging the analytical description across the GB thickness. This approach overcomes the limitations of previous finite element simulations (e.g., Refs.~\cite{Peng2024, Han2013}), where only the bulk domain is considered, leaving the diffusion mechanism within GBs unexplored. As a result, a more detailed understanding of local transport mechanisms and the corresponding overarching parametric regimes becomes accessible.
}

A significant impact of the GB properties on the overall diffusion behavior as well as on the effective properties has been found which is reflected in four characteristic parameter regimes: Starting from the case of idealized interfaces (\textit{neutral} regime), the overall transport is severely limited if the GB diffusivity in the normal direction is reduced (\textit{blocking} regime), effectively reducing flow across GBs. Opposed to that, the diffusion along the GB plays an enhancing role with respect to the overall transport by allowing for additional diffusion paths along the GBs (\textit{enhancing} regime). In some cases, a pronounced diffusion channeling effect was observed (\textit{connecting} regime). Accordingly, the effective diffusion coefficients are increased or reduced depending on the GB properties. All of the described effects are amplified for larger GB volume fractions, i.e., thicker GB regions. Interestingly, the normal and the tangential parts of the stiffness matrix and of the right-hand side vector scale differently with the GB width, leading to nontrivial changes in the related flow field.

While the presented model focuses on Fickian diffusion, it can be readily extended to incorporate electric driving forces without substantially modifying its formulation. Further, the presented work provides a foundation for extensions toward simulations of more realistic polycrystalline microstructures. This includes:
\begin{itemize}
    \item \textbf{Interplay of different GB types}: The parametrization of the model within this work can be extended to different GB types and anisotropic bulk behavior without additional effort or significant adaptations of the model. This enables the upscaling of realistic scenarios from atomistic to larger continuum scales, given the availability of information from atomistic simulations.
    \item \textbf{Sensitivity to geometrical features}: The present study has focused on the impact of material properties as well as the artificially introduced GB volume fraction. In reality, local geometry arrangement introduces additional variability of the material response \cite{Peng2024,Fritzen2011a}, be it through grain shape modifications or by studying multi-modal grain size distributions.
    \item \textbf{Extension of the range of admissible volume fractions}: The collapsed interface model relies on the assumption of the GB thickness being \textit{significantly smaller} than the grain size. For regimes of larger volume fractions, alternative approaches with potentially fully resolved GBs need to be considered, which could then be used to investigate the transition region.
\end{itemize}

{\color{rev}
By investigating the diffusivity of polycrystalline \Argyro{}, we have demonstrated that anisotropic GB diffusivities, different from the bulk, can notably affect effective diffusivities at larger scales. From an engineering perspective, the proposed model, on the one hand, provides accurate predictions in a forward (upscale) manner; on the other hand, it offers a foundation for solving inverse problems, such as estimating microstructural compositions, which are otherwise difficult to determine precisely from limited experimental data, as in the case for the argyrodite \Argyro{} that might contribute to improving all-solid-state batteries.

In general for polycrystalline materials, the present findings delineate the conditions under which GBs must be explicitly modeled in continuum simulations and when their influence is sufficiently minor to be neglected. Beyond GBs, other interface diffusion phenomena, such as transport along metal/ceramic interfaces, exhibit GB-like behavior (see, e.g., Ref.~\cite{Kumar2018}). Owing to its general formulation, the proposed model is extendable to a broad range of interface-driven transport applications.
}


\section*{CRediT authorship contribution statement}
\textbf{L. Scholz}: Conceptualization, Methodology, Software, Visualization, Writing - Original draft
\textbf{Y. Ou}: Conceptualization, Writing - Review and editing
\textbf{B. Grabowski}: Funding acquisition, Conceptualization, Writing - Review and editing
\textbf{F. Fritzen}: Funding acquisition, Conceptualization, Methodology, Writing - Original draft, Supervision

\section*{Declaration of competing interest}
The authors declare that they have no conflict of interest.

\section*{Acknowledgement}
This project is funded by the Deutsche Forschungsgemeinschaft (DFG, German Research Foundation) - EXC 2075 - 390740016 under Germany’s Excellence Strategy (L. Scholz and Y. Ou); FR 2702/10 - 517847245 (F. Fritzen). B. Grabowski is funded by the European Research Council (ERC) under the European Union’s Horizon 2020 Research and Innovation Program (Grant Agreement No. 865855). The authors acknowledge the support by the Stuttgart Center for Simulation Science (SimTech). 

\section*{Data availability}
The simulation results for the homogenized responses across the parameter space will be published on DaRUS, the data repository of the University of Stuttgart, upon acceptance.

\section*{Declaration of generative AI and AI-assisted technologies in the writing process}
During the preparation of this work the authors used ChatGPT (OpenAI o3 model) in order to refine the clarity of the manuscript. After using it, the authors reviewed and edited the content as needed and take full responsibility for the content of the published article.

\clearpage


\begin{appendix}
\counterwithin*{equation}{section}
\renewcommand\theequation{\thesection\arabic{equation}}

\section*{Appendix}
\section{FE discretization}
\subsection{3D bulk domain}
\label{app:T10ansatz}
\begin{figure}[h]
    \centering
    \begin{subfigure}[b]{0.25\textwidth}
        \centering
        \includegraphics[width=\textwidth]{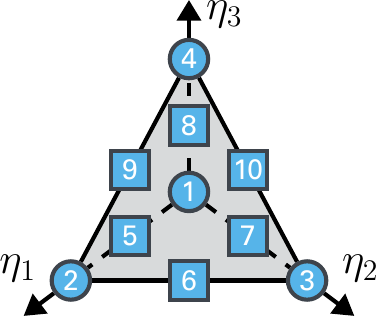}
        \caption{ten node tetrahedron}
        \label{fig:element_tet10}
    \end{subfigure}
    \hspace{5cm}
    \begin{subfigure}[b]{0.25\textwidth}
        \centering
        \includegraphics[width=\textwidth]{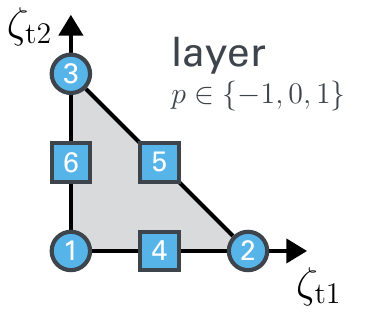}
        \caption{six node triangle layer}
        \label{fig:element_tet6s}
    \end{subfigure}
    \caption{\protect Reference elements for the FE discretization.}
\end{figure}

Within the bulk we use classical Lagrange elements with quadratic shape functions, that is, a ten-node tetrahedron as depicted in \cref{fig:element_tet10}. In the reference coordinates $\ul{\eta}=[\eta_1, \eta_2, \eta_3]$, the shape functions are
\begin{align}
    \ul{N}_\Omega(\ul{\eta}) = 
    \begin{bmatrix}
        (2(1-\eta_1-\eta_2-\eta_3)-1)(1-\eta_1-\eta_2-\eta_3) \\
        (2\eta_1-1)\eta_1 \\
        (2\eta_2-1)\eta_2 \\
        (2\eta_3-1)\eta_3 \\
        4(1-\eta_1-\eta_2-\eta_3)\eta_1 \\
        4\eta_1\eta_2 \\
        4(1-\eta_1-\eta_2-\eta_3)\eta_2 \\
        4(1-\eta_1-\eta_2-\eta_3)\eta_3 \\
        4\eta_1\eta_3 \\
        4\eta_2\eta_3
    \end{bmatrix}^\mathsf{T} \in\ffR^{1\times 10}.
\end{align}
The corresponding gradient with respect to the reference coordinates is
\begin{align}
    \ull{B}_\Omega^{\rm ref} (\ul{\eta}) = \begin{bmatrix}
        4(\eta_1+\eta_2+\eta_3) - 3 & 4(\eta_1+\eta_2+\eta_3)-3 & 4(\eta_1+\eta_2+\eta_3)-3 \\
        4\eta_1-1 & 0 & 0 \\
        0 & 4\eta_2-1 & 0 \\
        0 & 0 & 4\eta_3-1 \\
        -4(2\eta_1+\eta_2+\eta_3-1) & -4\eta_2 & -4\eta_3 \\
        4\eta_2 & 4\eta_1 & 0 \\
        -4\eta_1 & -4(2\eta_1+\eta_2+\eta_3-1) & -4\eta_3 \\
        -4\eta_1 & -4\eta_2 & -4(2\eta_1+\eta_2+\eta_3-1) \\
        4\eta_3 & 0 & 4\eta_1 \\
        0 & 4\eta_3 & 4\eta_2
    \end{bmatrix}^\mathsf{T} \in \ffR^{3\times 10}.
\end{align}

\subsection{2D GB interface domain}
\label{app:T6ansatz}
The proposed interface elements are described within the reference coordinates $\ul{\zeta} = [\zeta_{\rm t1}, \zeta_{\rm t2}, \zeta_{\rm n}]$ where $\ul{\zeta}_{\rm t}=[\zeta_{\rm t1}, \zeta_{\rm t2}]$ form the interface plane and $\zeta_n$ represents the interface normal of the prism element (see \Cref{Sec:fem:interface}). On the interface, only the $\ul{\zeta}_{\rm t}$ coordinates of the reference element are used to define the 6-node triangular element, as shown in \cref{fig:element_tet6s}. The corresponding shape functions are defined as:
\begin{align}
    \ul{N}_{\rm P2} (\ul{\zeta}_{\rm t}) =
    \begin{bmatrix}
        (1-\zeta_{\rm t1} - \zeta_{\rm t2})(1-2\zeta_{\rm t1}-2\zeta_{\rm t2}) \\
        \zeta_{\rm t1}(2\zeta_{\rm t1}-1) \\
        \zeta_{\rm t2}(2\zeta_{\rm t2}-1) \\
        4\zeta_{\rm t1}(1-\zeta_{\rm t1} - \zeta_{\rm t2}) \\
        4\zeta_{\rm t1}\zeta_{\rm t2} \\
        4\zeta_{\rm t2}(1-\zeta_{\rm t1}-\zeta_{\rm t2})
    \end{bmatrix}^\mathsf{T} \in \ffR^{1 \times 6}.
\end{align}
The gradient with respect to the reference coordinates is then
\begin{align}
    \ull{B}_{\rm P2}^{\rm ref}(\ul{\zeta}_{\rm t}) = 
    \begin{bmatrix}
        -3+4(\zeta_{\rm t1}+\zeta_{\rm t1}) & -3+4(\zeta_{\rm t1} + \zeta_{\rm t2}) \\
        -1+4\zeta_{\rm t1} & 0 \\
        0 & -1+4\zeta_{\rm t2}\\
        4(1-2\zeta_{\rm t1}-\zeta_{\rm t2}) & -4\zeta_{\rm t1} \\
        4\zeta_{\rm t2} & 4\zeta_{\rm t1} \\
        - 4\zeta_{\rm t2} & 4(1-\zeta_{\rm t1}-2\zeta_{\rm t2})
    \end{bmatrix}^\mathsf{T} \in \ffR^{2 \times 6}.
\end{align}
Note that this gradient is defined on a 2D submanifold and needs to be transformed to the 3D space as described in \Cref{Sec:fem:interface}.

\section{Affine parameter dependence}
\label{app:affine}
In the following a more detailed derivation of the affine dependencies for the parameter set given in \cref{eq:5d_param} is provided with respect to the reference quantities $D_0$ for the diffusion coefficients and $l_0$ for the geometry scaling. The superscript $0$ is used to indicate that the respective quantity depends on the reference quantities $D_0$ and $l_0$ alone and not on the characteristic values. 

\subsection{Bulk domain}
\label{app:bulk}
For the volume elements in the bulk domain, as introduced in \Cref{Sec:fem:bulk}, the integration weights $\omega_j$ and the gradient operator $\ull{B}_\Omega$ for a given element exhibit the following dependence:
\begin{align}
    \omega_j = \pi_l^3 \omega_j^0(l_0), && 
    \ull{B}_\Omega = \lb(\ull{B}_\Omega^{\rm ref} \pi_l\ull{X}_\Omega^0(l_0)\rb^{-1} \ull{B}_\Omega^{\rm ref} =\frac{1}{\pi_l} \ull{B}_\Omega^0(l_0).
\end{align}
The element stiffness matrix $\Kebulk$ (cf. \cref{eq:Kebulk}) exhibits a dependence on $\pi_\Omega$ and $\pi_l$
\begin{align}
    \Kebulk = \pi_\Omega D_0 \sum\limits_{j=1}^{n_{\rm GP}^\Omega} \pi_l^3\omega_j^0(l_0) \frac{1}{\pi_l}\lb\ull{B}_\Omega^0(\ul{\eta}_j,l_0)\rb^\mathsf{T} \frac{1}{\pi_l}\ull{B}_\Omega^0(\ul{\eta}_j,l_0) = \pi_\Omega\pi_l \KebulkO(D_0, l_0)
\end{align}
where $\KebulkO$ is independent of the characteristic values.

\subsection{GB domain}
The surface area of the individual interfaces scales according to
\begin{align}
    A_i = \pi_l^2 A_i^0(l_0).
\end{align}
The in-plane FE discretization using second-order triangles yields a dependence on $\pi_l$ for the integration weights, shape functions and their gradient operator:
\begin{align} 
    \nu_j &= \pi_l^2\nu_j^0(l_0), && 
    \ul{N}_{\rm P2} = \ul{N}_{\rm P2}^0, &&
    \ull{B}_{\rm P2} = \frac{1}{\pi_l}\ull{B}_{\rm P2}^0(l_0).
\end{align}
A dependence on the characteristic quantity accounting for the GB thickness is introduced via the analytical integration of the GB thickness for the stiffness matrices (cf. \cref{eq:Kepara} and \cref{eq:Keperp}):
\begin{align}
    \Kepara 
    &= \frac{\pi_\Vert D_0 \pi_h l_0}{15}
    \begin{bmatrix*}[r] 
        4[\ast] & 2[\ast] & -[\ast] \\ 
        2[\ast] & 16[\ast] & 2[\ast] \\ 
        -[\ast] & 2[\ast] & 4[\ast] 
    \end{bmatrix*} 
    = \pi_\Vert \pi_h \KeparaO(D_0,l_0) \\
    \Keperp
    &= \frac{\pi_\perp D_0}{6\pi_h l_0}
    \begin{bmatrix*}[r]
        7[\ast\ast] & -8[\ast\ast] & [\ast\ast] \\   
        -8[\ast\ast] & 16[\ast\ast] & -8[\ast\ast] \\ 
        [\ast\ast] & -8[\ast\ast] & 7[\ast\ast]
    \end{bmatrix*} 
    = \frac{\pi_\perp\pi_l^2}{\pi_h} \KeperpO(D_0, l_0)
\end{align}
with 
\begin{align}
    [\ast] &= \sum\limits_{j=1}^{n_{\rm GP}^{C18}}\pi_l^2\nu_j^0(l_0) \frac{1}{\pi_l} \lb\ull{B}_{\rm P2}^0(\ul{\zeta}_{\rm{t}(j)},l_0)\rb^\mathsf{T} \frac{1}{\pi_l}\ull{B}_{\rm P2}^0(\ul{\zeta}_{\rm{t}(j)}, l_0), \\
    [\ast\ast] &= \sum\limits_{j=1}^{n_{\rm GP}^{C18}}\pi_l^2\nu_j^0(l_0) \lb\ul{N}_{\rm P2}^0(\ul{\zeta}_{\rm{t}(j)})\rb^\mathsf{T}\ul{N}_{P2}^0(\ul{\zeta}_{\rm{t}(j)}).
\end{align}
This yields the global system as stated in \cref{eq:affine_system} which depends on the two intrinsic parameters $\Pi_\Vert$ and $\Pi_\perp$.

\subsection{Effective properties and analytical bounds}
\label{app:affine:eff}
The operators for integrating the gradient field over the respective domains $\mathbb{G}_{[\cdot]}$ (cf. \cref{eq:G_bulk}-\eqref{eq:G_perp}) follow a scaling with $\pi_h$ and $\pi_l$:
\begin{align}
    \Gbulk &= \sum\limits_{i=1}^{n_{\rm el}^\Omega} \sum\limits_{j=1}^{n_{\rm GP}^{\rm T10}} \pi_l^3\omega_j^0(l_0) \frac{1}{\pi_l}\ull{B}^0_\Omega(\ul{\eta}_j, l_0) \ull{L}_\Omega = \pi_l^2\GbulkO(l_0), \\
    \begin{split}
        \Gpara &= \frac{2\pi_h l_0}{6} \sum\limits_{i=1}^{n_{\rm el}^{\rm GB}} \sum\limits_{j=1}^{n_{\rm GP}^{\rm C18}} \pi_l^2\nu_j^0(l_0) \frac{1}{\pi_l}
        \begin{bmatrix}
            \ull{B}_{\rm P2}^0(\ul{\zeta}_{\rm{t}(j)}, l_0) & 4\ull{B}^0_{\rm P2}(\ul{\zeta}_{\rm{t}(j)}, l_0) & \ull{B}^0_{\rm P2}(\ul{\zeta}_{\rm{t}(j)},l_0)
        \end{bmatrix} \ull{L}_{\rm GB} \\
        &= \pi_h\pi_l\GparaO(l_0), 
    \end{split}\\
    \Gperp &= \sum\limits_{i=1}^{n_{\rm el}^{\rm GB}} \sum\limits_{j=1}^{n_{\rm GP}^{\rm C18}} \pi_l^2\nu_j^0(l_0)\ul{n}_i^0 \begin{bmatrix}
        -\ul{N}_{\rm P2}^0(\ul{\zeta}_{\rm{t}(j)}) & \ul{0} & \ul{N}_{\rm P2}^0(\ul{\zeta}_{\rm{t}(j)})
    \end{bmatrix} \ull{L}_{\rm GB} = \pi_l^2\GperpO.
\end{align}
This yields the following dependencies for the phase-wise gradient averages (cf. \cref{eq:int_grad}):
\begin{align}
    \langle \ul{g}_k \rangle_\Omega &= \frac{1}{\pi_l} \ol{\ul{g}}_k + \frac{1}{\pi_l v_\Omega^0(l_0)}\GbulkO \ulWT{c}_k, \\
    \langle \ul{g}_k \rangle_\Vert &= \frac{1}{\pi_l} (\ull{I} - \ull{S}_2)\ol{\ul{g}}_k + \frac{1}{\pi_l 2l_0A^0} \GparaO\ulWT{c}_k, &&
    \langle \ul{g}_k \rangle_\perp = \frac{1}{\pi_h 2l_0 A^0} \GperpO \ulWT{c}_k.
\end{align}

The volumes of bulk $v_\Omega$ and GB domain $v_{\rm GB}$ scale with
\begin{align}
    v_\Omega = \pi_l^3 v_\Omega^0(l_0), && v_{\rm GB} = \pi_h\pi_l^2 v_{\rm GB}^0(l_0)
\end{align}
yielding the volume fractions (\cref{eq:vol_fracs}) to be given as
\begin{align}
    \vfbulk = \frac{v_\Omega^0(l_0)}{v_\Omega^0(l_0) + \frac{\pi_h}{\pi_l}v_{\rm GB}^0(l_0)}, && 
    \vfgb = \frac{v_{\rm GB}^0(l_0)}{\frac{\pi_l}{\pi_h}v_\Omega^0(l_0) + v_{\rm GB}^0(l_0)}.
\end{align}

Since 5D parameter configurations $[\pi_\Omega, \pi_\Vert, \pi_\perp, \pi_h, \pi_l]$ mapping to one 2D configuration $[\Pi_\Vert, \Pi_\perp]$ yield identical fluctuation fields due to the lower intrinsic dimensionality, the parameter-free operators $\mathbb{G}_{[\cdot]}^0$ are identical as well and only need to be computed once for the entire set. However, the effective diffusion tensor (cf. \cref{eq:Deff}, that depends on the volume fractions (cf. \cref{eq:vol_fracs}), the volume-averaged gradients and the individual diffusion coefficients via the fluxes (cf. \cref{eq:Qk}), exhibits a dependence on the absolute values of the individual characteristic quantities. Hence, related 5D configurations yield the same fluctuation field but differing effective diffusion tensors. 

The same holds for the analytical Voigt (\cref{eq:voigt}) and Reuss bounds (\cref{eq:reuss}):
\begin{align}
    \ull{D}_{\rm V} &= D_0 \lb \vfbulk(\pi_h, \pi_l) \pi_\Omega \ull{I} + \vfgb(\pi_h, \pi_l) \pi_\Vert\lb \frac{\pi_\perp - \pi_\Vert}{\pi_\Vert} \ull{S}_2^0 + \ull{I} \rb \rb, \\
    \ull{D}_{\rm R} &= D_0 \lb \frac{\vfbulk(\pi_h, \pi_l)}{\pi_\Omega} \ull{I} + \frac{\vfgb(\pi_h, \pi_l)}{\pi_\Vert}\lb \frac{\pi_\Vert - \pi_\perp}{\pi_\perp} \ull{S}_2^0 +  \ull{I} \rb \rb^{-1}.
\end{align}
Note that the interface normals $\ul{n}_i$ as well as the structural tensor $\ull{S}_2=\ull{S}_2^0$ (cf. \cref{eq:S2}) are independent of the uniform scaling.

\section{Realistic setup for \texorpdfstring{Li\textsubscript{6}PS\textsubscript{5}Cl}{Li6PS5Cl}}
\label{app:realistic_setup}
\subsection{Parametrization}
Ou et al.~\cite{Ou2024} obtained the fitted parameters $D_{0}$ and the activation energies $E_{{\rm a}}$ for bulk and different GB types from Arrhenius fits. Together with the GB thickness, the values are given in \Cref{tab:argyro_param} for reference.

\begin{table}[h]
    \centering
    \begin{tabular}{l||c|c|c}
        & $D_0$ [$\text{10}^{-7}\text{cm}^2\text{/s}$] & $E_{\rm a}$ [meV] & $\delta$ [\AA] \\ \hline\hline
        anion-ordered bulk & $3.3 \times 10^4$ & 384 & - \\ \hline
        GB $\Sigma5(001)[001]$ ($\Vert$) & $1.1 \times 10^4$ & 270 & \multirow{2}{*}{26.5} \\ \cline{1-3}
        GB $\Sigma5(001)[001]$ ($\perp$) & $1.1 \times 10^4$ & 280 & \\ \hline
    \end{tabular}
    \caption{Overview of parameters for solid-state electrolyte {\Argyro}~\cite{Ou2024}. Diffusion parameters were obtained by fitting an Arrhenius relation to molecular dynamics simulation data over a temperature range of 500 to 800 K. }
    \label{tab:argyro_param}
\end{table}

For a given temperature $T$ the diffusion coefficients $D_\Omega$, $D_\Vert$ and $D_\perp$ (and their dedimensionalized equivalents $\pi_\Omega$, $\pi_\Vert$, $\pi_\perp$) can be recovered from these values via the Arrhenius relation in \cref{eq:arrhenius}.

In the proposed model, the GB volume fraction is represented implicitly by scaling the GB thickness and geometric sample size via the dimensionless parameters $\pi_h$ and $\pi_l$. Estimates for the GB thickness $\delta=2h$ were determined in Ref.~\cite{Ou2024} (cf. \Cref{tab:argyro_param}), and therefore, $\pi_h$ can be determined. For $\pi_l$, a more application-based perspective is used: Given the average grain volume $v_{\rm grain}=v_\Omega/N$ we define the average grain size $L_{\rm grain}$ as the diameter of the sphere of volume $v_{\rm grain}$\footnote{Sine $L_{\rm grain}$ is derived via the sphere volume the constant $\pi$ appears in \cref{eq:l_grain}. This should not be confused with the dimensionless scaling factors $\pi_\Omega$, $\pi_\Vert$, $\pi_\perp$, $\pi_h$ and $\pi_l$.},
{\color{rev}
\begin{align}
    L_{\rm grain} = 2 \lb\frac{v_{\rm grain}}{\frac{4}{3}\pi}\rb^{1/3}.
    \label{eq:l_grain}
\end{align}
The bulk volume $v_\Omega$ (and therefore the average grain volume $v_{\rm grain}$) directly scales with $\pi_l^3$, i.e. by scaling the overall size of the considered geometry with side length $l_0$ (see \Cref{Sec:affine} and \Cref{app:bulk}). Hence, if an average grain size $L_{\rm grain}$ is given, \cref{eq:l_grain} directly yields the corresponding scaling factor $\pi_l$ which is the last of the five input parameters in \cref{eq:5d_param}. The other four parameters are directly given as mentioned above.}

\subsection{Non-Arrhenius behavior and critical temperature}
To formalize the description of the non-Arrhenius behavior of the polycrystalline microstructure, we make use of the critical temperature $T_{\rm crit}$ defined by Heo et al.\footnote{Heo~et~al. used $T_{\rm c}$ when referring to the critical temperature. To avoid ambiguity with the concentration $c$, we use $T_{\rm crit}$ in this study.} \cite{Heo2021}. It denotes the transition temperature at which the effective diffusion coefficient switches from a bulk- to a GB-dominated behavior. For the construction of $T_{\rm crit}$, the intersection point of the tangents to the $\overline{D}$-curve in the Arrhenius plot is determined. Note that in the limit case of extreme temperatures, the tangents correspond to the shifted lines of $D_\Omega$ and $D_\Vert$, or $D_\perp$. For simplicity and in consistency with Heo.~et~al., we use the average of the GB diffusion coefficients $D_{\rm GB}(T) = (D_\Vert(T)+D_\perp(T))/2$ for the construction. The construction is visualized in \Cref{fig:T_Sigma5} by dashed lines.
\end{appendix}

\clearpage


\end{document}